\documentclass[journal]{IEEEtran}
\usepackage{amssymb}
\usepackage{algorithm}
\usepackage{algorithmic}
\usepackage{amsfonts}
\usepackage{amsmath}
\usepackage{amssymb}
\usepackage{amsthm}
\usepackage{array}
\usepackage{bbding}
\usepackage{bigints}
\usepackage{booktabs}
\usepackage{cite}
\usepackage{cleveref}
\usepackage{color}
\usepackage{diagbox}
\usepackage{epsfig,latexsym}
\usepackage{epstopdf}
\usepackage{graphicx}
\usepackage{fancyhdr}
\usepackage{float}
\usepackage{flushend}
\usepackage{indentfirst}
\usepackage{lastpage}
\usepackage{makecell}
\usepackage{mathtools}
\usepackage{multirow}
\usepackage{pifont}
\usepackage{psfrag}
\usepackage{setspace}
\usepackage{stfloats}
\usepackage{subfloat}
\usepackage{times}
\usepackage{subfig}
\newcommand{\cmark}{\ding{52}}

\theoremstyle{remark}
\newtheorem{theorem}{\quad \textbf{Theorem}}
\newtheorem{lemma}{\quad \textbf{Lemma}}
\newtheorem{Proposition}{\quad \textbf{Proposition}}

\allowdisplaybreaks[4]

\begin{document}
\title{Ergodic Spectral Efficiency Analysis of Intelligent Omni-Surface Aided Systems Suffering From Imperfect CSI and Hardware Impairments}

\author{Qingchao Li, \textit{Graduate Student Member, IEEE}, Mohammed El-Hajjar, \textit{Senior Member, IEEE},\\ and Lajos Hanzo, \textit{Life Fellow, IEEE}

\thanks{L. Hanzo would like to acknowledge the financial support of the Engineering and Physical Sciences Research Council projects EP/W016605/1, EP/X01228X/1, EP/Y026721/1 and EP/W032635/1 as well as of the European Research Council's Advanced Fellow Grant QuantCom (Grant No. 789028). \textit{(Corresponding author: Lajos Hanzo.)}

Qingchao Li, Mohammed El-Hajjar and Lajos Hanzo are with the Electronics and Computer Science, University of Southampton, Southampton SO17 1BJ, U.K. (e-mail: qingchao.li@soton.ac.uk; meh@ecs.soton.ac.uk; lh@ecs.soton.ac.uk).}}

\maketitle

\begin{abstract}
In contrast to the conventional reconfigurable intelligent surfaces (RIS), intelligent omni-surfaces (IOS) are capable of full-space coverage of smart radio environments by simultaneously transmitting and reflecting the incident signals. In this paper, we investigate the ergodic spectral efficiency of IOS-aided systems for transmission over random channel links, while considering both realistic imperfect channel state information (CSI) and transceiver hardware impairments (HWIs). Firstly, we formulate the linear minimum mean square error estimator of the equivalent channel spanning from the user equipments (UEs) to the access point (AP), where the transceiver HWIs are also considered. Then, we apply a two-timescale protocol for designing the beamformer of the IOS-aided system. Specifically, for the active AP beamformer, the minimum mean square error combining method is employed, which relies on the estimated equivalent channels, on the statistical information of the channel estimation error, on the inter-user interference as well as on the HWIs at the AP and UEs. By contrast, the passive IOS beamformer is designed based on the statistical CSI for maximizing the upper bound of the ergodic spectral efficiency. The theoretical analysis and simulation results show that the transceiver HWIs have a significant effect on the ergodic spectral efficiency, especially in the high transmit power region. Furthermore, we show that the HWIs at the AP can be effectively compensated by deploying more AP antennas.
\end{abstract}

\section{Introduction}
\IEEEPARstart{D}{riven} by the ever-increasing demand for capacity in the mobile networks, the massive multiple-input and multiple-output (MIMO) technique~\cite{li2021analog} has been employed in the fifth generation (5G) wireless systems for improving the spectral efficiency. However, the high energy consumption and hardware cost constitute major bottlenecks for the implementation of massive MIMO systems. As a potent technique, the reconfigurable intelligent surface (RIS) concept has been researched as a promising cost-efficient technique for wireless communications~\cite{basar2021present,pan2021reconfigurable,li2022reconfigurable_iot,pan2022overview,
li2023reconfigurable_tvt}.

The RIS is an artificial metamaterial surface comprised of numerous low-cost passive reflecting elements deployed at an elevated position between the access point (AP) and the user equipment (UE)~\cite{wu2019towards,gong2020toward,wu2021intelligent,bjornson2022reconfigurable,zheng2022survey}. By electronically tuning the phase shift of each RIS element, the signal propagation environment can be beneficially configured to realize increased information transmission reliability, especially when the direct AP-UE links are blocked~\cite{li2022reconfigurable_tvt,li2023achievable_tcom}.

\subsection{Prior Works}
Various aspects of the RIS have been investigated, such as its channel modeling~\cite{jiang2023reconfigurable}, channel estimation~\cite{zhang2023channel,li2023low}, resource allocation~\cite{xie2023performance}, performance analysis~\cite{li2023performance_tvt,yang2020outage}, and covert communications capacity~\cite{yang2023covert,chen2023active}. More particularly, by jointly designing the active beamformer of the AP and the passive beamformer constituted by the RIS, the transmission quality can be improved. In~\cite{wu2019intelligent}, Wu and Zhang investigated the problem of minimizing the transmission power by alternatively optimizing the transmit precoder (TPC) of the AP and the RIS phase shift matrix. Specifically, when the active AP TPC metrix is fixed, the RIS phase shift matrix is optimized by the semidefinite relaxation (SDR) algorithm. By contrast, when the RIS phase shift matrix is fixed, the active AP TPC matrix is designed based on the minimum mean squared error (MMSE) criterion. In~\cite{li2020weighted}, Li \textit{et al.} researched the problem of deploying multiple RISs for improving the weighted sum-rate of multiple users. The AP TPC and the RIS phase shift were alternately optimized, where the Lagrangian method and the Riemannian manifold conjugate gradient (RMCG) method were employed. The simulation results showed that the performance of multi-RIS aided wireless communication systems is better than that of single-RIS aided systems. In contrast to the RIS phase shift optimization based on instantaneous channel state information (CSI), Han \textit{et al.}~\cite{wu2019intelligent,li2020weighted} focused their attention on designing the phase shift matrix based on the statistics of the angle of departure (AoD) and angle of arrival (AoA), which substantially reduced the channel acquisition overhead~\cite{han2019large}. Specifically, the maximum-ratio transmission (MRT) criterion was employed for the design of the active beamformer at the AP, based on which a tight upper bound of the ergodic spectral efficiency was derived. Then, the RIS phase shift can be optimized for maximizing the ergodic spectral efficiency based on the statistical CSI.

\subsection{Motivations and Contributions}
Most of the existing treatises focus on the RIS family only capable of reflecting the impinging signals, but not transmitting them. This means that the RIS can work only when the AP and the UEs are at the same side of the RIS, which can only realize $180^\circ$ half-space coverage of the smart radio environment. To deal with this issue, the intelligent omni-surface (IOS), also called a simultaneously transmitting and reflecting (STAR) RIS, has been proposed to realize $360^\circ$ full coverage~\cite{xu2022simultaneously,zhang2022intelligent}. Explicitly, in the IOS-aided systems, the incident signals are not only reflected but also refracted by the metasurfaces for full-space coverage. In~\cite{mu2021simultaneously}, Mu \textit{et al.} proposed three practical operating protocols for the IOS, namely energy splitting (ES), time switching (TS) and mode stitching (MS). More specifically, in the ES protocol, the energy of signals incident on each IOS element is spitted into two parts, where one of them is configured for reflecting signals to support the receivers at the same side of the transmitter. The other one is configured for transmitting signals to support the receivers at the opposite side of the transmitter. By contrast, in the TS protocol, orthogonal time slots are employed for each IOS element switching between the reflecting mode and the transmitting mode. Finally, in the MS protocol the IOS elements are divided into two partitions, where the elements in one of the partitions is permanently configured to support the receivers at the same side of the transmitter, while the elements of the other partition are configured to support the receivers at the different side of the transmitter. The numerical results show that in terms of minimizing the power consumption, the ES protocol and the TS protocol are preferable for the multicast and unicast methods, respectively. Similar conclusion was also verified by Niu \textit{et al.} in~\cite{niu2021weighted}, where the block coordinate descent (BCD) algorithm was employed for jointly designing the AP and IOS beamformer to maximize the weighted sum-rate. In~\cite{zhang2020beyond}, Zhang \textit{et al.} solved a specific spectral efficiency maximization problem for the IOS-aided single-input and single-output (SISO) systems, where the closed-form expression and the branch-and-bound algorithm were employed for optimizing the continuous and discrete phase shift of the IOS elements. It was shown that the IOS-aided systems achieve considerably higher spectral efficiency than the state-of-the-art RIS-aided systems and the conventional cellular systems operating without IOS. In~\cite{zhang2021intelligent}, the above system model was extended to the case of the AP associated with multi-antennas supporting multiple users, where the active AP beamformer and the passive IOS beamformer were optimized in an iterative manner for maximizing the sum-rate. In~\cite{dhok2022rate}, Dhok \textit{et al.} considered the IOS-aided systems communicating over spatially-correlated Rician channels, where both the outage probability and channel capacity were derived for the infinite block-length transmissions. Furthermore, the block error rate, the system throughput and the channel capacity were also formulated for realistic finite block-length transmissions. In~\cite{wu2021coverage}, the sum coverage range maximization problem is formulated for IOS-aided systems, where the coefficients at the IOS elements and the resource allocation at the AP are jointly optimized for both a non-orthogonal multiple access (NOMA) scheme and an orthogonal multiple access (OMA) scheme. It was verified by the simulation results that the IOS promises wider coverage range than conventional RIS. In~\cite{wu2022resource,zhao2022ergodic}, an IOS was deployed in NOMA systems. It was shown that IOS-aided NOMA systems outperform the state-of-art RIS-aided systems both in terms of their sum-rate.

The IOS passive beamformer optimization reported in the above contributions are all based on the instantaneous CSI. To cut down the channel acquisition overhead, as well as the IOS phase shift update frequency, the authors of~\cite{papazafeiropoulos2022coverage,wu2023two} designed the IOS phase shift based on the statistical CSI, which changes slowly and may remain constant over numerous coherence time intervals. In~\cite{papazafeiropoulos2022coverage}, Papazafeiropoulos \textit{et al.} employed the projected gradient ascent method for finding the optimal IOS phase shift based on the statistical CSI for maximizing the coverage probability. In~\cite{wu2023two}, Wu \textit{et al.} employed the two-timescale protocol for IOS-aided systems. Specifically, the passive IOS beamformer is designed based on the statistical CSI, while the active AP beamformer relies on the instantaneous CSI of the equivalent AP-UE channel. This reduces the channel estimation overhead, at the expense of a passive beamforming gain erosion, especially in rich-scattering channel environments.

However, the beamformer design of the IOS-aided systems in the above treatises has the following limitations. Firstly, perfect CSI knowledge is assumed for both the beamformer at the AP and for the IOS elements, which is impractical for finite-length pilot sequences. Furthermore, the transceiver hardware is assumed to be ideal, where the signals can be transmitted and received without distortion, which is unrealistic, especially when a large number of antennas are deployed at the AP. More specifically, in the high transmit power region, the dominant factors determining the system performance are the channel estimation error and hardware impairment (HWI). Hence, beamformers must be designed for IOS-aided systems by considering both the hardware impairment, as well as imperfect CSI acquisition. To deal with the above issues, Table~\ref{Table_literature} explicitly contrasts our contributions to the literature.

\begin{table*}
\footnotesize
\begin{center}
\caption{Contrasting our contributions to the literature~\cite{mu2021simultaneously,niu2021weighted,zhang2020beyond,zhang2021intelligent,dhok2022rate,
wu2021coverage,wu2022resource,zhao2022ergodic,papazafeiropoulos2022coverage,wu2023two}.}
\label{Table_literature}
\begin{tabular}{*{12}{l}}
\toprule
     & \makecell[c]{Our paper} &~\cite{mu2021simultaneously} &~\cite{niu2021weighted} &~\cite{zhang2020beyond} &~\cite{zhang2021intelligent} &~\cite{dhok2022rate} &~\cite{wu2021coverage} &~\cite{wu2022resource} &~\cite{zhao2022ergodic} &~\cite{papazafeiropoulos2022coverage} &~\cite{wu2023two}\\
\midrule
\midrule
    IOS & \makecell[c]{\cmark} & \makecell[c]{\cmark} & \makecell[c]{\cmark} & \makecell[c]{\cmark} & \makecell[c]{\cmark} & \makecell[c]{\cmark} & \makecell[c]{\cmark} & \makecell[c]{\cmark} & \makecell[c]{\cmark} & \makecell[c]{\cmark} & \makecell[c]{\cmark} \\
\midrule
    Multiple AP antennas & \makecell[c]{\cmark} & \makecell[c]{\cmark} & \makecell[c]{\cmark} & & \makecell[c]{\cmark} &  &  & &  & \makecell[c]{\cmark} &  \\
\midrule
    IOS beamforming based on statistical CSI & \makecell[c]{\cmark} &  &  &  &  &  & &  &  & \makecell[c]{\cmark} & \makecell[c]{\cmark}  \\
\midrule
    Channel estimation & \makecell[c]{\cmark} &   &   &   &  &   &   &   &  &   &\\
\midrule
    Imperfect CSI & \makecell[c]{\cmark} &  &  &  &  &  &  & &  &  & \\
\midrule
    Transceiver hardware impairments & \makecell[c]{\cmark} &  &  &  &  &  &  &  &  &  &\\
\bottomrule
\end{tabular}
\end{center}
\end{table*}

\begin{itemize}
  \item Firstly, we formulate the linear minimum mean square error (LMMSE) estimator of the equivalent UE-AP channel for IOS-aided wireless systems, while considering the hardware impairments at both the UEs and the AP. Furthermore, the mean square error (MSE) of the LMMSE estimator is derived theoretically.
  \item Then, based on the estimated CSI of the equivalent channel, we employ a two-timescale protocol for jointly optimizing the active beamformer at the AP and the passive beamformer at the IOS for maximizing the ergodic spectral efficiency for transmission over random channel links. Specifically, the active beamformer at the AP is designed by the MMSE combination method based on the estimated instantaneous equivalent channel as well on the statistical channel estimation error covariance, the inter-user interference and the transceiver HWIs. Then, the ergodic spectral efficiency is formulated theoretically. Finally, the optimal closed-form IOS phase shift is derived for maximizing the upper bound of the ergodic spectral efficiency, based on the statistical CSI.
  \item The theoretical analysis and the numerical results show that the transceiver HWIs constrain the ergodic spectral efficiency improvement in the high transmit power region. We also show that employing more AP antennas is capable of compensating the HWI at the AP. By contrast, the HWI at the UEs cannot be compensated by harnessing more AP antennas.
\end{itemize}

\subsection{Organization and Notation}
The rest of this paper is organized as follows. In Section~\ref{System_Model}, we present the system model, while channel estimation is described in Section~\ref{Channel_Estimation}. Section~\ref{Beamforming_Design} presents our active AP beamformer design and the passive IOS phase shift optimization. Our simulation results are presented in Section~\ref{Numerical_and_Simulation_Results}, while we conclude in Section~\ref{Conclusion}.

\textit{Notations:} $\jmath=\sqrt{-1}$, vectors and matrices are denoted by boldface lower and upper case letters, respectively, $\left(\cdot\right)^{\dag}$ represents the conjugate operation, $\mathbb{C}^{m\times n}$ denotes the space of $m\times n$ complex-valued matrices, $a_n$ represents the $n$th element in vector $\mathbf{a}$, $\mathbf{0}_{N}$ is the $N\times1$ zero vector, $\mathbf{I}_{N}$ and $\mathbf{O}_{N_1,N_2}$ represents the $N\times N$ identity matrix and $N_1\times N_2$ zero matrix, respectively, $\mathbf{Diag}\left\{a_1,a_2,\cdots,a_N\right\}$ denotes a diagonal matrix having the elements of $a_1,a_2,\cdots,a_N$ in order, $\mathcal{CN}\left(\boldsymbol{\mu},\mathbf{\Sigma}\right)$ is a circularly symmetric complex Gaussian random vector with the mean $\boldsymbol{\mu}$ and the covariance matrix $\mathbf{\Sigma}$, $\mathbf{A}\succ\mathbf{0}$ indicates that $\mathbf{A}$ is a positive definite matrix, $\mathbf{A}\succeq\mathbf{0}$ represents $\mathbf{A}$ is a positive semi-definite matrix, $\mathbb{E}\left[\mathbf{x}\right]$ represents the mean of the random vector $\mathbf{x}$, the auto-covariance matrix of the random vector $\mathbf{x}$ is denoted as $\mathbf{C}_{\mathbf{x}\mathbf{x}}=\mathbb{E}\left[\left(\mathbf{x}-\mathbb{E}[\mathbf{x}]\right)
\left(\mathbf{x}-\mathbb{E}[\mathbf{x}]\right)^{\mathrm{H}}\right]$, the cross-covariance matrix between the random vectors $\mathbf{x}$ and $\mathbf{y}$ is expressed as $\mathbf{C}_{\mathbf{x}\mathbf{y}}=
\mathbb{E}\left[\left(\mathbf{x}-\mathbb{E}[\mathbf{x}]\right)
\left(\mathbf{y}-\mathbb{E}[\mathbf{y}]\right)^{\mathrm{H}}\right]$, the auto-correlation matrix of the random vector $\mathbf{x}$ is denoted as $\mathbb{E}\left[\mathbf{x}\mathbf{x}^{\mathrm{H}}\right]$, the cross-correlation matrix between the random vectors $\mathbf{x}$ and $\mathbf{y}$ is given by $\mathbb{E}\left[\mathbf{x}\mathbf{y}^{\mathrm{H}}\right]$, $[\mathbf{a}]_n$ represents the $n$th element in the vector $\mathbf{a}$ and $[\mathbf{A}]_{n_1,n_2}$ represents the $\left(n_1,n_2\right)$th element in matrix $\mathbf{A}$.

\section{System Model}\label{System_Model}
\begin{figure}[!t]
    \centering
    \includegraphics[width=2.8in]{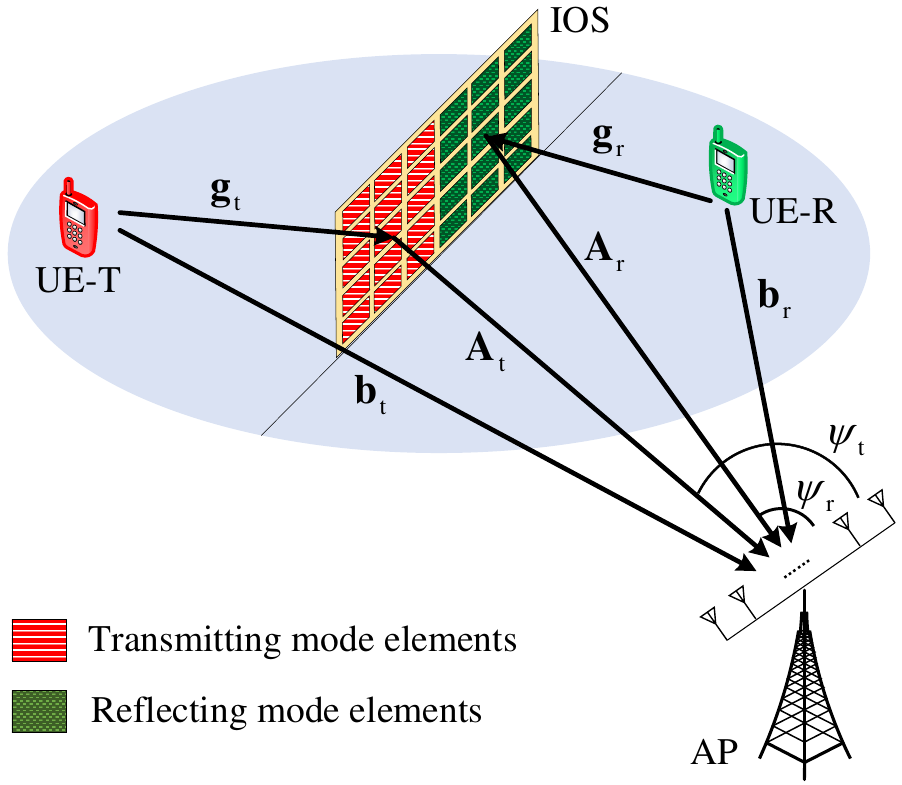}
    \caption{System model of the considered IOS-aided wireless communication system, including an AP equipped with $M$ ULA antennas, an IOS employing the mode stitching protocol with a total of $N$ elements, a user located at the same side of the IOS, denoted as UE-R, and a user located at the other side of the IOS, denoted as UE-T.}\label{Fig_System_model_IOS}
\end{figure}

The IOS-aided system model is shown in Fig.~\ref{Fig_System_model_IOS}, including an AP equipped with $M$ uniform linear array (ULA) antennas having an element spacing of $d_0$, an IOS with $N$ elements, and randomly positioned users. As shown in Fig.~\ref{Fig_System_model_IOS}, we denote the user located at the same side of the IOS by UE-R, and that at the other side of the IOS by UE-T\footnote{In this paper, we aim to characterize the fundamental ergodic spectral efficiency limits of an IOS aided dual-user communication network. However, it can be extended to a multi-user system by partitioning the IOS into multiple subsurfaces~\cite{aldababsa2023simultaneous}. Specifically, each subsurface is allocated to serve a specific user. Note that in this IOS partitioning method, we need to consider not only the interference between users but also the interference between different subsurfaces. This is a challenge set aside for our future work.}.

\subsection{IOS Architecture}
In the ES protocol, additional circuits are required to split the signal energy and some power leakage is inevitable. By contrast, in the TS protocol frequent switching is required between the reflecting and transmitting operation on each element, resulting in high hardware complexity. Fortunately, the MS protocol is easy to implement~\cite{liu2021star}. Thus, in this paper we focus on the IOS relying on the MS protocol. The part of the IOS configured for reflecting signals is a uniform rectangular planar array (URPA) containing $N_\mathrm{r}=N_\mathrm{r}^x\times N_\mathrm{r}^y$ elements, and that used for transmitting signals is a URPA containing $N_\mathrm{t}=N_\mathrm{t}^x\times N_\mathrm{t}^y$ elements, where $N_\mathrm{r}^x$ and $N_\mathrm{r}^y$ represent the numbers of reflecting elements in the horizontal and vertical direction respectively, while $N_\mathrm{t}^x$ and $N_\mathrm{t}^y$ represent the numbers of transmitting elements in the horizontal and vertical direction respectively. Furthermore, the spacing between adjacent IOS elements in the horizontal and vertical directions are represented by $\delta_x$ and $\delta_y$, respectively. Additionally, $\mathbf{\Theta}_\mathrm{r}$ and $\mathbf{\Theta}_\mathrm{t}$ denote the phase shift matrices of the reflective mode elements and transmit mode elements, given by
\begin{align}\label{IOS_Architecture_1}
    \mathbf{\Theta}_\mathrm{i}=\mathrm{diag}\left\{\mathrm{e}^{\jmath\theta_{\mathrm{i},1}},
    \mathrm{e}^{\jmath\theta_{\mathrm{i},2}},
    \cdots,\mathrm{e}^{\jmath\theta_{\mathrm{i},N_\mathrm{i}}}\right\},\quad \mathrm{i}\in\left\{\mathrm{r},\mathrm{t}\right\},
\end{align}
where $\theta_{\mathrm{r},n}$ represents the phase shift of the $n$th reflective mode element and $\theta_{\mathrm{t},n}$ is the phase shift of the $n$th transmit mode element.

\subsection{Channel Model}
Here we consider an uplink scenario, but emphasize that our performance analysis method proposed in this paper is also applicable to a downlink scenario. As shown in Fig.~\ref{Fig_System_model_IOS}, we denote the channel spanning from the UE-R to the AP as $\mathbf{b}_\mathrm{r}\in\mathbb{C}^{M\times1}$, from the UE-T to the AP as $\mathbf{b}_\mathrm{t}\in\mathbb{C}^{M\times1}$, from the UE-R to the reflective mode IOS elements as $\mathbf{g}_\mathrm{r}\in\mathbb{C}^{N_\mathrm{r}\times1}$, from the UE-T to the transmit mode IOS elements as $\mathbf{g}_\mathrm{t}\in\mathbb{C}^{N_\mathrm{t}\times1}$, and from the IOS to the AP as $\mathbf{A}=\left[\mathbf{A}_\mathrm{r},\mathbf{A}_\mathrm{t}\right]\in\mathbb{C}^{M\times N}$, where $\mathbf{A}_\mathrm{r}\in\mathbb{C}^{M\times N_{\mathrm{r}}}$ and $\mathbf{A}_\mathrm{t}^{M\times N_{\mathrm{t}}}$ represent the channel spanning from the reflective mode IOS elements and transmit mode IOS elements to the AP, respectively. Furthermore, $\varrho_{\mathbf{b}_\text{r}}=\text{C}_0d_{\mathbf{b}_\text{r}}^{-\alpha_{\mathbf{b}_\text{r}}}$, $\varrho_{\mathbf{b}_\text{t}}=\text{C}_0d_{\mathbf{b}_\text{t}}^{-\alpha_{\mathbf{b}_\text{t}}}$,
$\varrho_{\mathbf{g}_\text{r}}=\text{C}_0d_{\mathbf{g}_\text{r}}^{-\alpha_{\mathbf{g}_\text{r}}}$, $\varrho_{\mathbf{g}_\text{t}}=\text{C}_0d_{\mathbf{g}_\text{t}}^{-\alpha_{\mathbf{g}_\text{t}}}$,
$\varrho_{\mathbf{A}_\text{r}}=\text{C}_0d_{\mathbf{A}_\text{r}}^{-\alpha_{\mathbf{A}_\text{r}}}$
and $\varrho_{\mathbf{A}_\text{t}}=\text{C}_0d_{\mathbf{A}_\text{t}}^{-\alpha_{\mathbf{A}_\text{t}}}$ represent the path loss in the corresponding links, where $\text{C}_0$ is the path loss at the reference distance of 1 meter, $d_{\mathbf{b}_\text{r}}$, $d_{\mathbf{b}_\text{t}}$, $d_{\mathbf{g}_\text{r}}$, $d_{\mathbf{g}_\text{t}}$, $d_{\mathbf{A}_\text{r}}$ and $d_{\mathbf{A}_\text{t}}$ denote the distance in the corresponding links, and $\alpha_{\mathbf{b}_\text{r}}$, $\alpha_{\mathbf{b}_\text{t}}$, $\alpha_{\mathbf{g}_\text{r}}$, $\alpha_{\mathbf{g}_\text{t}}$, $\alpha_{\mathbf{A}_\text{r}}$ and $\alpha_{\mathbf{A}_\text{t}}$ represent the path loss exponent in the corresponding links.

In this paper, we consider the far-field channel model. However, our performance analysis approach is also valid, when the users are located in the near-field region of the IOS. Since the IOS is harnessed for creating additional paths for signal transmission among the AP and users, it is reasonable to assume that $\mathbf{g}_\mathrm{r}$, $\mathbf{g}_\mathrm{t}$, $\mathbf{A}_\mathrm{r}$ and $\mathbf{A}_\mathrm{t}$ experience Rician fading~\cite{yang2022performance}, while $\mathbf{b}_\mathrm{r}$ and $\mathbf{b}_\mathrm{t}$ face Rayleigh fading~\cite{an2022joint}. Thus, in the AP-UE links, we have $\mathbf{b}_\mathrm{r}\sim\mathcal{CN}\left(\mathbf{0}_M,\mathbf{I}_M\right)$ and $\mathbf{b}_\mathrm{t}\sim\mathcal{CN}\left(\mathbf{0}_M,\mathbf{I}_M\right)$. In the IOS-UE links, $\mathbf{g}_\mathrm{r}$ and $\mathbf{g}_\mathrm{t}$ are
\begin{align}\label{Channel_Model_2}
    \mathbf{g}_\mathrm{i}=\sqrt{\frac{\kappa_{\mathbf{g}_\mathrm{i}}}
    {1+\kappa_{\mathbf{g}_\mathrm{i}}}}
    \overline{\mathbf{g}}_\mathrm{i}
    +\sqrt{\frac{1}{1+\kappa_{\mathbf{g}_\mathrm{i}}}}\widetilde{\mathbf{g}}_\mathrm{i},\quad
    \mathrm{i}\in\left\{\mathrm{r},\mathrm{t}\right\},
\end{align}
where $\kappa_{\mathbf{g}_\mathrm{r}}$ and $\kappa_{\mathbf{g}_\mathrm{t}}$ are the Rician factor of $\mathbf{g}_\mathrm{r}$ and $\mathbf{g}_\mathrm{t}$ respectively. The NLoS component obeys $\widetilde{\mathbf{g}}_\mathrm{r}\sim\mathcal{CN}\left(\mathbf{0}_M,\mathbf{I}_M\right)$ and $\widetilde{\mathbf{g}}_\mathrm{t}\sim\mathcal{CN}\left(\mathbf{0}_M,\mathbf{I}_M\right)$. The LoS components $\overline{\mathbf{g}}_\mathrm{r}$ and $\overline{\mathbf{g}}_\mathrm{t}$ are formulated as~\cite{zhang2021reconfigurable}
\begin{align}\label{Channel_Model_3}
    \notag\overline{\mathbf{g}}_\mathrm{i}=&\left[1,\cdots,\mathrm{e}^{-\jmath\frac{2\pi}{\lambda}
    (\delta_{x}n_{x}\sin\phi_\mathrm{i}\cos\varphi_\mathrm{i}+\delta_{y}n_{y}\cos\phi_\mathrm{i})},
    \cdots,\right.\\
    \notag&\left.\mathrm{e}^{-\jmath\frac{2\pi}{\lambda}(\delta_{x}(N_\mathrm{i}^x-1)
    \sin\phi_\mathrm{i}\cos\varphi_\mathrm{i}
    +\delta_{y}(N_\mathrm{i}^y-1)\cos\phi_\mathrm{i})}\right]^{\mathrm{H}},\\
    &\quad
    \mathrm{i}\in\left\{\mathrm{r},\mathrm{t}\right\},
\end{align}
where $\lambda$ is the wavelength, $\phi_\mathrm{r}$ and $\varphi_\mathrm{r}$ are the elevation and azimuth AoA with respect to the reflective IOS elements respectively, while $\phi_\mathrm{t}$ and $\varphi_\mathrm{t}$ are the elevation and azimuth AoA with respect to the transmit IOS elements respectively. In the AP-IOS link, we have
\begin{align}\label{Channel_Model_5}
    \mathbf{A}_\mathrm{i}=\sqrt{\frac{\kappa_{\mathbf{A}_\mathrm{i}}}
    {1+\kappa_{\mathbf{A}_\mathrm{i}}}}
    \overline{\mathbf{A}}_\mathrm{i}+\sqrt{\frac{1}{1+\kappa_{\mathbf{A}_\mathrm{i}}}}
    \widetilde{\mathbf{A}}_\mathrm{i},\ \mathrm{i}\in\left\{\mathrm{r},\mathrm{t}\right\},
\end{align}
where $\kappa_{\mathbf{A}_\mathrm{i}}$ is the Rician factor of $\mathbf{A}_\mathrm{i}$. The NLoS component $\widetilde{\mathbf{A}}_\mathrm{i}=\left[\widetilde{\mathbf{a}}_{\mathrm{i},1}^{\mathrm{H}},
\widetilde{\mathbf{a}}_{\mathrm{i},2}^{\mathrm{H}},\cdots,
\widetilde{\mathbf{a}}_{\mathrm{i},M}^{\mathrm{H}}\right]^{\mathrm{H}}$, associated with $\widetilde{\mathbf{a}}_{\mathrm{i},m}\sim\mathcal{CN}\left(\mathbf{0}_N,\mathbf{I}_N\right)$ corresponds to the $m$th AP antenna. The LoS component $\overline{\mathbf{A}}_\mathrm{i}$ obeys:
\begin{align}\label{Channel_Model_6}
    \overline{\mathbf{A}}_\mathrm{i}=\overline{\mathbf{a}}^{(\mathrm{AP})}_\mathrm{i}
    \overline{\mathbf{a}}_\mathrm{i}^{(\mathrm{IOS})\mathrm{H}}.
\end{align}
In (\ref{Channel_Model_6}), $\overline{\mathbf{a}}^{(\mathrm{IOS})}_\mathrm{i}$ is given by~\cite{zhang2021reconfigurable}
\begin{align}\label{Channel_Model_7}
    \notag\overline{\mathbf{a}}_\mathrm{i}^{(\mathrm{IOS})}
    =&\left[1,\cdots,\mathrm{e}^{-\jmath\frac{2\pi}{\lambda}
    (\delta_{x}n_{x}\sin\omega_\mathrm{i}\cos\varpi_\mathrm{i}
    +\delta_{y}n_{y}\cos\omega_\mathrm{i})},\cdots,\right.\\
    &\left.\mathrm{e}^{-\jmath\frac{2\pi}{\lambda}(\delta_{x}(N_\mathrm{i}^x-1)
    \sin\omega_\mathrm{i}\cos\varpi_\mathrm{i}
    +\delta_{y}(N_\mathrm{i}^y-1)\cos\omega_\mathrm{i})}\right]^{\mathrm{H}},
\end{align}
where $\omega_\mathrm{i}$ and $\varpi_\mathrm{i}$ are the elevation and the azimuth AoD with respect to the IOS elements, respectively. Furthermore, in (\ref{Channel_Model_6}), $\overline{\mathbf{a}}
^{(\mathrm{AP})}_\mathrm{i}=\left[1,\mathrm{e}^{-\jmath\frac{2\pi}{\lambda}d_0\cos\psi_\mathrm{i}},
\cdots,\mathrm{e}^{-\jmath\frac{2\pi}{\lambda}d_0(M-1)\cos\psi_\mathrm{i}}\right]^{\mathrm{H}}$~\cite{zhang2021reconfigurable}, where $\psi_\mathrm{i}$ is the AoA of the signals spanning from the IOS elements to the AP antennas.

Therefore, the equivalent channel emerging from the UE-R to the AP is given by $\mathbf{h}_\mathrm{r}=\overline{\mathbf{h}}_\mathrm{r}+\widetilde{\mathbf{h}}_\mathrm{r}$, where $\overline{\mathbf{h}}_\mathrm{r}$ is the known component and $\widetilde{\mathbf{h}}_\mathrm{r}$ is the uncertainty. Similarly, the equivalent channel arriving from the UE-T to the AP is given by $\mathbf{h}_\mathrm{t}=\overline{\mathbf{h}}_\mathrm{t}+\widetilde{\mathbf{h}}_\mathrm{t}$, where $\overline{\mathbf{h}}_\mathrm{t}$ is the known component and $\widetilde{\mathbf{h}}_\mathrm{t}$ is the uncertainty. Hence, we have
\begin{align}\label{Channel_Model_11}
    \mathbf{h}_\mathrm{i}=\sqrt{\varrho_{\mathbf{A}_\text{i}}\varrho_{\mathbf{g}_\text{i}}}
    \mathbf{A}_\mathrm{i}\mathbf{\Theta}_\mathrm{i}\mathbf{g}_\mathrm{i}
    +\sqrt{\varrho_{\mathbf{b}_\text{i}}}\mathbf{b}_\mathrm{i},
\end{align}
\begin{align}\label{Channel_Model_12}
    \overline{\mathbf{h}}_\mathrm{i}=\sqrt{\varrho_{\mathbf{A}_\text{i}}
    \varrho_{\mathbf{g}_\text{i}}}
    \sqrt{\frac{\kappa_{{\mathbf{A}}_\mathrm{i}}\kappa_{{\mathbf{g}}_\mathrm{i}}}
    {(1+\kappa_{{\mathbf{A}}_\mathrm{i}})(1+\kappa_{{\mathbf{g}}_\mathrm{i}})}}
    \overline{\mathbf{A}}_\mathrm{i}\mathbf{\Theta}_\mathrm{i}\overline{\mathbf{g}}_\mathrm{i},
\end{align}
\begin{align}\label{Channel_Model_13}
    \notag\widetilde{\mathbf{h}}_\mathrm{i}=&\sqrt{\frac{\varrho_{\mathbf{A}_\text{i}}
    \varrho_{\mathbf{g}_\text{i}}}
    {(1+\kappa_{{\mathbf{A}}_\mathrm{i}})(1+\kappa_{{\mathbf{g}}_\mathrm{i}})}}
    \left(\sqrt{\kappa_{{\mathbf{A}}_\mathrm{i}}}
    \overline{\mathbf{A}}_\mathrm{i}\mathbf{\Theta}_\mathrm{i}\widetilde{\mathbf{g}}_\mathrm{i}+
    \sqrt{\kappa_{{\mathbf{g}}_\mathrm{i}}}
    \widetilde{\mathbf{A}}_\mathrm{i}\mathbf{\Theta}_\mathrm{i}\overline{\mathbf{g}}_\mathrm{i}\right.\\
    &\left.+\widetilde{\mathbf{A}}_\mathrm{i}\mathbf{\Theta}_\mathrm{i}\widetilde{\mathbf{g}}_\mathrm{i}
    \right)+\sqrt{\varrho_{\mathbf{b}_\text{i}}}\mathbf{b}_\mathrm{i},
\end{align}
where $\mathrm{i}\in\left\{\mathrm{r},\mathrm{t}\right\}$. Upon exploiting (\ref{Channel_Model_2}) and (\ref{Channel_Model_5}), we can formulate the mean and auto-covariance matrix of $\mathbf{h}_\mathrm{r}$ and $\mathbf{h}_\mathrm{t}$ as
\begin{align}\label{Channel_Model_14}
    \mathbb{E}[\mathbf{h}_\mathrm{i}]=\overline{\mathbf{h}}_\mathrm{i}
    =\sqrt{\varrho_{\mathbf{A}_\text{i}}\varrho_{\mathbf{g}_\text{i}}}
    \sqrt{\frac{\kappa_{{\mathbf{A}}_\mathrm{i}}\kappa_{{\mathbf{g}}_\mathrm{i}}}
    {(1+\kappa_{{\mathbf{A}}_\mathrm{i}})(1+\kappa_{{\mathbf{g}}_\mathrm{i}})}}
    \overline{\mathbf{A}}_\mathrm{i}\mathbf{\Theta}_\mathrm{i}\overline{\mathbf{g}}_\mathrm{i},
\end{align}
\begin{align}\label{Channel_Model_15}
    \notag\mathbf{C}_{\mathbf{h}_\mathrm{i}\mathbf{h}_\mathrm{i}}
    =\mathbf{C}_{\widetilde{\mathbf{h}}_\mathrm{i}\widetilde{\mathbf{h}}_\mathrm{i}}
    =&\varrho_{\mathbf{A}_\mathrm{i}}\varrho_{\mathbf{g}_\mathrm{i}}N_\mathrm{i}
    \frac{(1+\kappa_{{\mathbf{g}}_\mathrm{i}})\mathbf{I}_M+
    \kappa_{{\mathbf{A}}_\mathrm{i}}\overline{\mathbf{a}}^{(\mathrm{AP})}_\mathrm{i}
    \overline{\mathbf{a}}^{(\mathrm{AP})\mathrm{H}}_\mathrm{i}}
    {(1+\kappa_{{\mathbf{A}}_\mathrm{i}})(1+\kappa_{{\mathbf{g}}_\mathrm{i}})}\\
    &+\varrho_{\mathbf{b}_\mathrm{i}}\mathbf{I}_M,
\end{align}
where $\mathrm{i}\in\left\{\mathrm{r},\mathrm{t}\right\}$.

\subsection{Channel Estimation and Beamforming Protocol}
\begin{figure}[!t]
    \centering
    \includegraphics[width=3.4in]{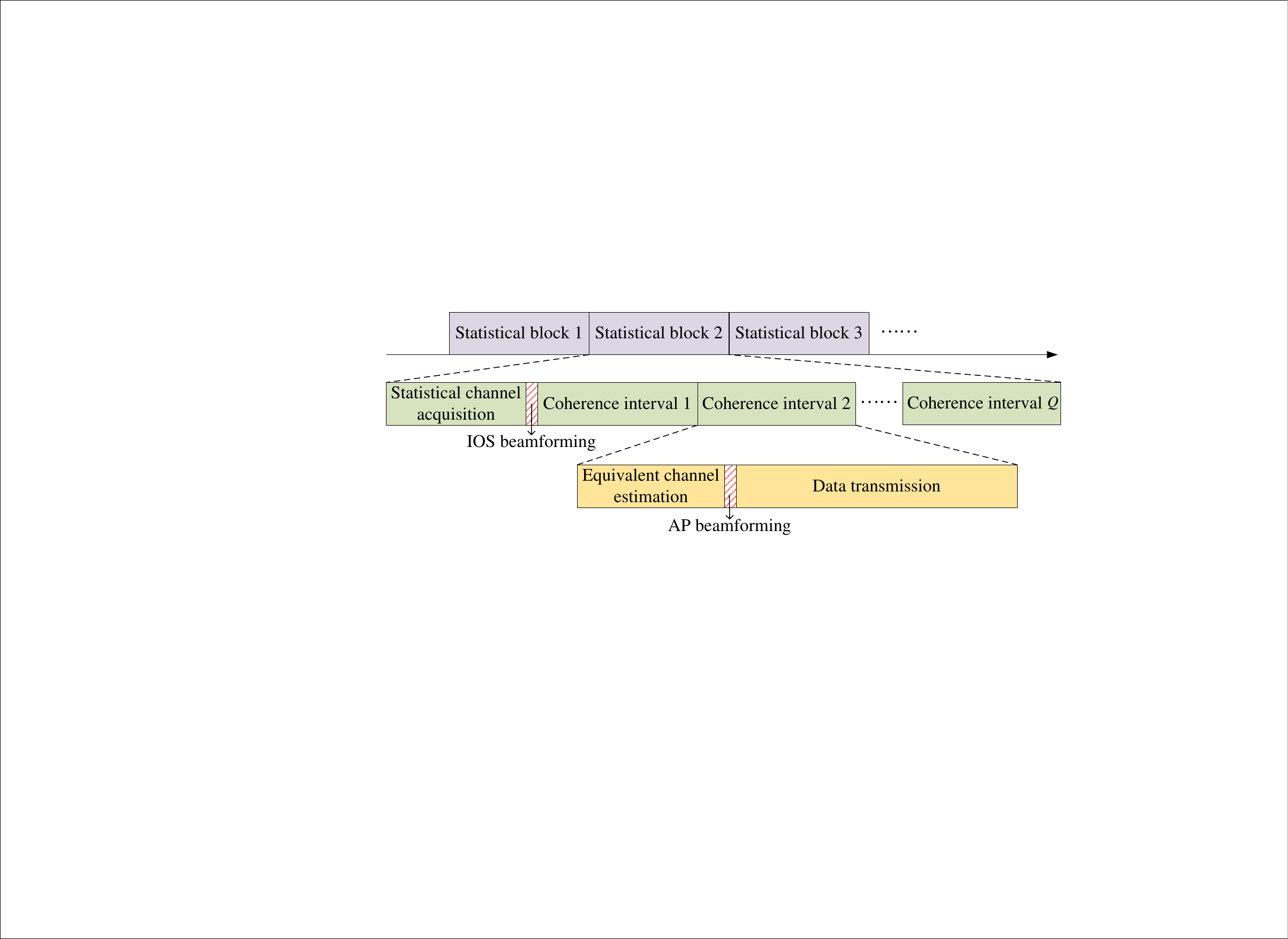}
    \caption{Illustration of the employed two-timescale channel estimation and beamforming protocol.}\label{Fig_diagram_two_timescale}
\end{figure}

The two-timescale channel estimation and beamforming protocol conceived is illustrated in Fig.~\ref{Fig_diagram_two_timescale}. In each statistical block, the statistical CSI, e.g. $\overline{\mathbf{g}}_\mathrm{r}$, $\overline{\mathbf{g}}_\mathrm{t}$, $\overline{\mathbf{a}}^{(\mathrm{AP})}_\mathrm{r}$, $\overline{\mathbf{a}}^{(\mathrm{AP})}_\mathrm{t}$,
$\overline{\mathbf{a}}_\mathrm{r}^{(\mathrm{IOS})\mathrm{H}}$ and $\overline{\mathbf{a}}_\mathrm{t}^{(\mathrm{IOS})\mathrm{H}}$, is fixed since they change slowly~\cite{pan2022overview}. The statistical CSI is estimated at the beginning of each statistical block, and the remaining part of each statistical block includes $Q$ coherence intervals. Since the statistical CSI changes slowly, the pilot overhead of the statistical CSI acquisition can be omitted~\cite{pan2022overview}. Each coherence interval is composed of several symbol slots, which are assumed to have the same instantaneous CSI. The IOS beamforming is designed based on the statistical CSI. Thus, the IOS phase shift is fixed within each statistical block, and the equivalent UE-AP channel is estimated at the beginning of each coherence interval. Then, the BS beamformer is designed based on the instantaneous equivalent UE-AP channel, followed by data transmission.

In the following, we formulate the LMMSE estimation of the equivalent channels $\mathbf{h}_\mathrm{r}$ and $\mathbf{h}_\mathrm{t}$ defined in (\ref{Channel_Model_11}), considering the signal distortions resulting from HWIs at both the APs and UEs.

\section{Channel Estimation}\label{Channel_Estimation}
In this section, we formulate the LMMSE estimation of the equivalent channel. For making efficient use of the large number of AP antennas, an uplink channel estimation scenario is considered. To eliminate the pilot contamination between the UE-R and UE-T, a pair of orthogonal pilot sequences are employed, denoted as $\left[\tau_{\mathrm{r}}^{(1)};\tau_{\mathrm{r}}^{(2)};\cdots;\tau_{\mathrm{r}}^{(K)}\right]$ and $\left[\tau_{\mathrm{t}}^{(1)};\tau_{\mathrm{t}}^{(2)};\cdots;\tau_{\mathrm{t}}^{(K)}\right]$, respectively. In previous IOS papers, the transceiver hardware is assumed to be ideal~\cite{mu2021simultaneously,
niu2021weighted,zhang2020beyond,zhang2021intelligent,dhok2022rate,wu2021coverage,wu2022resource,
zhao2022ergodic,papazafeiropoulos2022coverage,wu2023two}. However, practical circuits of the transceivers generally suffer from hardware impairments, including power amplifier non-linearities, amplitude/phase imbalance in the In-phase/Quadrature mixers, phase error in the local oscillator, sampling jitter and finite-resolution quantization in the analog-to-digital converters~\cite{bjornson2017massive}. To characterize the impact of transceiver hardware impairments, the non-ideal hardware circuits at the transmitter and the receiver can be modelled as non-linear memoryless filters~\cite{bjornson2014massive}. Specifically, the key modeling characteristics in this non-linear memoryless filter are that the desired signal is scaled by a deterministic factor and that an uncorrelated memoryless signal distortion term is added, which follows the Gaussian distribution in the worst case. In the $k$th ($k=1,2,\cdots,K$) symbol slot, the pilot $\tau_{\mathrm{r}}^{(k)}$ at the UE-R and the pilot $\tau_{\mathrm{t}}^{(k)}$ at the UE-T are transmitted simultaneously, leading to the received signal at the AP as~\cite{bjornson2017massive}
\begin{align}\label{Channel_estimation_1}
    \notag\mathbf{x}^{(k)}=&\underbrace{\sqrt{\rho_\mathrm{r}\varepsilon_\mathbf{v}
    \varepsilon_{u,\mathrm{r}}}\mathbf{h}_\mathrm{r}\tau_{\mathrm{r}}^{(k)}}
    _{\text{Desired UE-R pilot}}+\underbrace{\sqrt{\rho_\mathrm{t}\varepsilon_\mathbf{v}
    \varepsilon_{u,\mathrm{t}}}\mathbf{h}_\mathrm{t}\tau_{\mathrm{t}}^{(k)}}
    _{\text{Desired UE-T pilot}}\\
    \notag&+\underbrace{\sqrt{\rho_\mathrm{r}\varepsilon_\mathbf{v}\left(1-\varepsilon_{u,\mathrm{r}}\right)}
    \mathbf{h}_\mathrm{r}u_\mathrm{r}^{(k)}}_{\text{UE-R HWI distortion}}
    +\underbrace{\sqrt{\rho_\mathrm{t}\varepsilon_\mathbf{v}\left(1-\varepsilon_{u,\mathrm{t}}\right)}
    \mathbf{h}_\mathrm{t}u_\mathrm{t}^{(k)}}_{\text{UE-T HWI distortion}}\\
    \notag&+\underbrace{\sqrt{\rho_\mathrm{r}\left(1-\varepsilon_\mathbf{v}\right)}
    \mathbf{h}_\mathrm{r}\odot\mathbf{v}_\mathrm{r}^{(k)}+
    \sqrt{\rho_\mathrm{t}\left(1-\varepsilon_\mathbf{v}\right)}\mathbf{h}_\mathrm{t}
    \odot\mathbf{v}_\mathrm{t}^{(k)}}_{\text{AP HWI distortion}}\\
    &+\underbrace{\mathbf{w}^{(k)}}_{\text{Additive noise}},
\end{align}
where $\varepsilon_\mathbf{v}$, $\varepsilon_{u,\mathrm{r}}$ and $\varepsilon_{u,\mathrm{t}}$ represent the hardware quality factor of the AP, UE-R and UE-T, respectively, with $0\leq\varepsilon_\mathbf{v}\leq1$, $0\leq\varepsilon_{u,\mathrm{r}}\leq1$ and $0\leq\varepsilon_{u,\mathrm{t}}\leq1$. The hardware quality factor of 1 means that the hardware is ideal, while 0 means that the hardware is completely inadequate. Furthermore, $\rho_\mathrm{r}$ and $\rho_\mathrm{t}$ are the power of transmitted symbols at the UE-R and UE-T, respectively. We have $u_\mathrm{r}^{(k)}\sim\mathcal{CN}\left(0,1\right)$, $u_\mathrm{t}^{(k)}\sim\mathcal{CN}\left(0,1\right)$, $\mathbf{v}_\mathrm{r}^{(k)}\sim\mathcal{CN}\left(\mathbf{0}_M,\mathbf{I}_M\right)$, $\mathbf{v}_\mathrm{t}^{(k)}\sim\mathcal{CN}\left(\mathbf{0}_M,\mathbf{I}_M\right)$, and $\mathbf{w}^{(k)}\sim\mathcal{CN}\left(\mathbf{0}_{M},\sigma_w^2\mathbf{I}_M\right)$.

Firstly, to estimate $\mathbf{h}_\mathrm{r}$, the conjugate transpose of the UE-R pilot sequence, i.e. $\left[\tau_{\mathrm{r}}^{(1)\dag},\tau_{\mathrm{r}}^{(2)\dag},\cdots,
\tau_{\mathrm{r}}^{(K)\dag}\right]$, is employed to combine the AP observations $\mathbf{x}^{(1)},\mathbf{x}^{(2)},\cdots,\mathbf{x}^{(K)}$ and we have the processed received observation formulated as
\begin{align}\label{Channel_estimation_2}
    \notag\mathbf{x}_\mathrm{r}=&\frac{1}{\sqrt{K}}\sum_{k=1}^{K}\mathbf{x}^{(k)}
    \tau_{\mathrm{r}}^{(k)\dag}\\
    \notag\overset{(\text{a})}=&\sqrt{K\rho_\mathrm{r}\varepsilon_\mathbf{v}
    \varepsilon_{u,\mathrm{r}}}\mathbf{h}_\mathrm{r}+\sqrt{\rho_\mathrm{r}\varepsilon_\mathbf{v}
    \left(1-\varepsilon_{u,\mathrm{r}}\right)}\mathbf{h}_\mathrm{r}u_\mathrm{r}\\
    \notag&+\sqrt{\rho_\mathrm{t}\varepsilon_\mathbf{v}\left(1-\varepsilon_{u,\mathrm{t}}\right)}
    \mathbf{h}_\mathrm{t}u_\mathrm{t}
    +\sqrt{\rho_\mathrm{r}\left(1-\varepsilon_\mathbf{v}\right)}\mathbf{h}_\mathrm{r}\odot\mathbf{v}_\mathrm{r}\\
    &+\sqrt{\rho_\mathrm{t}\left(1-\varepsilon_\mathbf{v}\right)}\mathbf{h}_\mathrm{t}\odot\mathbf{v}_\mathrm{t}
    +\mathbf{w},
\end{align}
where  $u_\mathrm{r}=\frac{1}{\sqrt{K}}\sum_{k=1}^{K}u_\mathrm{r}^{(k)}\tau_{\mathrm{r}}^{(k)\dag}$, $u_\mathrm{t}=\frac{1}{\sqrt{K}}\sum_{k=1}^{K}u_\mathrm{t}^{(k)}\tau_{\mathrm{r}}^{(k)\dag}$, $\mathbf{v}_\mathrm{r}=\frac{1}{\sqrt{K}}\sum_{k=1}^{K}\mathbf{v}_\mathrm{r}^{(k)}
\tau_{\mathrm{r}}^{(k)\dag}$, $\mathbf{v}_\mathrm{t}=\frac{1}
{\sqrt{K}}\sum_{k=1}^{K}\mathbf{v}_\mathrm{t}^{(k)}\tau_{\mathrm{r}}^{(k)\dag}$, $\mathbf{w}=\frac{1}{\sqrt{K}}\sum_{k=1}^{K}\mathbf{w}^{(k)}\tau_{\mathrm{r}}^{(k)\dag}$, In (\ref{Channel_estimation_2}), (a) is based on the orthogonality of the pilot sequence, i.e. $\sum\nolimits_{k=1}^{K}\tau_\mathrm{r}^{(k)}\tau_\mathrm{r}^{(k)\dag}=
\sum\nolimits_{k=1}^{K}\tau_\mathrm{t}^{(k)}\tau_\mathrm{t}^{(k)\dag}=K$ and $\sum\nolimits_{k=1}^{K}\tau_\mathrm{r}^{(k)}\tau_\mathrm{t}^{(k)\dag}=
\sum\nolimits_{k=1}^{K}\tau_\mathrm{t}^{(k)}\tau_\mathrm{r}^{(k)\dag}=0$. Due to the independence of $u_\mathrm{r}^{(k)}$, $u_\mathrm{t}^{(k)}$, $\mathbf{v}_\mathrm{r}^{(k)}$, $\mathbf{v}_\mathrm{t}^{(k)}$ and $\mathbf{w}^{(k)}$ for $k=1,2,\cdots,K$, we have $u_\mathrm{r}\sim\mathcal{CN}\left(0,1\right)$, $u_\mathrm{t}\sim\mathcal{CN}\left(0,1\right)$, $\mathbf{v}_\mathrm{r}\sim\mathcal{CN}\left(\mathbf{0}_M,\mathbf{I}_M\right)$, $\mathbf{v}_\mathrm{t}\sim\mathcal{CN}\left(\mathbf{0}_M,\mathbf{I}_M\right)$ and $\mathbf{w}\sim\mathcal{CN}\left(\mathbf{0}_M,\sigma_w^2\mathbf{I}_M\right)$.

Based on (\ref{Channel_estimation_2}), we have
\begin{align}
    \mathbb{E}[\mathbf{x}_\mathrm{r}]=
    \sqrt{K\rho_\mathrm{r}\varepsilon_\mathbf{v}\varepsilon_{u,\mathrm{r}}}
    \overline{\mathbf{h}}_\mathrm{r},
\end{align}
\begin{align}
    \mathbf{C}_{\mathbf{h}_\mathrm{r}\mathbf{x}_\mathrm{r}}=
    \sqrt{K\rho_\mathrm{r}\varepsilon_\mathbf{v}\varepsilon_{u,\mathrm{r}}}
    \mathbf{C}_{\mathbf{h}_\mathrm{r}\mathbf{h}_\mathrm{r}}
\end{align}
and
\begin{align}
    \notag&\mathbf{C}_{\mathbf{x}_\mathrm{r}\mathbf{x}_\mathrm{r}}
    =\rho_\mathrm{r}\varepsilon_\mathbf{v}\left(1+\left(K-1\right)\varepsilon_{u,\mathrm{r}}\right)
    \mathbf{C}_{\mathbf{h}_\mathrm{r}\mathbf{h}_\mathrm{r}}
    +\rho_\mathrm{t}\varepsilon_\mathbf{v}\left(1-\varepsilon_{u,\mathrm{t}}\right)\cdot\\
    \notag&\mathbf{C}_{\mathbf{h}_\mathrm{t}\mathbf{h}_\mathrm{t}}
    +\rho_\mathrm{r}\left(1-\varepsilon_\mathbf{v}\right)
    \mathbf{C}_{\mathbf{h}_\mathrm{r}\mathbf{h}_\mathrm{r}}\odot\mathbf{I}_M+
    \rho_\mathrm{t}\left(1-\varepsilon_\mathbf{v}\right)
    \mathbf{C}_{\mathbf{h}_\mathrm{t}\mathbf{h}_\mathrm{t}}\odot\mathbf{I}_M\\
    \notag&+\rho_\mathrm{r}\varepsilon_\mathbf{v}\left(1-\varepsilon_{u,\mathrm{r}}\right)
    \overline{\mathbf{h}}_\mathrm{r}\overline{\mathbf{h}}_\mathrm{r}^{\mathrm{H}}
    +\rho_\mathrm{t}\varepsilon_\mathbf{v}\left(1-\varepsilon_{u,\mathrm{t}}\right)
    \overline{\mathbf{h}}_\mathrm{t}\overline{\mathbf{h}}_\mathrm{t}^{\mathrm{H}}\\
    \notag&+\left(\rho_\mathrm{r}\left(1-\varepsilon_\mathbf{v}\right)
    \varrho_{\mathbf{A}_\mathrm{r}}\varrho_{\mathbf{g}_\mathrm{r}}
    \frac{\kappa_{\mathbf{A}_\mathrm{r}}\kappa_{\mathbf{g}_\mathrm{r}}
    \left\|\overline{\mathbf{a}}_\mathrm{r}^{(\mathrm{IOS})\mathrm{H}}
    \mathbf{\Theta}_\mathrm{r}\overline{\mathbf{g}}_\mathrm{r}\right\|^2}
    {\left(1+\kappa_{\mathbf{A}_\mathrm{r}}\right)
    \left(1+\kappa_{\mathbf{g}_\mathrm{r}}\right)}\right.\\
    &\left.+\rho_\mathrm{t}\left(1-\varepsilon_\mathbf{v}\right)
    \varrho_{\mathbf{A}_\mathrm{t}}\varrho_{\mathbf{g}_\mathrm{t}}
    \frac{\kappa_{\mathbf{A}_\mathrm{t}}\kappa_{\mathbf{g}_\mathrm{t}}
    \|\overline{\mathbf{a}}_\mathrm{t}^{(\mathrm{IOS})\mathrm{H}}
    \mathbf{\Theta}_\mathrm{t}\overline{\mathbf{g}}_\mathrm{t}\|^2}
    {\left(1+\kappa_{\mathbf{A}_\mathrm{t}}\right)\left(1+\kappa_{\mathbf{g}_\mathrm{t}}\right)}+\sigma_w^2\right)\mathbf{I}_M.
\end{align}
Therefore, the LMMSE estimator of $\mathbf{h}_\mathrm{r}$ is given by
\begin{align}\label{Channel_estimation_5}
    \hat{\mathbf{h}}_\mathrm{r}=\overline{\mathbf{h}}_\mathrm{r}+
    \mathbf{C}_{\mathbf{h}_\mathrm{r}\mathbf{x}_\mathrm{r}}
    \mathbf{C}_{\mathbf{x}_\mathrm{r}\mathbf{x}_\mathrm{r}}^{-1}
    \left(\mathbf{x}_\mathrm{r}-\mathbb{E}[\mathbf{x}_\mathrm{r}]\right).
\end{align}
Hence, the estimation error is $\check{\mathbf{h}}_\mathrm{r}=\mathbf{h}_\mathrm{r}-\hat{\mathbf{h}}_\mathrm{r}$, and
the estimation covariance matrix and estimation error covariance matrix are
\begin{align}\label{Channel_estimation_6}
    \mathbf{C}_{\hat{\mathbf{h}}_\mathrm{r}\hat{\mathbf{h}}_\mathrm{r}}
    =\mathbf{C}_{\mathbf{h}_\mathrm{r}\mathbf{x}_\mathrm{r}}
    \mathbf{C}_{\mathbf{x}_\mathrm{r}\mathbf{x}_\mathrm{r}}^{-1}
    \mathbf{C}_{\mathbf{h}_\mathrm{r}\mathbf{x}_\mathrm{r}}^{\mathrm{H}},
\end{align}
\begin{align}\label{Channel_estimation_7}
    \mathbf{C}_{\check{\mathbf{h}}_\mathrm{r}\check{\mathbf{h}}_\mathrm{r}}
    =\mathbf{C}_{\mathbf{h}_\mathrm{r}\mathbf{h}_\mathrm{r}}
    -\mathbf{C}_{\mathbf{h}_\mathrm{r}\mathbf{x}_\mathrm{r}}
    \mathbf{C}_{\mathbf{x}_\mathrm{r}\mathbf{x}_\mathrm{r}}^{-1}
    \mathbf{C}_{\mathbf{h}_\mathrm{r}\mathbf{x}_\mathrm{r}}^{\mathrm{H}}.
\end{align}

Similarly, we can formulate the LMMSE estimator of $\mathbf{h}_\mathrm{t}$ and its corresponding estimation covariance matrix and estimation error covariance matrix as:
\begin{align}\label{Channel_estimation_8}
    \hat{\mathbf{h}}_\mathrm{t}=\overline{\mathbf{h}}_\mathrm{t}+
    \mathbf{C}_{\mathbf{h}_\mathrm{t}\mathbf{x}_\mathrm{t}}
    \mathbf{C}_{\mathbf{x}_\mathrm{t}\mathbf{x}_\mathrm{t}}^{-1}
    \left(\mathbf{x}_\mathrm{t}-\mathbb{E}[\mathbf{x}_\mathrm{t}]\right),
\end{align}
\begin{align}\label{Channel_estimation_9}
    \mathbf{C}_{\hat{\mathbf{h}}_\mathrm{t}\hat{\mathbf{h}}_\mathrm{t}}
    =\mathbf{C}_{\mathbf{h}_\mathrm{t}\mathbf{x}_\mathrm{t}}
    \mathbf{C}_{\mathbf{x}_\mathrm{t}\mathbf{x}_\mathrm{t}}^{-1}
    \mathbf{C}_{\mathbf{h}_\mathrm{t}\mathbf{x}_\mathrm{t}}^{\mathrm{H}},
\end{align}
\begin{align}\label{Channel_estimation_10}
    \mathbf{C}_{\check{\mathbf{h}}_\mathrm{t}\check{\mathbf{h}}_\mathrm{t}}
    =\mathbf{C}_{\mathbf{h}_\mathrm{t}\mathbf{h}_\mathrm{t}}
    -\mathbf{C}_{\mathbf{h}_\mathrm{t}\mathbf{x}_\mathrm{t}}
    \mathbf{C}_{\mathbf{x}_\mathrm{t}\mathbf{x}_\mathrm{t}}^{-1}
    \mathbf{C}_{\mathbf{h}_\mathrm{t}\mathbf{x}_\mathrm{t}}^{\mathrm{H}},
\end{align}
where $\mathbf{C}_{\mathbf{h}_\mathrm{t}\mathbf{x}_\mathrm{t}}$ and $\mathbf{C}_{\mathbf{x}_\mathrm{t}\mathbf{x}_\mathrm{t}}$ can be evaluated similarly to the case, when $\mathbf{h}_\mathrm{r}$ is estimated. According to (\ref{Channel_estimation_7}) and (\ref{Channel_estimation_10}), the normalized mean square error (N-MSE) is
\begin{align}
    \notag\epsilon=&\frac{1}{2}\left(\frac{\mathrm{Tr}\left[{\mathbf{C}_{\mathbf{h}_\mathrm{r}\mathbf{h}_\mathrm{r}}
    -\mathbf{C}_{\mathbf{h}_\mathrm{r}\mathbf{x}_\mathrm{r}}\mathbf{C}_{\mathbf{x}_\mathrm{r}
    \mathbf{x}_\mathrm{r}}^{-1}\mathbf{C}_{\mathbf{h}_\mathrm{r}\mathbf{x}_\mathrm{r}}^{\mathrm{H}}}\right]}
    {\mathrm{Tr}\left[\mathbf{C}_{\mathbf{h}_\mathrm{r}\mathbf{h}_\mathrm{r}}\right]}\right.\\
    &\left.+\frac{\mathrm{Tr}
    \left[{\mathbf{C}_{\mathbf{h}_\mathrm{t}\mathbf{h}_\mathrm{t}}-\mathbf{C}_{\mathbf{h}_\mathrm{t}
    \mathbf{x}_\mathrm{t}}\mathbf{C}_{\mathbf{x}_\mathrm{t}\mathbf{x}_\mathrm{t}}^{-1}
    \mathbf{C}_{\mathbf{h}_\mathrm{t}\mathbf{x}_\mathrm{t}}^{\mathrm{H}}}\right]}
    {\mathrm{Tr}\left[\mathbf{C}_{\mathbf{h}_\mathrm{t}\mathbf{h}_\mathrm{t}}\right]}\right).
\end{align}

\section{Beamforming Design}\label{Beamforming_Design}
In the section, we present the AP beamforming based on the estimated equivalent channels $\hat{\mathbf{h}}_\mathrm{r}$ and $\hat{\mathbf{h}}_\mathrm{t}$ defined in (\ref{Channel_estimation_5}) and (\ref{Channel_estimation_8}), respectively. Then, the ergodic spectral efficiency upper bound is theoretically derived with respect to the IOS design. Finally, we optimize the beamforming design of the IOS by maximizing the ergodic spectral efficiency upper bound.

\subsection{AP Beamforming}
We denote the information symbol at the UE-R and UE-T as $s_\mathrm{r}\sim\mathcal{CN}\left(0,1\right)$ and $s_\mathrm{t}\sim\mathcal{CN}\left(0,1\right)$, respectively. Then, the signal received at the AP is given by
\begin{align}\label{Channel_Model_24}
    \notag\mathbf{y}=&\underbrace{\sqrt{\rho_\mathrm{r}\varepsilon_\mathbf{v}
    \varepsilon_{u,\mathrm{r}}}\hat{\mathbf{h}}_\mathrm{r}s_\mathrm{r}}
    _{\text{$s_\mathrm{r}$ over estimated channel}}
    +\underbrace{\sqrt{\rho_\mathrm{t}\varepsilon_\mathbf{v}
    \varepsilon_{u,\mathrm{t}}}\hat{\mathbf{h}}_\mathrm{t}s_\mathrm{t}}
    _{\text{$s_\mathrm{t}$ over estimated channel}}\\
    \notag&+\underbrace{\sqrt{\rho_\mathrm{r}\varepsilon_\mathbf{v}
    \varepsilon_{u,\mathrm{r}}}\check{\mathbf{h}}_\mathrm{r}s_\mathrm{r}}
    _{\text{$s_\mathrm{r}$ over unknown channel}}
    +\underbrace{\sqrt{\rho_\mathrm{t}\varepsilon_\mathbf{v}
    \varepsilon_{u,\mathrm{t}}}\check{\mathbf{h}}_\mathrm{t}s_\mathrm{t}}
    _{\text{$s_\mathrm{t}$ over unknown channel}}\\
    \notag&+\underbrace{\sqrt{\rho_\mathrm{r}\varepsilon_\mathbf{v}
    \left(1-\varepsilon_{u,\mathrm{r}}\right)}\mathbf{h}_\mathrm{r}u_\mathrm{r}}
    _{\text{UE-R HWI distortion}}
    +\underbrace{\sqrt{\rho_\mathrm{t}\varepsilon_\mathbf{v}\left(1-\varepsilon_{u,\mathrm{t}}\right)}
    \mathbf{h}_\mathrm{t}u_\mathrm{t}}_{\text{UE-T HWI distortion}}\\
    \notag&+\underbrace{\sqrt{\rho_\mathrm{r}\left(1-\varepsilon_\mathbf{v}\right)}
    \mathbf{h}_\mathrm{r}\odot\mathbf{v}_\mathrm{r}+
    \sqrt{\rho_\mathrm{t}\left(1-\varepsilon_\mathbf{v}\right)}\mathbf{h}_\mathrm{t}
    \odot\mathbf{v}_\mathrm{t}}_{\text{AP HWI distortion}}\\
    &+\underbrace{\mathbf{w}}_{\text{Additive noise}},
\end{align}
where $u_\mathrm{r}\sim\mathcal{CN}\left(0,1\right)$, $u_\mathrm{t}\sim
\mathcal{CN}\left(0,1\right)$, $\mathbf{v}_\mathrm{r}\sim\mathcal{CN}(\mathbf{0}_M,\mathbf{I}_M)$, $\mathbf{v}_\mathrm{t}\sim\mathcal{CN}\left(\mathbf{0}_M,\mathbf{I}_M\right)$, and $\mathbf{w}\sim\mathcal{CN}\left(\mathbf{0}_{M},\sigma_w^2\mathbf{I}_M\right)$.

The MMSE combiner is employed at the AP to recover the information $s_\mathrm{r}$ and $s_\mathrm{t}$ based on the observation $\mathbf{y}$.
\begin{theorem}\label{theorem_1}
The MMSE combining vectors $\mathbf{q}_\mathrm{r}$ and $\mathbf{q}_\mathrm{t}$ of the UE-R and UE-T, respectively, are given by~\cite{bjornson2017massive}
\begin{align}\label{Beamforming_Design_1}
    \mathbf{q}_\mathrm{r}=\rho_\mathrm{r}\varepsilon_{\mathbf{v}}\varepsilon_{u,\mathrm{r}}
    \left(\rho_\mathrm{r}\varepsilon_{\mathbf{v}}
    \hat{\mathbf{h}}_\mathrm{r}\hat{\mathbf{h}}_\mathrm{r}^{\mathrm{H}}
    +\rho_\mathrm{t}\varepsilon_{\mathbf{v}}\hat{\mathbf{h}}_\mathrm{t}
    \hat{\mathbf{h}}_\mathrm{t}^{\mathrm{H}}+\mathbf{R}\right)^{-1}\hat{\mathbf{h}}_\mathrm{r},
\end{align}
\begin{align}\label{Beamforming_Design_2}
    \mathbf{q}_\mathrm{t}=\rho_\mathrm{t}\varepsilon_{\mathbf{v}}\varepsilon_{u,\mathrm{t}}
    \left(\rho_\mathrm{r}\varepsilon_{\mathbf{v}}
    \hat{\mathbf{h}}_\mathrm{r}\hat{\mathbf{h}}_\mathrm{r}^{\mathrm{H}}
    +\rho_\mathrm{t}\varepsilon_{\mathbf{v}}\hat{\mathbf{h}}_\mathrm{t}
    \hat{\mathbf{h}}_\mathrm{t}^{\mathrm{H}}+\mathbf{R}\right)^{-1}\hat{\mathbf{h}}_\mathrm{t},
\end{align}
where we have:
\begin{align}\label{Beamforming_Design_2_1}
    \notag\mathbf{R}=&\rho_\mathrm{r}\varepsilon_\mathbf{v}(1-\varepsilon_{u,\mathrm{r}})
    \mathbf{C}_{\check{\mathbf{h}}_\mathrm{r}\check{\mathbf{h}}_\mathrm{r}}
    +\rho_\mathrm{t}\varepsilon_\mathbf{v}\left(1-\varepsilon_{u,\mathrm{t}}\right)
    \mathbf{C}_{\check{\mathbf{h}}_\mathrm{t}\check{\mathbf{h}}_\mathrm{t}}\\
    \notag&+\rho_\mathrm{r}\left(1-\varepsilon_\mathbf{v}\right)
    \left(\varrho_{\mathbf{A}_\mathrm{r}}\varrho_{\mathbf{g}_\mathrm{r}}
    \frac{\left(1+\kappa_{\mathbf{g}_\mathrm{r}}+\kappa_{\mathbf{A}_\mathrm{r}}\right)N_\mathrm{r}}
    {\left(1+\kappa_{\mathbf{A}_\mathrm{r}}\right)\left(1+\kappa_{\mathbf{g}_\mathrm{r}}\right)}
    +\varrho_{\mathbf{b}_\mathrm{r}}\right.\\
    \notag&\left.+\varrho_{\mathbf{A}_\mathrm{r}}\varrho_{\mathbf{g}_\mathrm{r}}
    \frac{\kappa_{\mathbf{A}_\mathrm{r}}\kappa_{\mathbf{g}_\mathrm{r}}
    \|\overline{\mathbf{a}}_\mathrm{r}^{(\mathrm{IOS})\mathrm{H}}
    \mathbf{\Theta}_\mathrm{r}\overline{\mathbf{g}}_\mathrm{r}\|^2}
    {\left(1+\kappa_{\mathbf{A}_\mathrm{r}}\right)\left(1+\kappa_{\mathbf{g}_\mathrm{r}}\right)}\right)\mathbf{I}_M\\
    \notag&+\rho_\mathrm{t}\left(1-\varepsilon_\mathbf{v}\right)
    \left(\varrho_{\mathbf{A}_\mathrm{t}}\varrho_{\mathbf{g}_\mathrm{t}}
    \frac{\left(1+\kappa_{\mathbf{g}_\mathrm{t}}+\kappa_{\mathbf{A}_\mathrm{t}}\right)N_\mathrm{t}}
    {\left(1+\kappa_{\mathbf{A}_\mathrm{t}}\right)\left(1+\kappa_{\mathbf{g}_\mathrm{t}}\right)}
    +\varrho_{\mathbf{b}_\mathrm{t}}\right.\\
    &\left.+\varrho_{\mathbf{A}_\mathrm{t}}\varrho_{\mathbf{g}_\mathrm{t}}
    \frac{\kappa_{\mathbf{A}_\mathrm{t}}\kappa_{\mathbf{g}_\mathrm{t}}
    \|\overline{\mathbf{a}}_\mathrm{t}^{(\mathrm{IOS})\mathrm{H}}
    \mathbf{\Theta}_\mathrm{t}\overline{\mathbf{g}}_\mathrm{t}\|^2}
    {\left(1+\kappa_{\mathbf{A}_\mathrm{t}}\right)\left(1+\kappa_{\mathbf{g}_\mathrm{t}}\right)}\right)\mathbf{I}_M
    +\sigma_w^2\mathbf{I}_M.
\end{align}
Based on the MMSE combiner at the AP, the instantaneous SINR for the received symbols of the UE-R and UE-T are
\begin{align}\label{Beamforming_Design_3}
    \notag\gamma_\mathrm{r}=&\rho_\mathrm{r}\varepsilon_{\mathbf{v}}
    \varepsilon_{u,\mathrm{r}}\hat{\mathbf{h}}_\mathrm{r}^{\mathrm{H}}
    \left(\rho_\mathrm{r}\varepsilon_{\mathbf{v}}\left(1-\varepsilon_{u,\mathrm{r}}\right)
    \hat{\mathbf{h}}_\mathrm{r}\hat{\mathbf{h}}_\mathrm{r}^{\mathrm{H}}
    +\rho_\mathrm{t}\varepsilon_{\mathbf{v}}\hat{\mathbf{h}}_\mathrm{t}
    \hat{\mathbf{h}}_\mathrm{t}^{\mathrm{H}}\right.\\
    &\left.+\mathbf{R}\right)^{-1}\hat{\mathbf{h}}_\mathrm{r},
\end{align}
\begin{align}\label{Beamforming_Design_4}
    \notag\gamma_\mathrm{t}=&\rho_\mathrm{t}\varepsilon_{\mathbf{v}}
    \varepsilon_{u,\mathrm{t}}\hat{\mathbf{h}}_\mathrm{t}^{\mathrm{H}}
    \left(\rho_\mathrm{t}\varepsilon_{\mathbf{v}}\left(1-\varepsilon_{u,\mathrm{t}}\right)
    \hat{\mathbf{h}}_\mathrm{t}\hat{\mathbf{h}}_\mathrm{t}^{\mathrm{H}}
    +\rho_\mathrm{r}\varepsilon_{\mathbf{v}}\hat{\mathbf{h}}_\mathrm{r}
    \hat{\mathbf{h}}_\mathrm{r}^{\mathrm{H}}\right.\\
    &\left.+\mathbf{R}\right)^{-1}\hat{\mathbf{h}}_\mathrm{t}.
\end{align}
\end{theorem}
\begin{IEEEproof}
    See Appendix~\ref{Appendix_A}.
\end{IEEEproof}

\begin{theorem}\label{theorem_2}
Upon using the MMSE combiner at the AP, the instantaneous spectral efficiency of the UE-R and UE-T, denoted as $\mathcal{R}_{\mathrm{ins},\mathrm{r}}$ and $\mathcal{R}_{\mathrm{ins},\mathrm{t}}$, respectively, becomes
\begin{align}\label{Beamforming_Design_5}
    \mathcal{R}_{\mathrm{ins},\mathrm{i}}=\log_2\left(1+\frac{\rho_\mathrm{i}
    \varepsilon_\mathbf{v}\varepsilon_{u,\mathrm{i}}\zeta_\mathrm{i}}
    {1+\rho_\mathrm{i}\varepsilon_\mathbf{v}\left(1-\varepsilon_{u,\mathrm{i}}\right)\zeta_\mathrm{i}}\right),
    \ \mathrm{i}\in\left\{\mathrm{r},\mathrm{t}\right\},
\end{align}
where $\zeta_\mathrm{r}=\hat{\mathbf{h}}_\mathrm{r}^{\mathrm{H}}\mathbf{R}^{-1}
\hat{\mathbf{h}}_\mathrm{r}-\frac{\rho_\mathrm{t}\varepsilon_\mathbf{v}
\hat{\mathbf{h}}_\mathrm{r}^{\mathrm{H}}\mathbf{R}^{-1}\hat{\mathbf{h}}_\mathrm{t}
\hat{\mathbf{h}}_\mathrm{t}^{\mathrm{H}}\mathbf{R}^{-1}
\hat{\mathbf{h}}_\mathrm{r}}{1+\rho_\mathrm{t}\varepsilon_\mathbf{v}
\hat{\mathbf{h}}_\mathrm{t}^{\mathrm{H}}\mathbf{R}^{-1}\hat{\mathbf{h}}_\mathrm{t}}$ and $\zeta_\mathrm{t}=\hat{\mathbf{h}}_\mathrm{t}^{\mathrm{H}}\mathbf{R}^{-1}
\hat{\mathbf{h}}_\mathrm{t}-\frac{\rho_\mathrm{r}\varepsilon_\mathbf{v}
\hat{\mathbf{h}}_\mathrm{t}^{\mathrm{H}}
\mathbf{R}^{-1}\hat{\mathbf{h}}_\mathrm{r}\hat{\mathbf{h}}_\mathrm{r}^{\mathrm{H}}\mathbf{R}^{-1}
\hat{\mathbf{h}}_\mathrm{t}}{1+\rho_\mathrm{r}\varepsilon_\mathbf{v}
\hat{\mathbf{h}}_\mathrm{r}^{\mathrm{H}}\mathbf{R}^{-1}\hat{\mathbf{h}}_\mathrm{r}}$.
\end{theorem}
\begin{IEEEproof}
    See Appendix~\ref{Appendix_B}.
\end{IEEEproof}

Therefore, according to \textit{Theorem}~\ref{theorem_2}, we can express the ergodic spectral efficiency of UE-R and UE-T as $\mathcal{R}_{\mathrm{erg},\mathrm{i}}=\mathbb{E}\left[\mathcal{R}_{\mathrm{ins},\mathrm{i}}\right]
=\mathbb{E}\left[\log_2\left(1+\frac{\rho_\mathrm{i}\varepsilon_\mathbf{v}
\varepsilon_{u,\mathrm{i}}\zeta_\mathrm{i}}{1+\rho_\mathrm{i}\varepsilon_\mathbf{v}
\left(1-\varepsilon_{u,\mathrm{i}}\right)\zeta_\mathrm{i}}\right)\right]$, with $\mathrm{i}\in\left\{\mathrm{r},\mathrm{t}\right\}$.

\begin{theorem}\label{theorem_3}
When $\cos\psi_\mathrm{t}\neq\cos\psi_\mathrm{r}$, the ergodic spectral efficiency upper bound of UE-R and UE-T for transmission over random channel links, denoted as $\ddot{\mathcal{R}}_{\mathrm{erg},\mathrm{r}}$ and $\ddot{\mathcal{R}}_{\mathrm{erg},\mathrm{t}}$, respectively, can be written as
\begin{align}\label{Beamforming_Design_9}
    \ddot{\mathcal{R}}_{\mathrm{erg},\mathrm{i}}=\log_2\left(1+\frac{\rho_\mathrm{i}
    \varepsilon_\mathbf{v}\varepsilon_{u,\mathrm{i}}\ddot{\zeta}_\mathrm{i}}
    {1+\rho_\mathrm{i}\varepsilon_\mathbf{v}\left(1-\varepsilon_{u,\mathrm{i}}\right)
    \ddot{\zeta}_\mathrm{i}}\right),\ \mathrm{i}\in\{\mathrm{r},\mathrm{t}\},
\end{align}
where $\ddot{\zeta}_\mathrm{i}$ can be derived as follows:
\begin{align}\label{Beamforming_Design_10_1}
    \notag\ddot{\zeta}_\mathrm{i}
    =&\mathbb{E}\left[\hat{\mathbf{h}}_\mathrm{i}^{\mathrm{H}}\mathbf{R}^{-1}
    \hat{\mathbf{h}}_\mathrm{i}\right]\\
    \notag=&\text{Tr}\left[\mathbf{R}^{-1}\left(\overline{\mathbf{h}}_\mathrm{i}
    \overline{\mathbf{h}}_\mathrm{i}^{\mathrm{H}}
    +\mathbf{C}_{\hat{\mathbf{h}}_\mathrm{i}\hat{\mathbf{h}}_\mathrm{i}}\right)\right]\\
    \notag=&\frac{\varrho_{\mathbf{A}_\mathrm{i}}
    \varrho_{\mathbf{g}_\mathrm{i}}\kappa_{\mathbf{A}_\mathrm{i}}
    \kappa_{\mathbf{g}_\mathrm{i}}\|\overline{\mathbf{a}}^{(\mathrm{IOS})\mathrm{H}}_\mathrm{i}
    \mathbf{\Theta}_\mathrm{i}\overline{\mathbf{g}}_\mathrm{i}\|^2}
    {\left(1+\kappa_{\mathbf{A}_\mathrm{i}}\right)\left(1+\kappa_{\mathbf{g}_\mathrm{i}}\right)}
    \cdot\overline{\mathbf{a}}^{(\mathrm{AP})\mathrm{H}}_\mathrm{i}\mathbf{R}^{-1}
    \overline{\mathbf{a}}^{(\mathrm{AP})}_\mathrm{i}\\
    &+\text{Tr}\left[\mathbf{R}^{-1}
    \mathbf{C}_{\hat{\mathbf{h}}_\mathrm{i}\hat{\mathbf{h}}_\mathrm{i}}\right].
\end{align}
The upper bound can be attained by letting $M\rightarrow\infty$.
\end{theorem}
\begin{IEEEproof}
    See Appendix~\ref{Appendix_C}.
\end{IEEEproof}

\subsection{IOS Beamforming}
In this section, we design the reflective IOS beamformer matrix $\mathbf{\Theta}_\mathrm{r}$ based on the statistical CSI of UE-R, i.e. $\overline{\mathbf{A}}_\mathrm{r}$, $\overline{\mathbf{g}}_\mathrm{r}$, and the transmit IOS beamformer matrix $\mathbf{\Theta}_\mathrm{t}$ based on the statistical CSI of UE-T, i.e. $\overline{\mathbf{A}}_\mathrm{t}$, $\overline{\mathbf{g}}_\mathrm{t}$. This is carried out by maximizing the ergodic spectral efficiency upper bound of the UE-R, i.e. $\ddot{\mathcal{R}}_{\mathrm{erg},\mathrm{r}}$, and that of the UE-T, i.e. $\ddot{\mathcal{R}}_{\mathrm{erg},\mathrm{t}}$, respectively. Thus, we formulate the IOS beamforming problem as
\begin{align}\label{Beamforming_Design_12}
    \notag\text{(P1)}\quad &\max_{\mathbf{\Theta}_\mathrm{i}}\ \ddot{\mathcal{R}}_{\mathrm{erg},\mathrm{i}}\\
    \text{s.t.}&\quad \theta_{\mathrm{i},n}\in[0,2\pi),\ n=1,2,\cdots,N_\mathrm{i},\ \mathrm{i}\in\{\mathrm{r},\mathrm{t}\}.
\end{align}
\begin{theorem}\label{theorem_4}
In practical systems, we have $\frac{\varepsilon_{u,\mathrm{r}}}{1-\varepsilon_{u,\mathrm{r}}}\gg1$, $\frac{\varepsilon_{u,\mathrm{t}}}{1-\varepsilon_{u,\mathrm{t}}}\gg1$ and $\frac{\varepsilon_\mathbf{v}}{1-\varepsilon_\mathbf{v}}\gg1$. Hence, the problem (P1) can be derived as
\begin{align}\label{Beamforming_Design_13}
    \notag\text{(P2)}\quad &\max_{\mathbf{\Theta}_\mathrm{i}}\ \left\|\overline{\mathbf{A}}_\mathrm{i}\mathbf{\Theta}_\mathrm{i}
    \overline{\mathbf{g}}_\mathrm{i}\right\|^2\\
    \text{s.t.}&\quad \theta_{\mathrm{i},n}\in[0,2\pi),\ n=1,2,\cdots,N_\mathrm{i},\ \mathrm{i}\in\left\{\mathrm{r},\mathrm{t}\right\}.
\end{align}
\end{theorem}
\begin{IEEEproof}
    See Appendix~\ref{Appendix_E}.
\end{IEEEproof}

In problem (P2), based on (\ref{Channel_Model_6}) we can exploit that $\left\|\overline{\mathbf{A}}_\mathrm{i}
\mathbf{\Theta}_\mathrm{i}\overline{\mathbf{g}}_\mathrm{i}\right\|^2
=M\left\|\overline{\mathbf{a}}_\mathrm{i}^{(\mathrm{IOS})\mathrm{H}}\mathbf{\Theta}_\mathrm{i}
\overline{\mathbf{g}}_\mathrm{i}\right\|^2\overset{(\text{a})}{\leq}MN_\mathrm{i}^2$, where the relationship in (a) is established when $\left[\mathbf{\Theta}_\mathrm{r}\right]_{n,n}=\left[\overline{\mathbf{a}}
_\mathrm{r}^{(\mathrm{IOS})}\right]_n
\cdot\left[\overline{\mathbf{g}}_\mathrm{r}^{\mathrm{H}}\right]_n$ and $\left[\mathbf{\Theta}_\mathrm{t}\right]_{n,n}
=\left[\overline{\mathbf{a}}_\mathrm{t}^{(\mathrm{IOS})}\right]_n
\cdot\left[\overline{\mathbf{g}}_\mathrm{t}^{\mathrm{H}}\right]_n$. According to (\ref{IOS_Architecture_1}), (\ref{Channel_Model_3}) and (\ref{Channel_Model_7}), we can express the optimal phase shift of the IOS elements as
\begin{align}\label{Beamforming_Design_16}
    \notag\theta_{\mathrm{i},n}=&\frac{2\pi}{\lambda}\left(\delta_{x}n_\mathrm{i}^{x}
    \left(\sin\phi_\mathrm{i}\cos\varphi_\mathrm{i}-\sin\omega_\mathrm{i}\cos\varpi_\mathrm{i}\right)\right.\\
    &\left.+\delta_{y}n_\mathrm{i}^{y}\left(\cos\phi_\mathrm{i}-\cos\omega_\mathrm{i}\right)\right),
    \ \mathrm{i}\in\left\{\mathrm{r},\mathrm{t}\right\},
\end{align}
with $n^x_\mathrm{i}\in\left\{1,2,\cdots,N^x_\mathrm{i}\right\}$, $n^y_\mathrm{i}\in\left\{1,2,\cdots,N^y_\mathrm{i}\right\}$, and $n=\left(n_\mathrm{i}^{y}-1\right)N^x_\mathrm{i}+n_\mathrm{i}^{x}$. Therefore, $\ddot{\zeta}_\mathrm{i}$ in (\ref{Beamforming_Design_9}) is given by
\begin{align}\label{Beamforming_Design_18}
    \ddot{\zeta}_\mathrm{i}=\frac{\varrho_{\mathbf{A}_\mathrm{i}}\varrho_{\mathbf{g}_\mathrm{i}}
    \kappa_{\mathbf{A}_\mathrm{i}}\kappa_{\mathbf{g}_\mathrm{i}}N_\mathrm{i}^2}
    {\left(1+\kappa_{\mathbf{A}_\mathrm{i}}\right)\left(1+\kappa_{\mathbf{g}_\mathrm{i}}\right)}
    \overline{\mathbf{a}}^{(\mathrm{AP})\mathrm{H}}_\mathrm{i}\mathbf{R}^{-1}
    \overline{\mathbf{a}}^{(\mathrm{AP})}_\mathrm{i}
    +\text{Tr}\left[\mathbf{R}^{-1}\mathbf{C}_{\hat{\mathbf{h}}_\mathrm{i}\hat{\mathbf{h}}_\mathrm{i}}\right],
\end{align}
where $\left\|\overline{\mathbf{a}}^{(\mathrm{IOS})\mathrm{H}}_\mathrm{i}
\mathbf{\Theta}_\mathrm{i}\overline{\mathbf{g}}_\mathrm{i}\right\|^2=N_\mathrm{i}^2$ in the matrices $\mathbf{R}$ and $\mathbf{C}_{\hat{\mathbf{h}}_\mathrm{i}\hat{\mathbf{h}}_\mathrm{i}}$.

\begin{theorem}\label{theorem_5}
To characterize the scaling law of the transceiver HWIs versus the number of AP antennas $M$, and the number of IOS elements $N_\mathrm{r}$ and $N_\mathrm{t}$, we present a loose upper bound of the ergodic spectral efficiency of the UE-R and UE-T, denoted as $\ddot{\mathcal{R}}_{\mathrm{erg},\mathrm{r}}'$ and $\ddot{\mathcal{R}}_{\mathrm{erg},\mathrm{t}}'$ respectively, as
\begin{align}\label{Beamforming_Design_19}
    \notag&\ddot{\mathcal{R}}_{\mathrm{erg},\mathrm{i}}'=\log_2\left(1+\right.\\
    &\left.\frac{\rho_\mathrm{i}\varepsilon_\mathbf{v}\varepsilon_{u,\mathrm{i}}M\eta_\mathrm{i}}
    {\rho_\mathrm{i}\varepsilon_\mathbf{v}\left(1-\varepsilon_{u,\mathrm{i}}\right)M\eta_\mathrm{i}
    +\rho_\mathrm{r}\left(1-\varepsilon_{\mathbf{v}}\right)\eta_\mathrm{r}
    +\rho_\mathrm{t}\left(1-\varepsilon_{\mathbf{v}}\right)\eta_\mathrm{t}+\sigma_w^2}\right),
\end{align}
where $\mathrm{i}\in\left\{\mathrm{r},\mathrm{t}\right\}$ and $\eta_\mathrm{i}=\frac{\varrho_{\mathbf{A}_\mathrm{i}}\varrho_{\mathbf{g}_\mathrm{i}}
\kappa_{\mathbf{A}_\mathrm{i}}\kappa_{\mathbf{g}_\mathrm{i}}N_\mathrm{i}^2+(1+
\kappa_{\mathbf{g}_\mathrm{i}}+\kappa_{\mathbf{A}_\mathrm{i}})N_\mathrm{i}}{(1+
\kappa_{\mathbf{A}_\mathrm{i}})(1+\kappa_{\mathbf{g}_\mathrm{i}})}+\varrho_{\mathbf{b}_\mathrm{i}}$.
\end{theorem}
\begin{IEEEproof}
    See Appendix~\ref{Appendix_F}.
\end{IEEEproof}

Observe Theorem \ref{theorem_5} that increasing the number of AP antennas $M$ or the number of IOS elements $N_\mathrm{r}$ and $N_\mathrm{t}$ is beneficial for improving the ergodic spectral efficiency in the low signal-to-noise ratio (SNR) region. By contrast, there is a performance floor in the high SNR region, i.e. the ergodic spectral efficiency is limited by the hardware quality of the UEs. This means that deploying more AP antennas or IOS elements improves the ergodic spectral efficiency performance when the SNR is low. However, any ergodic spectral efficiency enhancement in the high SNR region requires high-quality hardware at the UEs.

\subsection{Hardware-Quality Scaling Law}
In this section, we present the scaling law of transceiver HWIs, with respect to the number of AP antennas $M$, and the number of IOS elements $N_\mathrm{r}$ and $N_\mathrm{t}$, to gather valuable insights on the hardware efficiency.

\subsubsection{AP HWI}
Observe in (\ref{Beamforming_Design_19}) that in the low transmit power region, the egordic spectral efficiency degradation resulting from the AP HWI can be compensated by harnessing an additional $\frac{\left(1-\varepsilon_\mathbf{v}\right)M}{\varepsilon_\mathbf{v}}$ number of AP antennas.

\subsubsection{UE HWI}
According to (\ref{Beamforming_Design_19}), when UE-R and UE-T have HWIs, the loss of ergodic spectral efficiency cannot be completely compensated by increasing either the number of AP antennas or that of the IOS elements. When the transmit power obeys $\rho_\mathrm{i}\rightarrow\infty$, or $M\rightarrow\infty$, or $N_\mathrm{i}\rightarrow\infty$, the loose egordic spectral efficiency upper bound $\ddot{\mathcal{R}}_{\mathrm{erg},\mathrm{i}}'$ tends to saturate according to $\ddot{\mathcal{R}}_{\mathrm{erg},\mathrm{i}}'\rightarrow\log_2
\left(1+\frac{\varepsilon_{u,\mathrm{i}}}{1-\varepsilon_{u,\mathrm{i}}}\right)$.

\subsubsection{Number of IOS elements}
We consider the scenario when the direct UE-AP link is completely blocked, i.e. $\varrho_{\mathbf{b}_\mathrm{i}}=\varrho_{\mathbf{b}_\mathrm{i}}=0$. If the signals suffer from Rayleigh fading, i.e. $\kappa_{\mathbf{A}_\mathrm{i}}=\kappa_{\mathbf{g}_\mathrm{i}}=0$, doubling the number of IOS elements $N_\mathrm{i}$ is equivalent to a $3\text{dB}$ enhancement of the transmit power $\rho_\mathrm{i}$. By contrast, if the channel is perfectly known, i.e. $\kappa_{\mathbf{A}_\mathrm{i}}=\kappa_{\mathbf{g}_\mathrm{i}}\rightarrow\infty$, doubling the number of IOS elements $N_\mathrm{i}$ is equivalent to a $6\text{dB}$ enhancement of the transmit power $\rho_\mathrm{i}$.

It is worth noting that, according to (\ref{Beamforming_Design_19}), increasing the transmit power $\rho_\mathrm{i}$ or the number of IOS elements $N_\mathrm{i}$ cannot compensate for the effect of the AP HWI or UE HWI since the signal distortion is exacerbated upon increasing of the transmit power and the number of IOS elements used.

\section{Numerical and Simulation Results}\label{Numerical_and_Simulation_Results}
In this section, our theoretical analysis and simulation results characterizing the ergodic sum-rate of the UE-R and UE-T are presented in the face of transceiver HWIs. Referring to the RIS-aided system setup in~\cite{kang2023active}, the distance between the nodes of the AP, the UEs and the RIS ranges from tens to one hundred meters. Therefore, in the system model of Fig.~\ref{Fig_System_model_IOS}, the AP, UE-R, UE-T, reflective IOS elements and transmit IOS elements are located at the Cartesian coordinates of (0m, -100m, 20m), (0m, -20m, 5m), (0m, 20m, 5m), (2m, 0m, 15m) and (-2m, 0m, 15m), respectively. Furthermore, $\phi_\mathrm{r}$, $\phi_\mathrm{t}$, $\varphi_\mathrm{r}$, $\varphi_\mathrm{t}$, $\omega_\mathrm{r}$, $\omega_\mathrm{t}$, $\varpi_\mathrm{r}$, $\varpi_\mathrm{t}$ $\psi_\mathrm{r}$ and $\psi_\mathrm{t}$ are randomly generated following a uniform distribution. According to~\cite{kang2023active}, we set $\mathrm{C}_0=-30\text{dB}$, $\sigma_w^2=-90\text{dBm}$, $\alpha_{\mathbf{A}_\mathrm{r}}=\alpha_{\mathbf{A}_\mathrm{t}}=
\alpha_{\mathbf{g}_\mathrm{r}}=\alpha_{\mathbf{g}_\mathrm{t}}=2.2$. Unless otherwise specified, the other simulation parameters are: $\alpha_{\mathbf{b}_\mathrm{r}}=\alpha_{\mathbf{b}_\mathrm{t}}=\alpha_{\mathbf{b}}=4.8$, $K=16$, $M=100$, $N_\mathrm{r}=N_\mathrm{t}=20\times20$, $\delta_x=\delta_y=\frac{\lambda}{2}$, $d_0=\frac{\lambda}{2}$, $\Delta\psi=\psi_\mathrm{r}-\psi_\mathrm{t}=0.1\pi$, $\rho_\mathrm{r}=\rho_\mathrm{t}=\rho=20\text{dBm}$, $\varepsilon_{u,\mathrm{r}}=\varepsilon_{u,\mathrm{t}}=\varepsilon_{u}$, $\kappa_{\mathbf{A}_\mathrm{r}}=\kappa_{\mathbf{A}_\mathrm{t}}=
\kappa_{\mathbf{g}_\mathrm{r}}=\kappa_{\mathbf{g}_\mathrm{t}}=\kappa=0\text{dB}$, and the ergodic sum-rate performance variation with the change of these parameters will be evaluated in the following simulation results.

\begin{figure}[!t]
    \centering
    \includegraphics[width=2.8in]{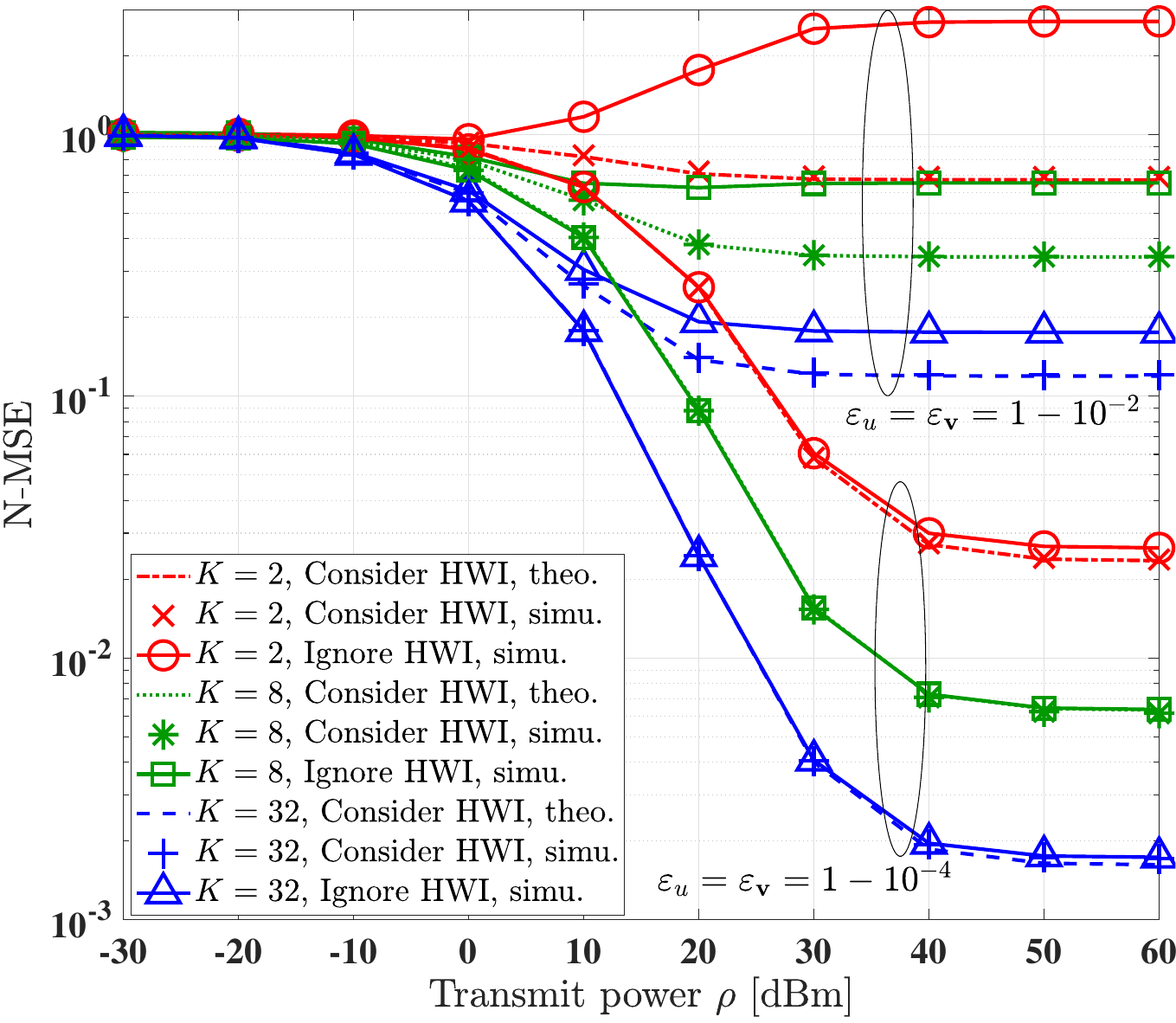}
    \caption{Comparison of the N-MSE versus the transmit power.}\label{Fig_simu_CE}
\end{figure}

Fig. \ref{Fig_simu_CE} compares the theoretical analysis (theo.) and the simulation results (simu.) of the N-MSE versus the transmit power $\rho$ under different pilot overhead $K$ with the number of AP antennas $M=50$, where the legend `Consider HWI' means the hardware impairment is considered in the LMMSE estimator, while the legend `Ignore HWI' means the hardware impairment is ignored in the design of the LMMSE estimator. Fig. \ref{Fig_simu_CE} shows that increasing the pilot overhead can effectively improve the N-MSE performance, while at the cost of transmitting more pilot sequences. With the increase of the transmit power, the N-MSE tends to a constant, since the channel estimation error dominantly results from the effect of HWI instead of the additive noise at the AP. Furthermore, Fig. \ref{Fig_simu_CE} shows that ignoring the effect of HWI exacerbates the N-MSE performance especially in the low hardware quality factor region and high transmit power region, which demonstrates the necessity of considering the effect of HWI in the design of channel estimator.

\begin{figure}[!t]
    \centering
    \subfloat[The UE hardware quality factor $\varepsilon_{u}=1$]
    {\begin{minipage}{1\linewidth}
        \centering
        \includegraphics[width=2.8in]{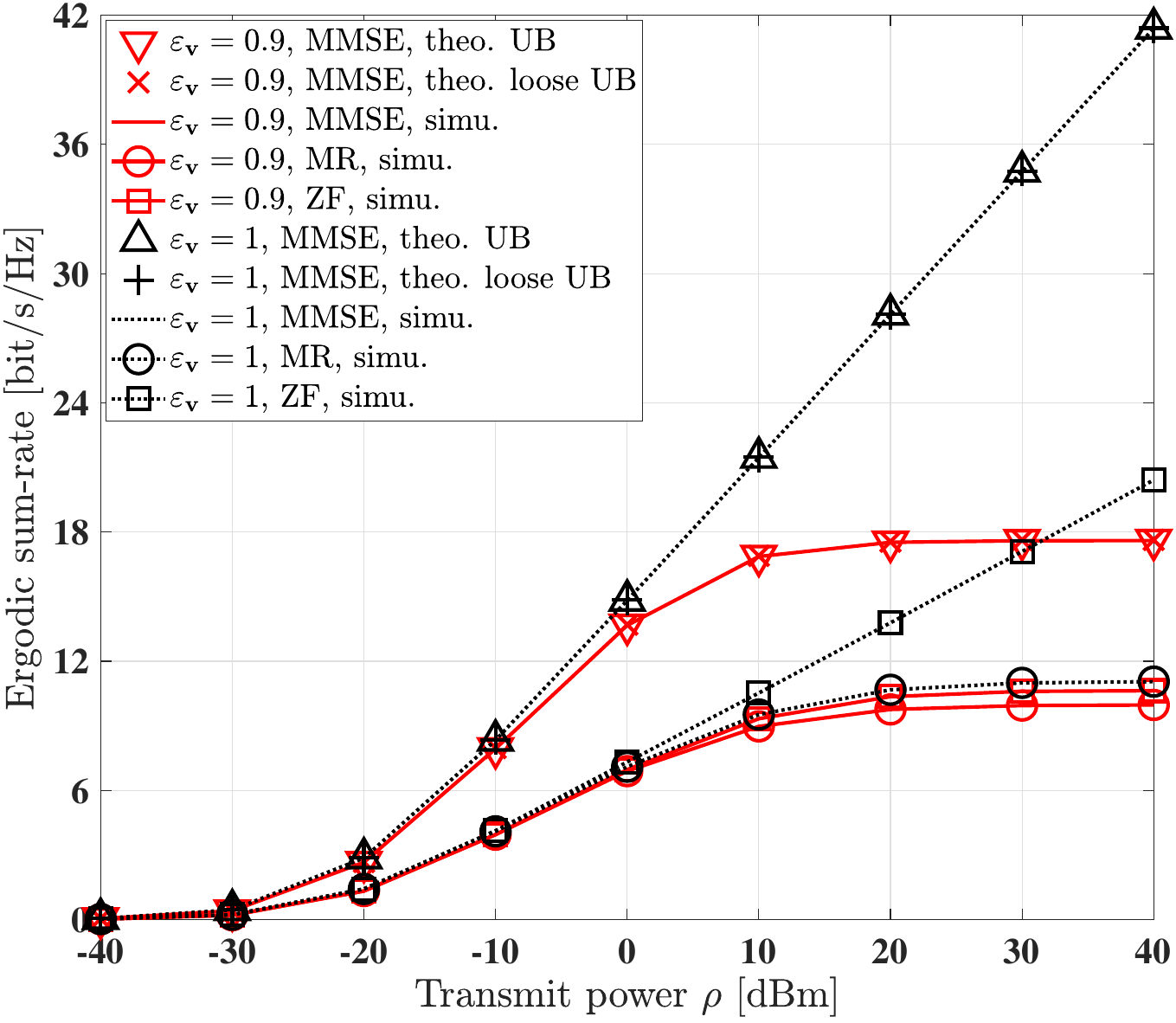}
    \end{minipage}}\\
    \subfloat[The AP hardware quality factor $\varepsilon_\mathbf{v}=1$]
    {\begin{minipage}{1\linewidth}
        \centering
        \includegraphics[width=2.8in]{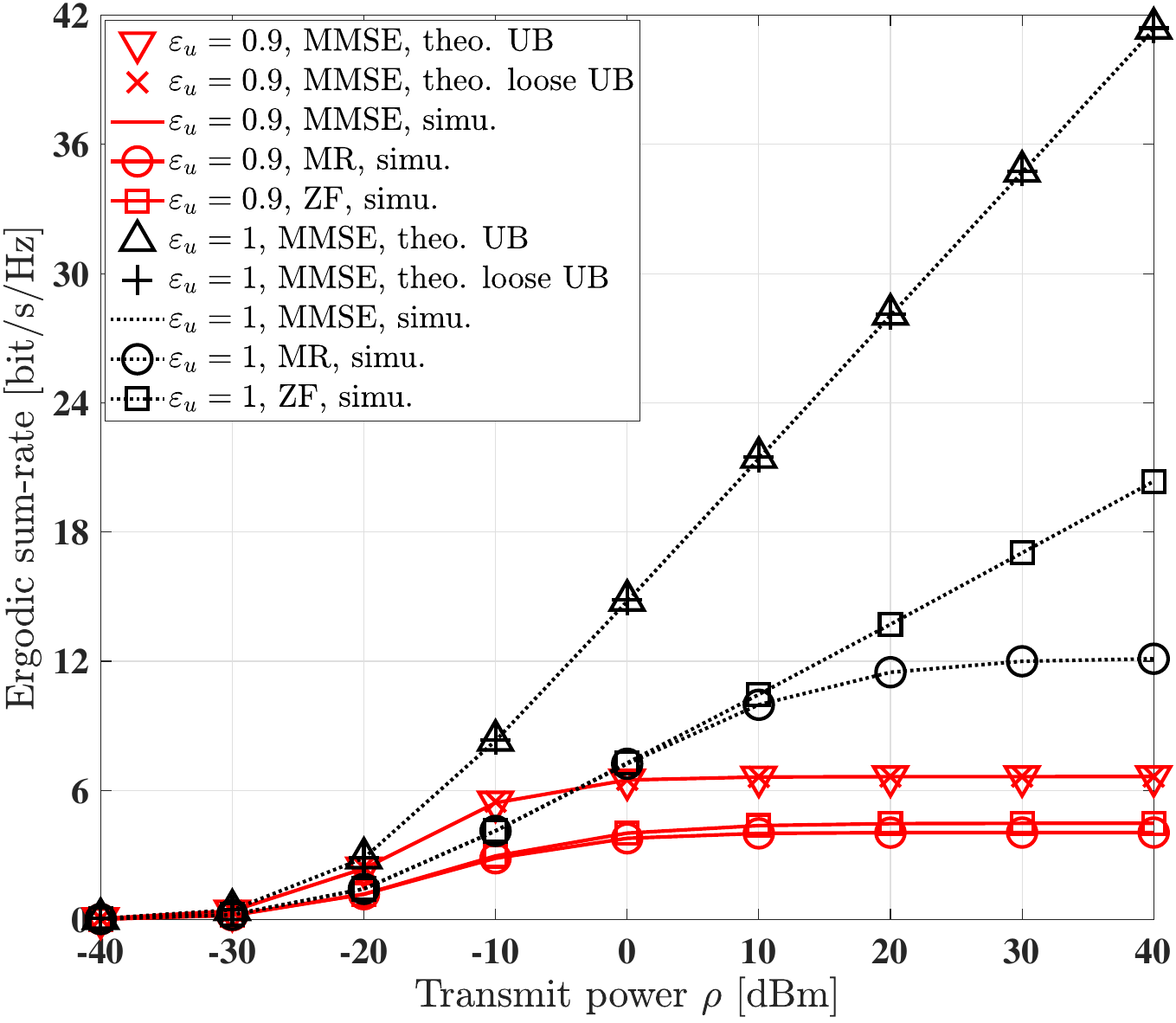}
    \end{minipage}}
    \caption{Theoretical (\ref{Beamforming_Design_9}), (\ref{Beamforming_Design_19}) and simulation comparison of the ergodic sum-rate versus the transmit power in various combining methods.}\label{Fig_simu_SE_12}
\end{figure}

Fig.~\ref{Fig_simu_SE_12} compares the ergodic sum-rate versus the transmit power $\rho$ for various combining methods, including the MMSE, maximum ratio (MR) and zero-forcing (ZF) techniques. Apart from the simulation results, the theoretical upper bound (theo. UB) in (\ref{Beamforming_Design_9}) and the loose theoretical upper bound (theo. loose UB) in (\ref{Beamforming_Design_19}) of the MMSE combining method are also presented for comparison. Fig.~\ref{Fig_simu_SE_12} (a) shows that the MMSE combiner promises higher ergodic sum-rate than the MR and ZF combiners. Observe that for the MR method, the design of combining vector for the UE-R only employs the information of the estimated channel, but ignores all other information, including the statistical channel estimation error, the interference arriving from the UE-T, the effect of HWIs and the additive noise at the AP. On the other hand, in the design of the combining vector for UE-R based on the ZF method, the hostile interference emanating from the UE-T is eliminated, but the effects of HWIs, of the channel estimation error and of the additive noise at the AP are amplified. By contrast, the MMSE method comprehensively considers the statistical information concerning the channel estimation error, the inter-user interference, the HWIs and the additive noise at the AP in the design of combining vectors. Observe that when $\varepsilon_u=\varepsilon_\mathbf{v}=1$, the ergodic sum-rate increase linearly with the transmit power, while when the AP hardware quality factor obeys $\varepsilon_\mathbf{v}<1$, the ergodic sum-rate tends to a constant in the high transmit power region, since the signal distortion at the AP is amplified upon increasing of the transmit power. Fig.~\ref{Fig_simu_SE_12} (b) compares the ergodic sum-rate versus the transmit power $\rho$ for the MMSE, MR and ZF, under different UE hardware quality factors $\varepsilon_{u}$, along with an ideal AP hardware, i.e. for $\varepsilon_\mathbf{v}=1$. Observe that upon increasing the transmit power, the ergodic sum-rate of the MMSE method approximately equals $\log_2\left(1+\frac{\varepsilon_{u,\mathrm{r}}}{1-\varepsilon_
{u,\mathrm{r}}}\right)+\log_2\left(1+\frac{\varepsilon_{u,\mathrm{t}}}{1-\varepsilon_
{u,\mathrm{t}}}\right)=\log_2\left(1+\frac{0.9}{1-0.9}\right)+\log_2\left(1+\frac{0.9}{1-0.9}\right)\thickapprox6.644$, which verifies our theoretical analysis.

\begin{figure}[!t]
    \centering
    \includegraphics[width=2.8in]{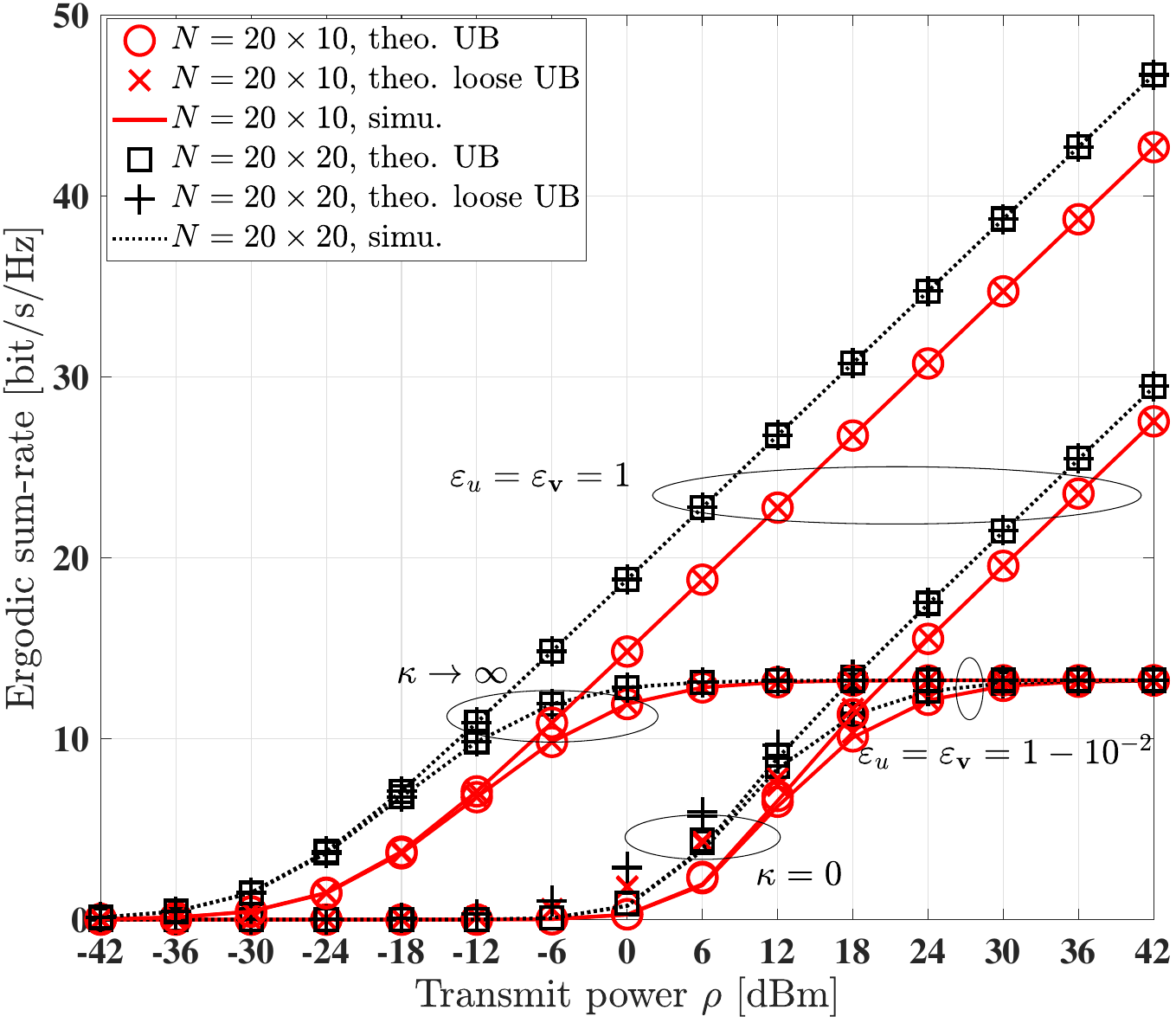}
    \caption{Theoretical (\ref{Beamforming_Design_9}), (\ref{Beamforming_Design_19}) and simulation comparison of the ergodic sum-rate versus the transmit power under different number of IOS elements, hardware quality factor and Rician factor.}\label{Fig_simu_SE_3}
\end{figure}

\begin{figure}[!t]
    \centering
    \includegraphics[width=2.8in]{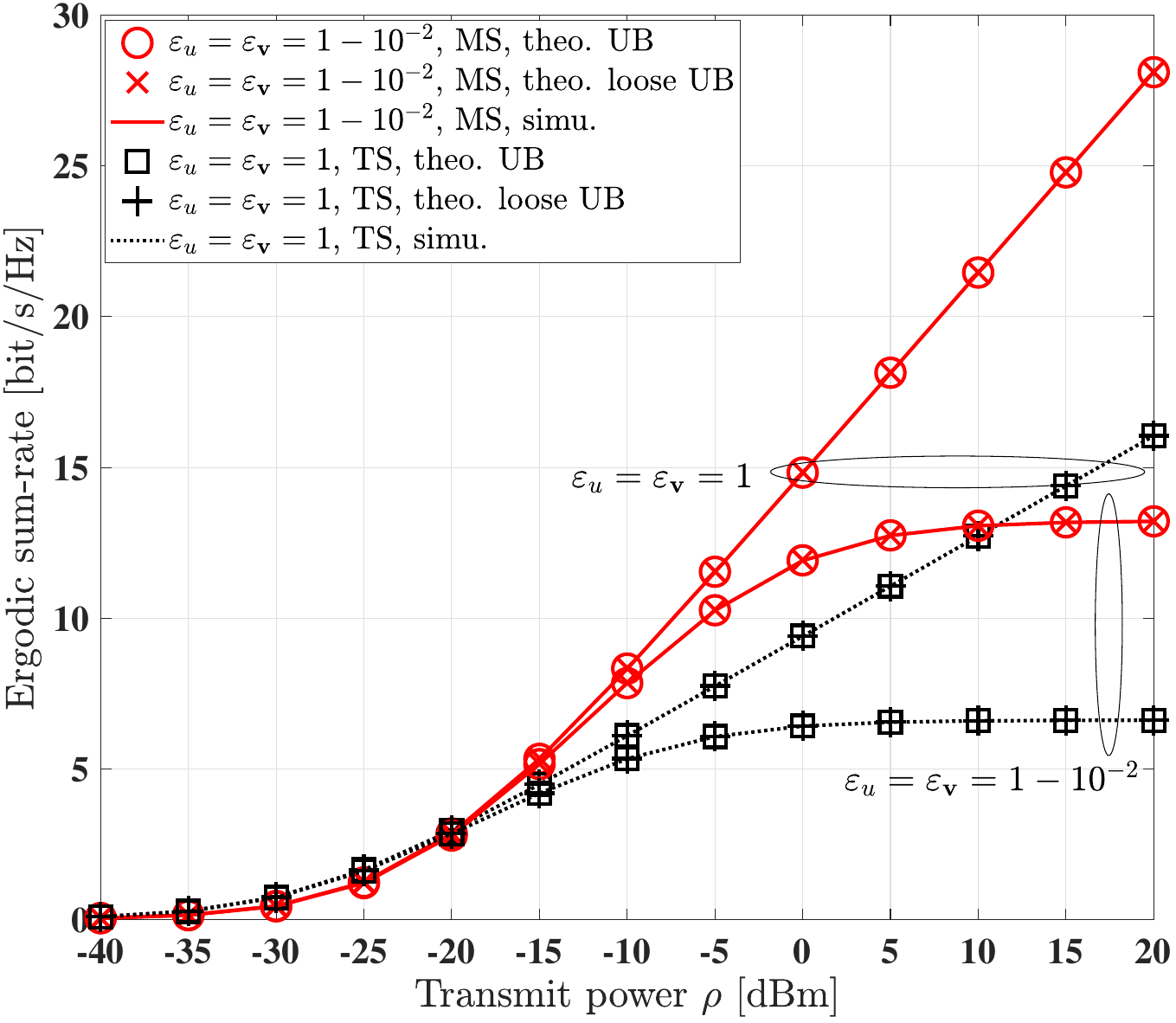}
    \caption{Theoretical (\ref{Beamforming_Design_9}), (\ref{Beamforming_Design_19}) and simulation comparison of the ergodic sum-rate versus the transmit power under different hardware quality factor for the IOS-assisted systems based on the MS protocol and the TS protocol, respectively.}\label{Fig_simu_SE_4}
\end{figure}

\begin{figure}[!t]
    \centering
    \includegraphics[width=2.8in]{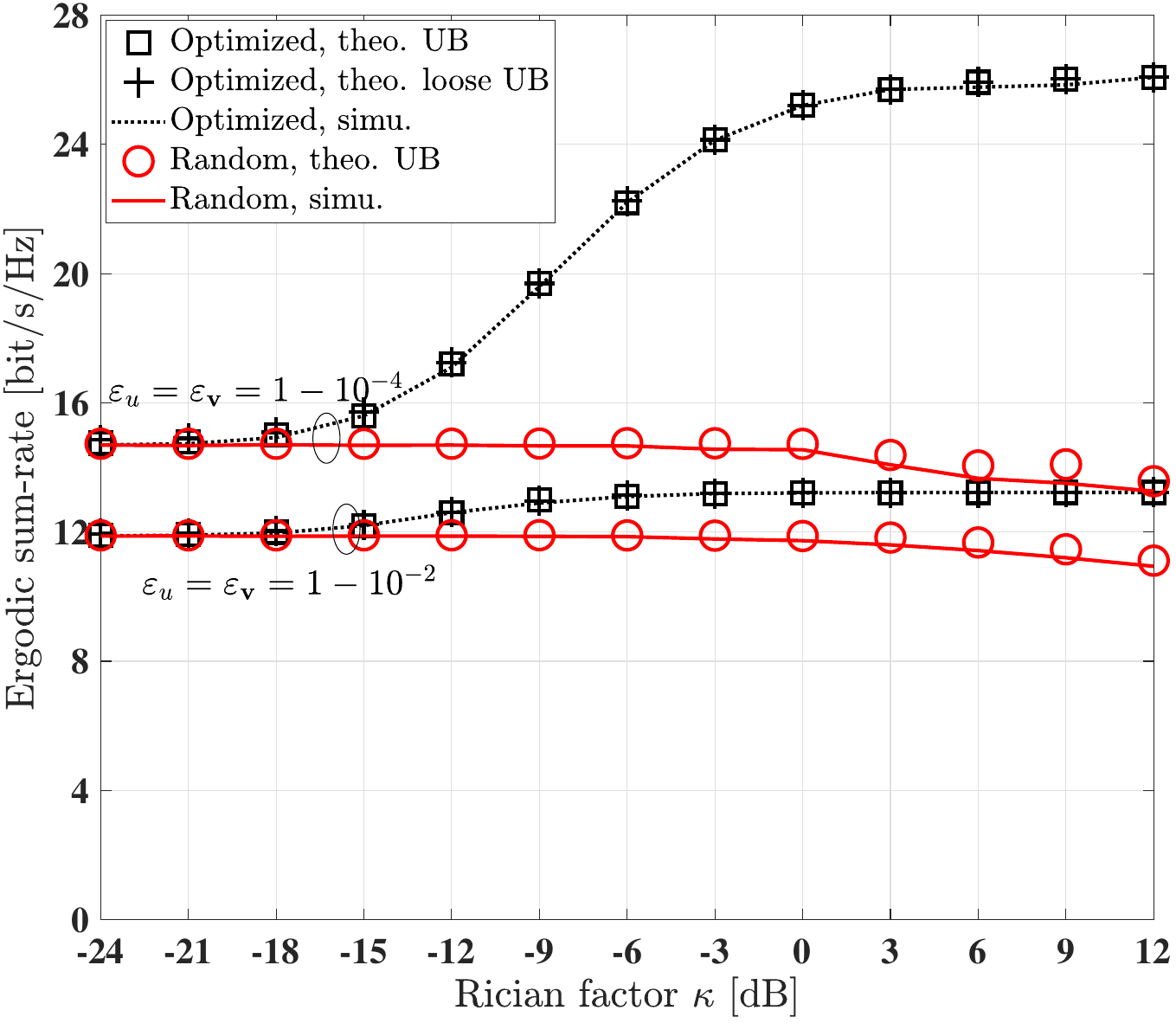}
    \caption{Theoretical (\ref{Beamforming_Design_9}), (\ref{Beamforming_Design_19}) and simulation comparison of the ergodic sum-rate versus the Rician factor.}\label{Fig_simu_SE_5}
\end{figure}

Fig.~\ref{Fig_simu_SE_3} compares the ergodic sum-rate versus the transmit power $\rho$ for different number of the IOS elements, hardware quality factors and Rician factors, with the phase shift of IOS elements being optimally designed based on (\ref{Beamforming_Design_16}). Observe that when $\varepsilon_{\mathbf{v}}=\varepsilon_{u}=1$, the ergodic sum-rate tends to infinity upon increasing the transmit power. However, when $\varepsilon_{\mathbf{v}}=\varepsilon_{u}=1-10^{-2}$, the ergodic sum-rate saturates at $2\log_2\left(1+\frac{1-10^{-2}}{10^{-2}}\right)\thickapprox13.29$. Furthermore, it is worth noting that for realistic HWIs, increasing the number of IOS elements cannot increase the ergodic sum-rate in the high transmit power region since it is limited by the hardware quality of UE-R and UE-T. Besides, when the channel is dominated by LoS components, i.e. $\kappa\rightarrow\infty$, doubling the number of IOS elements can bring approximately $6\text{dB}$ transmit power gain. By contrast, when the channel is dominated by NLoS components, i.e. $\kappa=0$, doubling the number of IOS elements can bring approximately $3\text{dB}$ transmit power gain.

Fig.~\ref{Fig_simu_SE_4} compares the ergodic sum-rate versus the transmit power $\rho$ under different hardware quality factors for the IOS-assisted systems based on the MS protocol and the TS protocol, respectively. The ergodic sum-rate based on the MS protocol is almost 2 times that based on the TS protocol. This is due to the fact that the UE-T and the UE-R are allocated to the orthogonal time resources of the TS protocol, while they are instantaneously supported within the same time resource in the MS protocol.

Fig.~\ref{Fig_simu_SE_5} compares the ergodic sum-rate versus the Rician factor $\kappa$ under different hardware quality factors. Observe that in the low Rician factor region, i.e. when the NLoS components are dominant, the ergodic sum-rate associated with the optimal IOS phase shift is almost the same as that for random IOS phase shifts. By contrast, in the high Rician factor region, i.e. when the LoS components are dominant, the ergodic sum-rate associated with optimal IOS phase shifts is significantly better than that for random IOS phase shift, since the IOS phase shift is purely designed based on the LoS components.

\begin{figure}[!t]
    \centering
    \subfloat[The UE hardware quality factor $\varepsilon_{u}=1$]
    {\begin{minipage}{1\linewidth}
        \centering
        \includegraphics[width=2.8in]{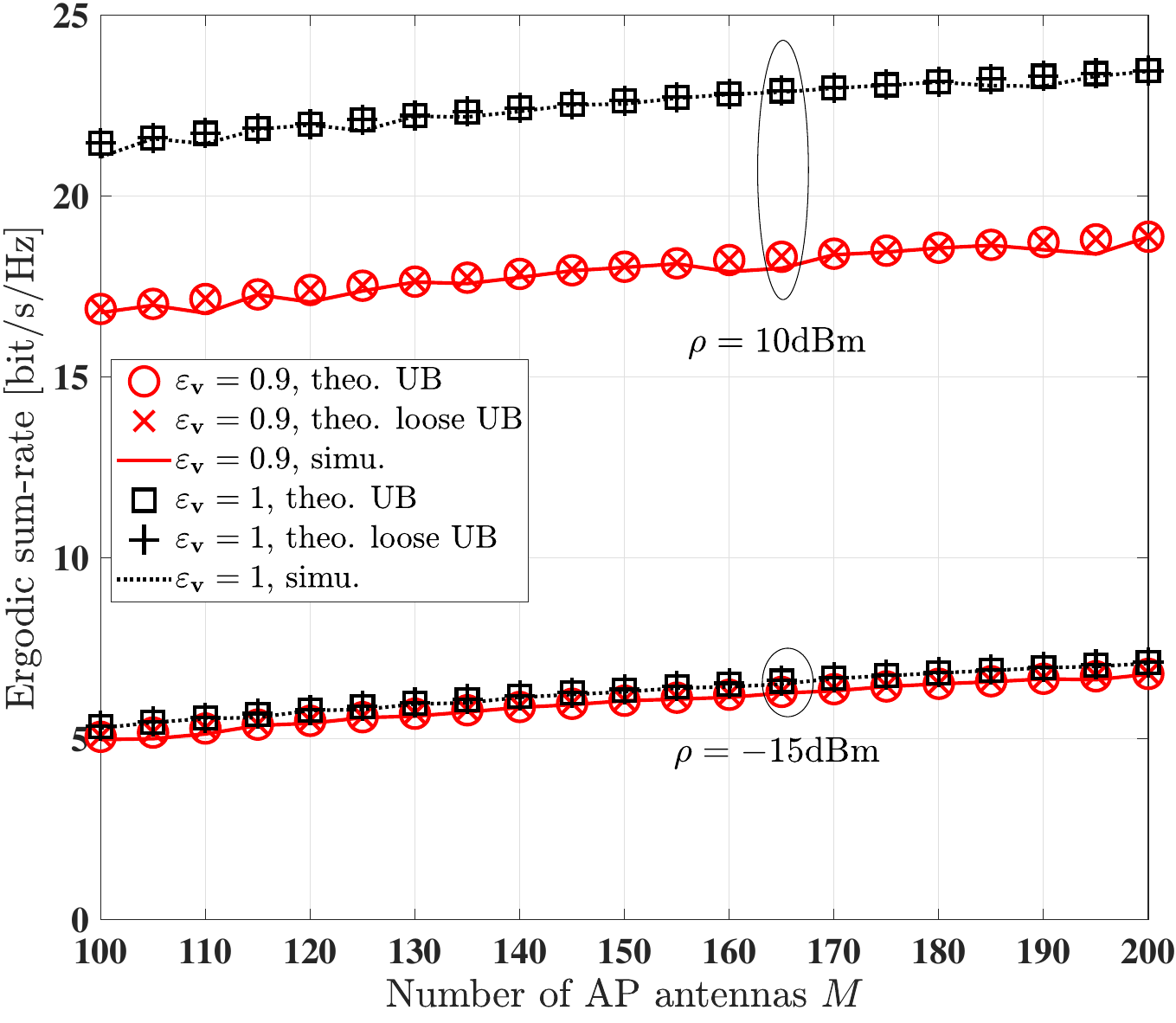}
    \end{minipage}}\\
    \subfloat[The AP hardware quality factor $\varepsilon_\mathbf{v}=1$]
    {\begin{minipage}{1\linewidth}
        \centering
        \includegraphics[width=2.8in]{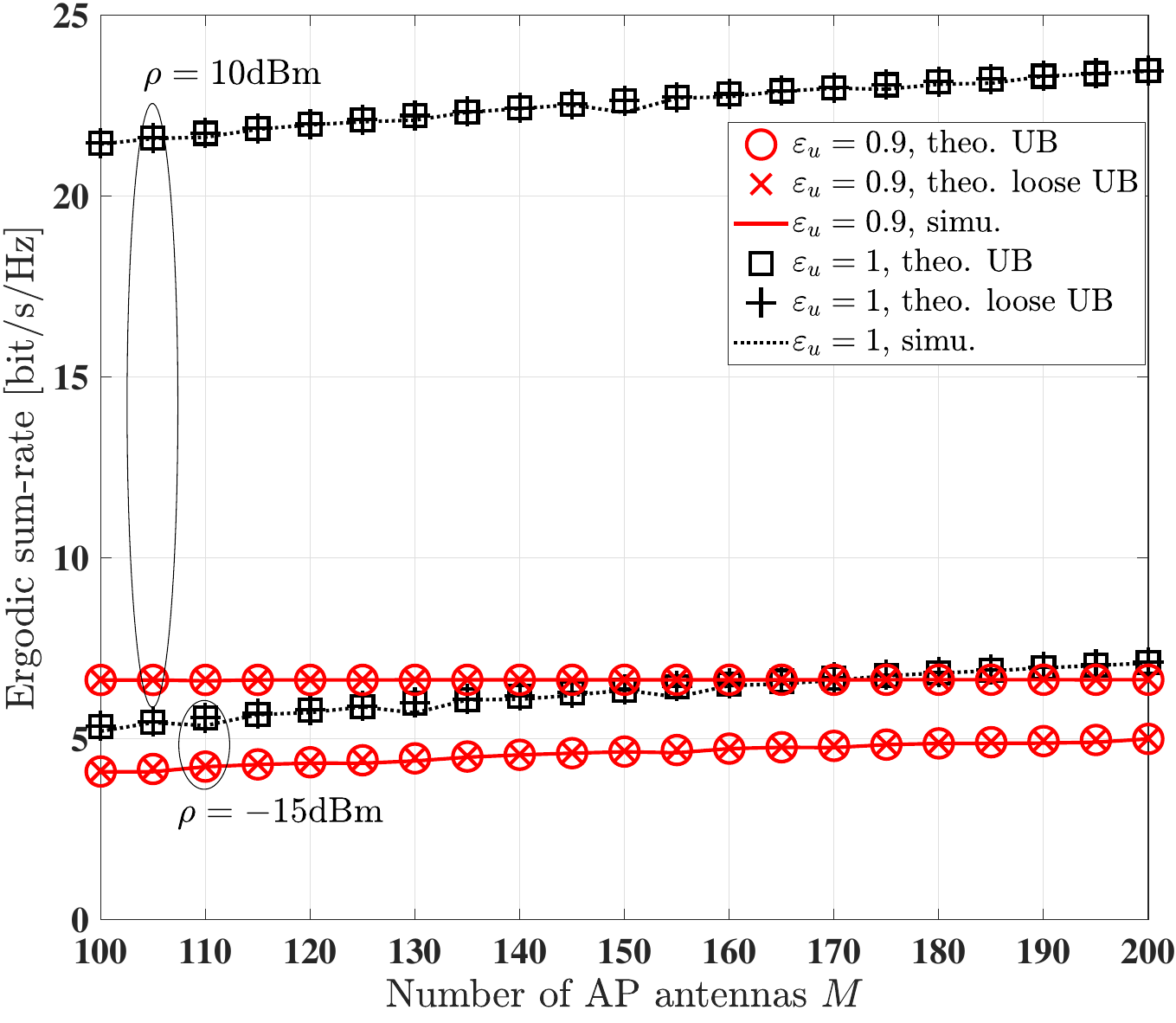}
    \end{minipage}}
    \caption{Theoretical (\ref{Beamforming_Design_9}), (\ref{Beamforming_Design_19}) and simulation comparison of the ergodic sum-rate versus the number of AP antennas $M$.}\label{Fig_simu_SE_67}
\end{figure}

Fig.~\ref{Fig_simu_SE_67} (a) compares the ergodic sum-rate versus the number of AP antennas $M$ for different transmit powers and AP hardware quality factors, with the UE hardware quality factor set to $\varepsilon_{u}=1$. Observe that in the low transmit power region, such as $\rho=-15\text{dBm}$, employing slightly more AP antennas efficiently compensates the spectral efficiency degradation resulting from HWIs. For example, the ergodic sum-rate for $M=110$ AP antennas with the AP hardware quality factor of $\varepsilon_{\mathbf{v}}=0.9$ is almost the same as that for $M=100$ AP antennas with ideal AP hardware. In the high transmit power region, such as $\rho=10\text{dBm}$, using more AP antennas can also compensate the spectral efficiency degradation resulting from HWIs, but more AP antennas are required than in the low transmit power region. Fig.~\ref{Fig_simu_SE_67} (b) compares the ergodic sum-rate versus the number of AP antennas $M$ for different transmit power regions ($\rho=10\text{dBm}$ and $\rho=-15\text{dBm}$) and UE hardware quality factors, along with the AP hardware quality factor of $\varepsilon_{\mathbf{v}}=1$. In contrast to the systems in Fig.~\ref{Fig_simu_SE_67} (a) having non-ideal AP hardware and ideal UE hardware, employing more AP antennas has limited benefit in terms of compensating for the UE hardware impairments. For example, in the low transmit power region of $\rho=-15\text{dBm}$, the ergodic sum-rate for $M=200$ AP antennas and the UE hardware quality factor of $\varepsilon_{u}=0.9$ is still lower than that for $M=100$ AP antennas and ideal UE hardware. Furthermore, in the high transmit power region of $\rho=10\text{dBm}$, the ergodic sum-rate for the UE hardware quality factor of $\varepsilon_{u}=0.9$ has a large gap compared to that associated with ideal UE hardware. As a further observation, using more AP antennas fails to compensate for the UE HWIs.

\begin{figure}[!t]
    \centering
    \includegraphics[width=2.8in]{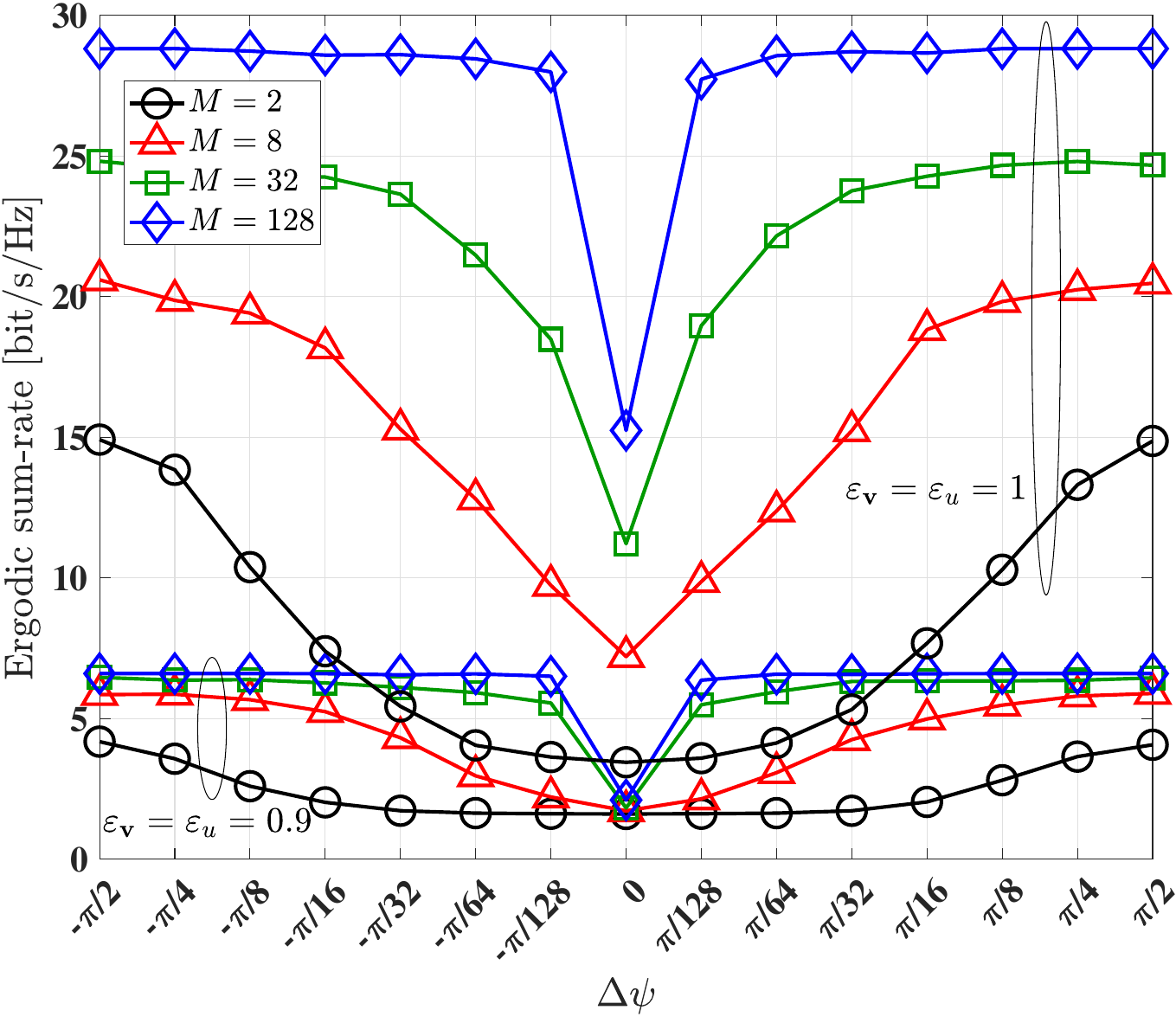}
    \caption{Simulation comparison of the ergodic sum-rate versus the AoA difference at the AP $\Delta\psi=\psi_\mathrm{r}-\psi_\mathrm{t}$.}\label{Fig_simu_SE_8}
\end{figure}

Fig.~\ref{Fig_simu_SE_8} presents the simulation results of the ergodic sum-rate versus the AoA difference $\Delta\psi=\psi_\mathrm{r}-\psi_\mathrm{t}$ at the AP for different number of AP antennas and UE hardware quality factors. Observe that the ergodic sum-rate is improved upon increasing the AoA difference at the AP. Furthermore, employing more AP antennas allows the IOS-based wireless system to accommodate a wider AoA difference range. For example, although when the AoA difference at the AP is small, such as $\Delta\psi=\pm\pi/128$, the ergodic sum-rate can be improved upon using more AP antennas.

\begin{figure}[!t]
    \centering
    \includegraphics[width=2.8in]{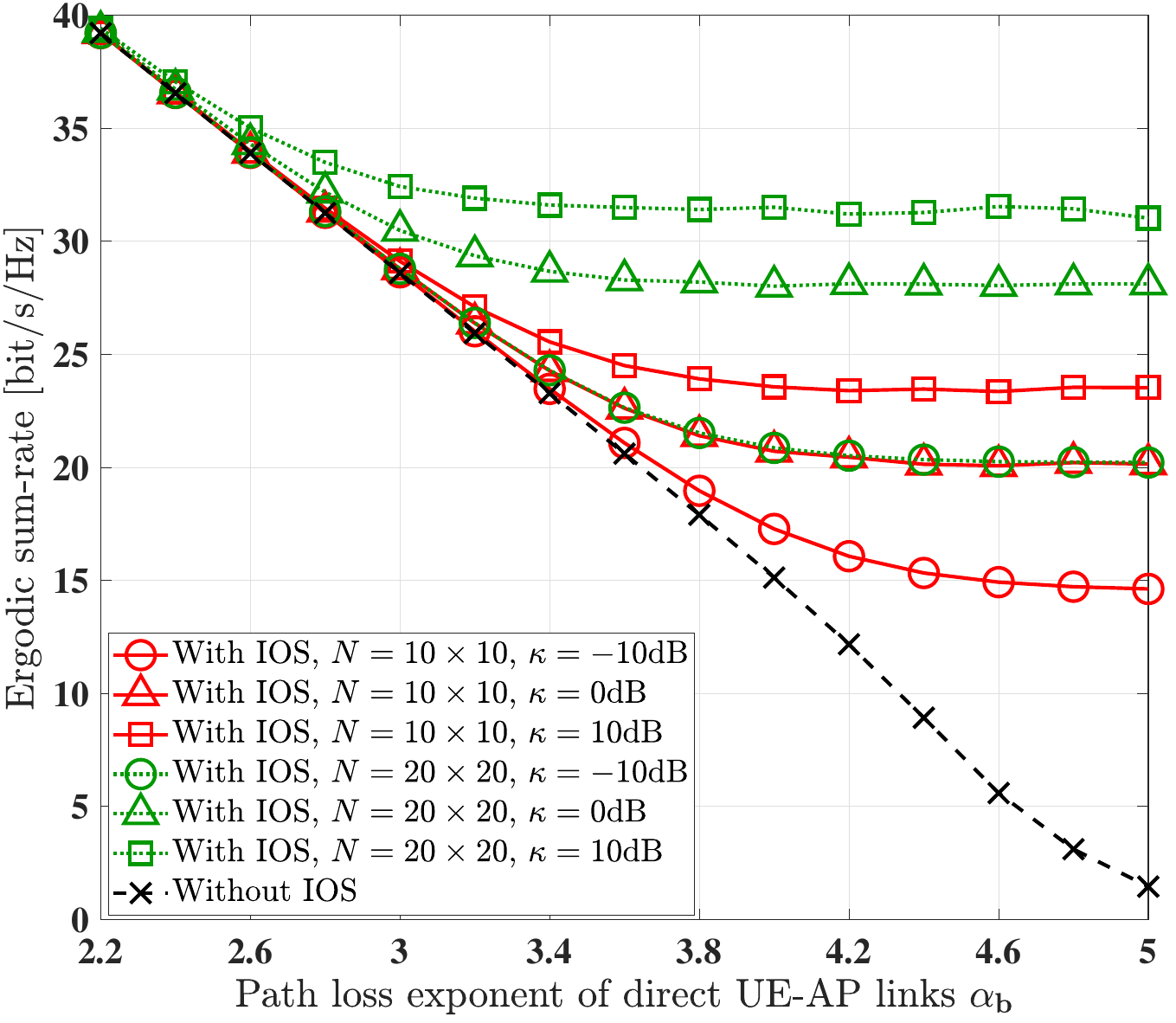}
    \caption{Simulation comparison of the ergodic sum-rate versus the path loss exponent of direct UE-AP links $\alpha_{\mathbf{b}}$ for the IOS-assisted system and the conventional system without IOS, respectively.}\label{Fig_simu_SE_9}
\end{figure}

Fig.~\ref{Fig_simu_SE_9} presents the simulation results of the ergodic sum-rate versus the path loss exponent of the direct UE-AP links $\alpha_{\mathbf{b}}$ for the IOS-assisted system and the conventional system without IOS, respectively. Observe that when the IOS is absent, the ergodic sum-rate degrades with the increase of the path loss exponent of the direct UE-AP links. Furthermore, when the path loss exponent of the direct UE-AP links is small, the ergodic sum-rate performance cannot be substantially improved by deploying the IOS elements, which is due to the fact that the direct UE-AP channel links are more reliable than that created by the IOS. By contrast, when the path loss exponent of direct UE-AP links is high, deploying the IOS elements is beneficial for the improvement of the ergodic sum-rate, since reliable channel links are created by the IOS elements, even though the signals suffer from severe attenuation in the direct UE-AP links.

\section{Conclusions}\label{Conclusion}
The ergodic spectral efficiency of the IOS-aided MIMO uplink associated with imperfect CSI and transceiver HWIs was characterized. Firstly the LMMSE estimator of the equivalent channel spanning from the UEs to the AP was derived, in the face of transceiver HWIs. Next, a two-timescale protocol was conceived for the joint beamformer design of IOS-aided systems. At the AP, an MMSE combiner was employed based on the estimated equivalent channels, as well as on the statistical channel estimation error, on the inter-user interference and on transceiver HWIs. Then, we derived the closed-form expression of the optimal IOS phase shift based on the statistical CSI for maximizing the upper bound of the ergodic spectral efficiency. Our theoretical analysis and numerical results show that the transceiver HWIs have a grave deleterious impact on the ergodic spectral efficiency. Employing more AP antennas can compensate for the HWIs at the AP, but the HWI at the UEs cannot be compensated by using more AP antennas or by increasing the transmit power.

\appendices
\section{Proof of Theorem~\ref{theorem_1}}\label{Appendix_A}
In (\ref{Channel_Model_24}), based on the independence of $s_\mathrm{r}$, $s_\mathrm{t}$, $u_\mathrm{r}$, $u_\mathrm{t}$, $\mathbf{v}_\mathrm{r}$, $\mathbf{v}_\mathrm{t}$ and $\mathbf{w}$, the auto-correlation matrix of $\mathbf{y}$ can be formulated as
\begin{align}\label{Proof_A_1}
    \notag\mathbb{E}\left[\mathbf{y}\mathbf{y}^{\mathrm{H}}\right]
    &=\rho_\mathrm{r}\varepsilon_\mathbf{v}\varepsilon_{u,\mathrm{r}}
    \hat{\mathbf{h}}_\mathrm{r}\hat{\mathbf{h}}_\mathrm{r}^{\mathrm{H}}
    +\rho_\mathrm{t}\varepsilon_\mathbf{v}\varepsilon_{u,\mathrm{t}}
    \hat{\mathbf{h}}_\mathrm{t}\hat{\mathbf{h}}_\mathrm{t}^{\mathrm{H}}
    +\rho_\mathrm{r}\varepsilon_\mathbf{v}\varepsilon_{u,\mathrm{r}}
    \mathbf{C}_{\check{\mathbf{h}}_\mathrm{r}\check{\mathbf{h}}_\mathrm{r}}\\
    \notag&+\rho_\mathrm{t}\varepsilon_\mathbf{v}\varepsilon_{u,\mathrm{t}}
    \mathbf{C}_{\check{\mathbf{h}}_\mathrm{t}\check{\mathbf{h}}_\mathrm{t}}
    +\rho_\mathrm{r}\varepsilon_\mathbf{v}\left(1-\varepsilon_{u,\mathrm{r}}\right)
    (\hat{\mathbf{h}}_\mathrm{r}\hat{\mathbf{h}}_\mathrm{r}^{\mathrm{H}}
    +\mathbf{C}_{\check{\mathbf{h}}_\mathrm{r}\check{\mathbf{h}}_\mathrm{r}})\\
    \notag&+\rho_\mathrm{t}\varepsilon_\mathbf{v}\left(1-\varepsilon_{u,\mathrm{t}}\right)
    \left(\hat{\mathbf{h}}_\mathrm{t}\hat{\mathbf{h}}_\mathrm{t}^{\mathrm{H}}
    +\mathbf{C}_{\check{\mathbf{h}}_\mathrm{t}\check{\mathbf{h}}_\mathrm{t}}\right)
    +\left(\rho_\mathrm{r}\left(1-\varepsilon_\mathbf{v}\right)\cdot\right.\\
    \notag&\left.\mathbf{E}\left[\mathbf{h}_\mathrm{r}\mathbf{h}_\mathrm{r}^{\mathrm{H}}\right]
    +\rho_\mathrm{t}\left(1-\varepsilon_\mathbf{v}\right)
    \mathbf{E}\left[\mathbf{h}_\mathrm{t}\mathbf{h}_\mathrm{t}^{\mathrm{H}}\right]\right)\odot\mathbf{I}_M
    +\sigma_w^2\mathbf{I}_M\\
    &\overset{(\text{a})}=\rho_\mathrm{r}\varepsilon_\mathbf{v}\varepsilon_{u,\mathrm{r}}
    \hat{\mathbf{h}}_\mathrm{r}\hat{\mathbf{h}}_\mathrm{r}^{\mathrm{H}}
    +\rho_\mathrm{t}\varepsilon_\mathbf{v}\varepsilon_{u,\mathrm{t}}
    \hat{\mathbf{h}}_\mathrm{t}\hat{\mathbf{h}}_\mathrm{t}^{\mathrm{H}}+\mathbf{R},
\end{align}
where (a) is based on (\ref{Channel_Model_14}), (\ref{Channel_Model_15}) and (\ref{Beamforming_Design_2_1}). The MMSE combining vector and the SINR of UE-R are given by
\begin{align}\label{Proof_A_8}
    \mathbf{q}_\mathrm{r}=\rho_\mathrm{r}\varepsilon_{\mathbf{v}}\varepsilon_{u,\mathrm{r}}
    \left(\mathbb{E}\left[\mathbf{y}\mathbf{y}^\mathrm{H}\right]\right)^{-1}\hat{\mathbf{h}}_\mathrm{r},
\end{align}
\begin{align}\label{Proof_A_9}
    \gamma_\mathrm{r}=\rho_\mathrm{r}\varepsilon_{\mathbf{v}}\varepsilon_{u,\mathrm{r}}
    \hat{\mathbf{h}}_\mathrm{r}^{\mathrm{H}}\left(\mathbb{E}[\mathbf{y}\mathbf{y}^\mathrm{H}]
    -\rho_\mathrm{r}\varepsilon_\mathbf{v}\varepsilon_{u,\mathrm{r}}
    \hat{\mathbf{h}}_\mathrm{r}\hat{\mathbf{h}}_\mathrm{r}^{\mathrm{H}}\right)^{-1}
    \hat{\mathbf{h}}_\mathrm{r}.
\end{align}
According to (\ref{Proof_A_1}), (\ref{Proof_A_8}) and (\ref{Proof_A_9}), we can arrive at (\ref{Beamforming_Design_1}) and (\ref{Beamforming_Design_3}). Furthermore, the MMSE combining vector and the SINR of UE-T in (\ref{Beamforming_Design_2}) and (\ref{Beamforming_Design_4}) can be similarly evaluated.

\section{Proof of Theorem~\ref{theorem_2}}\label{Appendix_B}
\begin{lemma}\label{lemma_1}
For the matrices $\mathbf{D}\in\mathbb{C}^{L_1{\times}L_1}$, $\mathbf{F}_1\in\mathbb{C}^{L_1{\times}L_2}$, $\mathbf{F}_2\in\mathbb{C}^{L_2{\times}L_2}$, $\mathbf{F}_3\in\mathbb{C}^{L_2{\times}L_1}$, the inverse of $\mathbf{D}+\mathbf{F}_1\mathbf{F}_2\mathbf{F}_3$ is $\mathbf{D}^{-1}-\mathbf{D}^{-1}\mathbf{F}_1\left(\mathbf{F}_3\mathbf{D}^{-1}
\mathbf{F}_1+\mathbf{F}_2^{-1}\right)^{-1}\mathbf{F}_3\mathbf{D}^{-1}$~\cite{zhang2017matrix}.
\end{lemma}

Based on Lemma~\ref{lemma_1}, letting $L_1=M$, $L_2=1$, $L_3=M$, $\mathbf{D}=\rho_\mathrm{t}\varepsilon_{\mathbf{v}}\hat{\mathbf{h}}_\mathrm{t}
\hat{\mathbf{h}}_\mathrm{t}^{\mathrm{H}}+\mathbf{R}$, $\mathbf{F}_1=\sqrt{\rho_\mathrm{r}\varepsilon_{\mathbf{v}}
\left(1-\varepsilon_{u,\mathrm{r}}\right)}\hat{\mathbf{h}}_\mathrm{r}$, $\mathbf{F}_2=1$ and $\mathbf{F}_3=\sqrt{\rho_\mathrm{r}\varepsilon_{\mathbf{v}}
\left(1-\varepsilon_{u,\mathrm{r}}\right)}\hat{\mathbf{h}}_\mathrm{r}^{\mathrm{H}}$ for (\ref{Beamforming_Design_3}), we can formulate $\gamma_\mathrm{r}$ as
\begin{align}\label{Proof_B_2}
    \gamma_\mathrm{r}=&\frac{\rho_\mathrm{r}\varepsilon_{\mathbf{v}}
    \varepsilon_{u,\mathrm{r}}\hat{\mathbf{h}}_\mathrm{r}^{\mathrm{H}}
    \left(\rho_\mathrm{t}\varepsilon_{\mathbf{v}}\hat{\mathbf{h}}_\mathrm{t}
    \hat{\mathbf{h}}_\mathrm{t}^{\mathrm{H}}+\mathbf{R}\right)^{-1}\hat{\mathbf{h}}_\mathrm{r}}
    {1+\rho_\mathrm{r}\varepsilon_{\mathbf{v}}\left(1-\varepsilon_{u,\mathrm{r}}\right)
    \hat{\mathbf{h}}_\mathrm{r}^{\mathrm{H}}
    \left(\rho_\mathrm{t}\varepsilon_{\mathbf{v}}\hat{\mathbf{h}}_\mathrm{t}
    \hat{\mathbf{h}}_\mathrm{t}^{\mathrm{H}}
    +\mathbf{R}\right)^{-1}\hat{\mathbf{h}}_\mathrm{r}}.
\end{align}
Again, based on Lemma~\ref{lemma_1}, letting $L_1=M$, $L_2=1$, $L_3=M$, $\mathbf{D}=\mathbf{R}$, $\mathbf{F}_1=\sqrt{\rho_\mathrm{t}\varepsilon_{\mathbf{v}}}\hat{\mathbf{h}}_\mathrm{t}$, $\mathbf{F}_2=1$ and $\mathbf{F}_3=\sqrt{\rho_\mathrm{t}\varepsilon_{\mathbf{v}}}
\hat{\mathbf{h}}_\mathrm{t}^{\mathrm{H}}$ for (\ref{Proof_B_2}), we have $\gamma_\mathrm{r}=\frac{\rho_\mathrm{r}\varepsilon_\mathbf{v}
\varepsilon_{u,\mathrm{r}}\zeta_\mathrm{r}}
{1+\rho_\mathrm{r}\varepsilon_\mathbf{v}\left(1-\varepsilon_{u,\mathrm{r}}\right)\zeta_\mathrm{r}}$. Similarly, we can get $\gamma_\mathrm{t}=\frac{\rho_\mathrm{t}\varepsilon_\mathbf{v}
\varepsilon_{u,\mathrm{t}}\zeta_\mathrm{t}}{1+\rho_\mathrm{t}\varepsilon_\mathbf{v}
\left(1-\varepsilon_{u,\mathrm{t}}\right)\zeta_\mathrm{t}}$. Therefore, we can arrive at the instantaneous ergodic spectral efficiency of UE-R and UE-T in (\ref{Beamforming_Design_5}).

\section{Proof of Theorem~\ref{theorem_3}}\label{Appendix_C}
Since $f(\zeta_\mathrm{i})=\log_2\left(1+\frac{\rho_\mathrm{i}\varepsilon_\mathbf{v}
\varepsilon_{u,\mathrm{i}}\zeta_\mathrm{i}}{1+\rho_\mathrm{i}\varepsilon_\mathbf{v}
\left(1-\varepsilon_{u,\mathrm{i}}\right)\zeta_\mathrm{i}}\right)$ is a concave function, according to Jensen's inequality we can get
\begin{align}\label{Beamforming_Design_11_1}
    \mathcal{R}_{\mathrm{erg},\mathrm{i}}
    \leq\log_2\left(1+\frac{\rho_\mathrm{i}\varepsilon_\mathbf{v}\varepsilon_{u,\mathrm{i}}
    \mathbb{E}[\zeta_\mathrm{i}]}
    {1+\rho_\mathrm{i}\varepsilon_\mathbf{v}\left(1-\varepsilon_{u,\mathrm{i}}\right)
    \mathbb{E}[\zeta_\mathrm{i}]}\right).
\end{align}
In (\ref{Beamforming_Design_11_1}), $\mathbb{E}[\zeta_\mathrm{r}]$ and $\mathbb{E}[\zeta_\mathrm{t}]$ are given by
\begin{align}\label{Beamforming_Design_11_2}
    \notag\mathbb{E}[\zeta_\mathrm{r}]
    =&\mathbb{E}\left[\hat{\mathbf{h}}_\mathrm{r}^{\mathrm{H}}\mathbf{R}^{-1}\hat{\mathbf{h}}_\mathrm{r}\right]
    -\mathbb{E}\left[\frac{\rho_\mathrm{t}\varepsilon_\mathbf{v}
    \hat{\mathbf{h}}_\mathrm{r}^{\mathrm{H}}
    \mathbf{R}^{-1}\hat{\mathbf{h}}_\mathrm{t}\mathbf{R}^{-1}
    \hat{\mathbf{h}}_\mathrm{t}^{\mathrm{H}}\hat{\mathbf{h}}_\mathrm{r}}
    {1+\rho_\mathrm{t}\varepsilon_\mathbf{v}\hat{\mathbf{h}}_\mathrm{r}^{\mathrm{H}}\mathbf{R}^{-1}
    \hat{\mathbf{h}}_\mathrm{r}}\right]\\
    \overset{(\text{a})}{\leq}&\mathbb{E}\left[\hat{\mathbf{h}}_\mathrm{r}^{\mathrm{H}}\mathbf{R}^{-1}
    \hat{\mathbf{h}}_\mathrm{r}\right],
\end{align}
\begin{align}\label{Beamforming_Design_11_3}
    \notag\mathbb{E}[\zeta_\mathrm{t}]
    =&\mathbb{E}\left[\hat{\mathbf{h}}_\mathrm{t}^{\mathrm{H}}\mathbf{R}^{-1}\hat{\mathbf{h}}_\mathrm{t}\right]
    -\mathbb{E}\left[\frac{\rho_\mathrm{r}\varepsilon_\mathbf{v}
    \hat{\mathbf{h}}_\mathrm{t}^{\mathrm{H}}
    \mathbf{R}^{-1}\hat{\mathbf{h}}_\mathrm{r}\mathbf{R}^{-1}
    \hat{\mathbf{h}}_\mathrm{r}^{\mathrm{H}}\hat{\mathbf{h}}_\mathrm{t}}
    {1+\rho_\mathrm{r}\varepsilon_\mathbf{v}\hat{\mathbf{h}}_\mathrm{t}^{\mathrm{H}}\mathbf{R}^{-1}
    \hat{\mathbf{h}}_\mathrm{t}}\right]\\
    \overset{(\text{b})}{\leq}&\mathbb{E}\left[\hat{\mathbf{h}}_\mathrm{t}^{\mathrm{H}}\mathbf{R}^{-1}
    \hat{\mathbf{h}}_\mathrm{t}\right],
\end{align}
where the equalities in (a) and (b) are established when $M\rightarrow\infty$ in conjunction with $\cos\psi_\mathrm{t}\neq\cos\psi_\mathrm{r}$ and the proof is presented in Appendix~\ref{Appendix_D}. Furthermore, it can be shown that when $M\rightarrow\infty$, the derivative of the function $f(\zeta_\mathrm{i})$ with respect to $\zeta_\mathrm{i}$ tends to 0. Therefore, (\ref{Beamforming_Design_9}) is the upper bound of the ergodic spectral efficiency, and upon increasing of $M$, the ergodic spectral efficiency tends to its upper bound. Furthermore, $\mathbb{E}\left[\hat{\mathbf{h}}_\mathrm{r}^{\mathrm{H}}\mathbf{R}^{-1}
\hat{\mathbf{h}}_\mathrm{r}\right]$ in (\ref{Beamforming_Design_11_2}) and $\mathbb{E}\left[\hat{\mathbf{h}}_\mathrm{t}^{\mathrm{H}}\mathbf{R}^{-1}
\hat{\mathbf{h}}_\mathrm{t}\right]$ in (\ref{Beamforming_Design_11_3}) can be exploited to arrive at (\ref{Beamforming_Design_10_1}) based on (\ref{Channel_Model_12}), (\ref{Channel_estimation_6}) and (\ref{Channel_estimation_9}).

\section{Proof of (\ref{Beamforming_Design_11_1}), (\ref{Beamforming_Design_11_2}) and (\ref{Beamforming_Design_11_3})}\label{Appendix_D}
The proof of $\mathbb{E}\left[\hat{\mathbf{h}}_\mathrm{r}^{\mathrm{H}}\mathbf{R}^{-1}
\hat{\mathbf{h}}_\mathrm{r}\right]-\mathbb{E}\left[\frac{\rho_\mathrm{t}\varepsilon_\mathbf{v}
\hat{\mathbf{h}}_\mathrm{r}^{\mathrm{H}}\mathbf{R}^{-1}\hat{\mathbf{h}}_\mathrm{t}
\hat{\mathbf{h}}_\mathrm{t}^{\mathrm{H}}\mathbf{R}^{-1}\hat{\mathbf{h}}_\mathrm{r}}
{1+\rho_\mathrm{t}\varepsilon_\mathbf{v}\hat{\mathbf{h}}_\mathrm{t}^{\mathrm{H}}\mathbf{R}^{-1}
\hat{\mathbf{h}}_\mathrm{t}}\right]$ tends to $\mathbb{E}\left[\hat{\mathbf{h}}
_\mathrm{r}^{\mathrm{H}}\mathbf{R}^{-1}\hat{\mathbf{h}}_\mathrm{r}\right]$ when $M\rightarrow\infty$ and $\cos\psi_\mathrm{t}\neq\cos\psi_\mathrm{r}$ is equivalent to the following proposition:

\begin{Proposition}\label{proposition_1}
    When $M\rightarrow\infty$ and $\cos\psi_\mathrm{t}\neq\cos\psi_\mathrm{r}$, we have
    \begin{align}\label{Proof_C_1}
        \frac{\mathbb{E}\left[\frac{\rho_\mathrm{t}\varepsilon_\mathbf{v}
        \hat{\mathbf{h}}_\mathrm{r}^{\mathrm{H}}
        \mathbf{R}^{-1}\hat{\mathbf{h}}_\mathrm{t}\hat{\mathbf{h}}_\mathrm{t}^{\mathrm{H}}
        \mathbf{R}^{-1}
        \hat{\mathbf{h}}_\mathrm{r}}{1+\rho_\mathrm{t}\varepsilon_\mathbf{v}
        \hat{\mathbf{h}}_\mathrm{t}^{\mathrm{H}}\mathbf{R}^{-1}\hat{\mathbf{h}}_\mathrm{t}}\right]}
        {\mathbb{E}\left[\hat{\mathbf{h}}_\mathrm{r}^{\mathrm{H}}\mathbf{R}^{-1}
        \hat{\mathbf{h}}_\mathrm{r}\right]}=0.
    \end{align}
\end{Proposition}

In (\ref{Proof_C_1}), let $\mathbf{R}=\mathbf{P}\mathbf{\Lambda}\mathbf{P}^{-1}$ represent the eigenvalue decomposition of $\mathbf{R}$ with $\mathbf{P}$ being a unitary matrix and $\mathbf{\Lambda}=\mathbf{Diag}\left\{\lambda_1,\lambda_2,\cdots,\lambda_M\right\}$. Then, upon defining $\hat{\mathbf{h}}_\mathrm{r}'=\sqrt{\mathbf{\Lambda}^{-1}}\mathbf{P}^{-1}
\hat{\mathbf{h}}_\mathrm{r}$, $\hat{\mathbf{h}}_\mathrm{t}'=\sqrt{\mathbf{\Lambda}^{-1}}
\mathbf{P}^{-1}\hat{\mathbf{h}}_\mathrm{t}$, $\overline{\mathbf{h}}_\mathrm{r}'
=\sqrt{\mathbf{\Lambda}^{-1}}\mathbf{P}^{-1}\overline{\mathbf{h}}_\mathrm{r}$ and $\overline{\mathbf{h}}_\mathrm{t}'=\sqrt{\mathbf{\Lambda}^{-1}}\mathbf{P}^{-1}
\overline{\mathbf{h}}_\mathrm{t}$, the numerator of (\ref{Proof_C_1}) can be expressed as
\begin{align}\label{Proof_C_2}
    \notag&\mathbb{E}\left[\frac{\rho_\mathrm{t}\varepsilon_\mathbf{v}
    \hat{\mathbf{h}}_\mathrm{r}^{\mathrm{H}}\mathbf{R}^{-1}\hat{\mathbf{h}}_\mathrm{t}
    \hat{\mathbf{h}}_\mathrm{t}^{\mathrm{H}}\mathbf{R}^{-1}
    \hat{\mathbf{h}}_\mathrm{r}}{1+\rho_\mathrm{t}\varepsilon_\mathbf{v}
    \hat{\mathbf{h}}_\mathrm{t}^{\mathrm{H}}\mathbf{R}^{-1}\hat{\mathbf{h}}_\mathrm{t}}\right]\\
    \notag<&\mathbb{E}\left[\frac{\hat{\mathbf{h}}_\mathrm{r}'^{\mathrm{H}}
    \hat{\mathbf{h}}_\mathrm{t}'\hat{\mathbf{h}}_\mathrm{t}'^{\mathrm{H}}
    \hat{\mathbf{h}}_\mathrm{r}'}
    {\hat{\mathbf{h}}_\mathrm{t}'^{\mathrm{H}}\hat{\mathbf{h}}_\mathrm{t}'}\right]\\
    \notag=&\mathbb{E}
    \left[\left|(\hat{\mathbf{h}}_\mathrm{r}'-\overline{\mathbf{h}}_\mathrm{r}')^{\mathrm{H}}
    \frac{\hat{\mathbf{h}}_\mathrm{t}'}{\left\|\hat{\mathbf{h}}_\mathrm{t}'\right\|}
    +\overline{\mathbf{h}}_\mathrm{r}'^{\mathrm{H}}
    \frac{\hat{\mathbf{h}}_\mathrm{t}'-\overline{\mathbf{h}}_\mathrm{t}'}
    {\left\|\hat{\mathbf{h}}_\mathrm{t}'\right\|}
    +\overline{\mathbf{h}}_\mathrm{r}'^{\mathrm{H}}
    \frac{\overline{\mathbf{h}}_\mathrm{t}'}{\left\|\hat{\mathbf{h}}_\mathrm{t}'\right\|}\right|^2\right]\\
    \notag\overset{(\text{a})}{=}&\mathbb{E}
    \left[\left|\left(\hat{\mathbf{h}}_\mathrm{r}'-\overline{\mathbf{h}}_\mathrm{r}'\right)
    ^{\mathrm{H}}
    \frac{\hat{\mathbf{h}}_\mathrm{t}'}{\left\|\hat{\mathbf{h}}_\mathrm{t}'\right\|}\right|^2\right]
    +\mathbb{E}\left[\left|\overline{\mathbf{h}}_\mathrm{r}'^{\mathrm{H}}
    \frac{\hat{\mathbf{h}}_\mathrm{t}'-\overline{\mathbf{h}}_\mathrm{t}'}
    {\left\|\hat{\mathbf{h}}_\mathrm{t}'\right\|}\right|^2\right]\\
    &+\mathbb{E}\left[\left|\overline{\mathbf{h}}_\mathrm{r}'^{\mathrm{H}}
    \frac{\overline{\mathbf{h}}_\mathrm{t}'}{\|\hat{\mathbf{h}}_\mathrm{t}'\|}\right|^2\right],
\end{align}
where (a) is based on $\mathbb{E}\left[\left(\hat{\mathbf{h}}_\mathrm{r}'-\overline{\mathbf{h}}_\mathrm{r}'\right)^{\mathrm{H}}
\frac{\hat{\mathbf{h}}_\mathrm{t}'}{\|\hat{\mathbf{h}}_\mathrm{t}'\|}
\frac{\hat{\mathbf{h}}_\mathrm{t}'^{\mathrm{H}}}{\left\|\hat{\mathbf{h}}_\mathrm{t}'\right\|}
\overline{\mathbf{h}}_\mathrm{r}'\right]=0$ and $\mathbb{E}\left[\overline{\mathbf{h}}_\mathrm{r}'^{\mathrm{H}}\frac{\hat{\mathbf{h}}_\mathrm{t}'
-\overline{\mathbf{h}}_\mathrm{t}'}{\left\|\hat{\mathbf{h}}_\mathrm{t}'\right\|}
\frac{\overline{\mathbf{h}}_\mathrm{t}'^{\mathrm{H}}}{\left\|\hat{\mathbf{h}}_\mathrm{t}'\right\|}
\overline{\mathbf{h}}_\mathrm{r}'\right]=0$. Since $\left\|\frac{\hat{\mathbf{h}}_\mathrm{t}'}{\|\hat{\mathbf{h}}_\mathrm{t}'\|}\right\|=1$ and the elements in $\hat{\mathbf{h}}_\mathrm{r}'-\overline{\mathbf{h}}_\mathrm{r}'$ have identical covariance, we may write:
\begin{align}\label{Proof_C_4}
    \notag\mathbb{E}\left[\left|(\hat{\mathbf{h}}_\mathrm{r}'-\overline{\mathbf{h}}_\mathrm{r}')
    ^{\mathrm{H}}\frac{\hat{\mathbf{h}}_\mathrm{t}'}{\|\hat{\mathbf{h}}_\mathrm{t}'\|}\right|^2\right]
    \leq&\frac{\lambda_{\text{min}}^{-2}}{M}\mathbb{E}\left[\|\hat{\mathbf{h}}_\mathrm{r}'
    -\overline{\mathbf{h}}_\mathrm{r}'\|^2\right]\\
    \overset{(\text{a})}\leq&\frac{\lambda_{\text{min}}^{-2}}{M}
    \mathbb{E}\left[\|\hat{\mathbf{h}}_\mathrm{r}'\|^2\right],
\end{align}
where $\lambda_{\text{min}}$ represents the minimum value of $\lambda_{1},\lambda_{2},\cdots,
\lambda_{M}$, and (a) is based on $\left\|\hat{\mathbf{h}}_\mathrm{r}'\right\|^2=\left\|\overline{\mathbf{h}}
_\mathrm{r}'\right\|^2+\left\|\hat{\mathbf{h}}_\mathrm{r}'-\overline{\mathbf{h}}_\mathrm{r}'\right\|^2$. In $\mathbb{E}\left[|\overline{\mathbf{h}}_\mathrm{r}'^{\mathrm{H}}\frac{\hat{\mathbf{h}}
_\mathrm{t}'-\overline{\mathbf{h}}_\mathrm{t}'}{\|\hat{\mathbf{h}}_\mathrm{t}'\|}|^2\right]$, since $\left\|\frac{\overline{\mathbf{h}}_\mathrm{r}'}{\|\overline{\mathbf{h}}_\mathrm{r}'\|}
\right\|=1$ and the elements in $\hat{\mathbf{h}}_\mathrm{t}'-\overline{\mathbf{h}}_\mathrm{t}'$ have the identical covariance, we can get
\begin{align}\label{Proof_C_5}
    \notag\mathbb{E}\left[\left|\overline{\mathbf{h}}_\mathrm{r}'^{\mathrm{H}}
    \frac{\hat{\mathbf{h}}_\mathrm{t}'-\overline{\mathbf{h}}_\mathrm{t}'}
    {\|\hat{\mathbf{h}}_\mathrm{t}'\|}\right|^2\right]
    =&\mathbb{E}\left[\left|\frac{\overline{\mathbf{h}}_\mathrm{r}'^{\mathrm{H}}}
    {\left\|\overline{\mathbf{h}}_\mathrm{r}'\right\|}(\hat{\mathbf{h}}_\mathrm{t}'
    -\overline{\mathbf{h}}_\mathrm{t}')
    \frac{\left\|\overline{\mathbf{h}}_\mathrm{r}'\right\|}
    {\left\|\hat{\mathbf{h}}_\mathrm{t}'\right\|}\right|^2\right]\\
    \overset{(\text{a})}{\leq}&\frac{\lambda_{\text{min}}^{-2}}{M}
    \mathbb{E}\left[\left\|\hat{\mathbf{h}}_\mathrm{r}'\right\|^2\right],
\end{align}
where (a) is based on exploiting that $\left\|\hat{\mathbf{h}}_\mathrm{t}'\right\|^2=\left\|\overline{\mathbf{h}}_\mathrm{t}'\right\|^2
+\left\|\hat{\mathbf{h}}_\mathrm{t}'-\overline{\mathbf{h}}_\mathrm{t}'\right\|^2
\geq\left\|\hat{\mathbf{h}}_\mathrm{t}'-\overline{\mathbf{h}}_\mathrm{t}'\right\|^2$ and $\left\|\hat{\mathbf{h}}_\mathrm{r}'\|^2=\|\overline{\mathbf{h}}_\mathrm{r}'\right\|^2
+\left\|\hat{\mathbf{h}}_\mathrm{r}'-\overline{\mathbf{h}}_\mathrm{r}'\right\|^2
\geq\left\|\overline{\mathbf{h}}_\mathrm{r}'\right\|^2$. Furthermore, $\mathbb{E}[\left|\overline{\mathbf{h}}
_\mathrm{r}'^{\mathrm{H}}\frac{\overline{\mathbf{h}}_\mathrm{t}'}
{\left\|\hat{\mathbf{h}}_\mathrm{t}'\right\|}\right|^2]$ can be derived as
\begin{align}\label{Proof_C_6}
    \notag&\mathbb{E}\left[\left|\overline{\mathbf{h}}_\mathrm{r}'^{\mathrm{H}}
    \frac{\overline{\mathbf{h}}_\mathrm{t}'}
    {\|\hat{\mathbf{h}}_\mathrm{t}'\|}\right|^2\right]\\
    \notag=&\left\|\overline{\mathbf{a}}^{(\mathrm{AP})\mathrm{H}}_\mathrm{t}\mathbf{R}^{-1}
    \overline{\mathbf{a}}^{(\mathrm{AP})}_\mathrm{r}\right\|^2\cdot
    \frac{\varrho_{\mathbf{A}_\mathrm{t}}\varrho_{\mathbf{g}_\mathrm{t}}
    \kappa_{\mathbf{A}_\mathrm{t}}\kappa_{\mathbf{g}_\mathrm{t}}
    \left\|\overline{\mathbf{a}}_\mathrm{t}^{(\mathrm{IOS})\mathrm{H}}
    \mathbf{\Theta}_\mathrm{t}\overline{\mathbf{g}}_\mathrm{t}\right\|^2}
    {(1+\kappa_{\mathbf{A}_\mathrm{t}})\left(1+\kappa_{\mathbf{g}_\mathrm{t}}\right)}\cdot\\
    \notag&\frac{\varrho_{\mathbf{A}_\mathrm{r}}\varrho_{\mathbf{g}_\mathrm{r}}
    \kappa_{\mathbf{A}_\mathrm{r}}\kappa_{\mathbf{g}_\mathrm{r}}
    \left\|\overline{\mathbf{a}}_\mathrm{r}^{(\mathrm{IOS})\mathrm{H}}
    \mathbf{\Theta}_\mathrm{r}\overline{\mathbf{g}}_\mathrm{r}\right\|^2}
    {\left(1+\kappa_{\mathbf{A}_\mathrm{r}}\right)\left(1+\kappa_{\mathbf{g}_\mathrm{r}}\right)}\cdot
    \mathbb{E}\left[\frac{1}{\|\hat{\mathbf{h}}_\mathrm{t}'\|^2}\right]\\
    \notag=&\frac{1}{M^2}\left\|\overline{\mathbf{h}}_\mathrm{r}'\right\|^2
    \left\|\overline{\mathbf{h}}_\mathrm{t}'\right\|^2
    \sum_{m=1}^{M}\lambda_m^{-1}\mathrm{e}^{\jmath2\pi{d_0}
    m\left(\cos\psi_\mathrm{t}-\cos\psi_\mathrm{r}\right)}\cdot
    \mathbb{E}\left[\frac{1}{\|\hat{\mathbf{h}}_\mathrm{t}'\|^2}\right]\\
    \overset{(\text{a})}{\leq}&\frac{1}{M^2}\sum_{m=1}^{M}\lambda_m^{-1}
    \mathrm{e}^{\jmath2\pi{d_0}m\left(\cos\psi_\mathrm{t}-\cos\psi_\mathrm{r}\right)}\cdot
    \mathbb{E}\left[\|\hat{\mathbf{h}}_\mathrm{r}'\|^2\right],
\end{align}
where (a) is based on $\left\|\hat{\mathbf{h}}_\mathrm{r}'\right\|^2=\left\|\overline{\mathbf{h}}_\mathrm{r}'\right\|^2+\left\|\hat{\mathbf{h}}
_\mathrm{r}'-\overline{\mathbf{h}}_\mathrm{r}'\right\|^2\geq\left\|\overline{\mathbf{h}}_\mathrm{r}'\right\|^2$, $\left\|\hat{\mathbf{h}}_\mathrm{t}'\right\|^2=\left\|\overline{\mathbf{h}}_\mathrm{t}'\right\|^2+\left\|\hat{\mathbf{h}}
_\mathrm{t}'-\overline{\mathbf{h}}_\mathrm{t}'\right\|^2\geq\left\|\overline{\mathbf{h}}_\mathrm{t}'\right\|^2$.

The denominator in (\ref{Proof_C_1}) can be further expressed as
\begin{align}\label{Proof_C_8}
    \mathbb{E}\left[\hat{\mathbf{h}}_\mathrm{r}^{\mathrm{H}}\mathbf{R}^{-1}\hat{\mathbf{h}}_\mathrm{r}\right]
    =\mathbb{E}\left[\left\|\hat{\mathbf{h}}_\mathrm{t}'\right\|^2\right].
\end{align}

According to (\ref{Proof_C_2}), (\ref{Proof_C_4}), (\ref{Proof_C_5}), (\ref{Proof_C_6}) and (\ref{Proof_C_8}), we can get $\frac{\mathbb{E}\left[\frac{\rho_\mathrm{t}\varepsilon_\mathbf{v}
\hat{\mathbf{h}}_\mathrm{r}^{\mathrm{H}}\mathbf{R}^{-1}\hat{\mathbf{h}}_\mathrm{t}
\hat{\mathbf{h}}_\mathrm{t}^{\mathrm{H}}\mathbf{R}^{-1}
\hat{\mathbf{h}}_\mathrm{r}}{1+\rho_\mathrm{t}\varepsilon_\mathbf{v}
\hat{\mathbf{h}}_\mathrm{t}^{\mathrm{H}}\mathbf{R}^{-1}\hat{\mathbf{h}}_\mathrm{t}}\right]}
{\mathbb{E}\left[\hat{\mathbf{h}}_\mathrm{r}^{\mathrm{H}}\mathbf{R}^{-1}\hat{\mathbf{h}}_\mathrm{r}\right]}
<\frac{\lambda_{\text{min}}^{-2}}{M}\left(2+\frac{\sin^2\left(\pi{d_0}M\left(\cos\psi_\mathrm{t}
-\cos\psi_\mathrm{r}\right)\right)}{M\sin^2\left(\pi{d_0}\left(\cos\psi_\mathrm{t}-\cos\psi_\mathrm{r}\right)\right)}\right)$, which tends to 0 when $M\rightarrow\infty$ and $\cos\psi_\mathrm{t}\neq\cos\psi_\mathrm{r}$.

\section{Proof of Theorem~\ref{theorem_4}}\label{Appendix_E}
When $\frac{\varepsilon_{u,\mathrm{r}}}{1-\varepsilon_{u,\mathrm{r}}}\gg1$, $\frac{\varepsilon_{u,\mathrm{t}}}{1-\varepsilon_{u,\mathrm{t}}}\gg1$ and $\frac{\varepsilon_\mathbf{v}}{1-\varepsilon_\mathbf{v}}\gg1$, based on (\ref{Channel_estimation_5}), $\mathbf{C}_{\mathbf{x}_\mathrm{i}\mathbf{x}_\mathrm{i}}$ is approximately given by $\mathbf{C}_{\mathbf{x}_\mathrm{i}\mathbf{x}_\mathrm{i}}
=\rho_\mathrm{i}\varepsilon_\mathbf{v}(1+(K-1)\varepsilon_{u,\mathrm{i}})
\mathbf{C}_{\mathbf{h}_\mathrm{i}\mathbf{h}_\mathrm{i}}+\sigma_w^2\mathbf{I}_M$. Therefore, $\mathbf{C}_{\hat{\mathbf{h}}_\mathrm{i}\hat{\mathbf{h}}_\mathrm{i}}$ and $\mathbf{C}_{\check{\mathbf{h}}_\mathrm{i}\check{\mathbf{h}}_\mathrm{i}}$ can be approximated as
\begin{align}\label{Proof_D_2_1}
    \notag\mathbf{C}_{\hat{\mathbf{h}}_\mathrm{i}\hat{\mathbf{h}}_\mathrm{i}}
    \notag=&K\rho_\mathrm{i}\varepsilon_\mathbf{v}\varepsilon_{u,\mathrm{i}}
    \mathbf{C}_{\mathbf{h}_\mathrm{i}\mathbf{h}_\mathrm{i}}
    \left(\rho_\mathrm{i}\varepsilon_\mathbf{v}\left(1+\left(K-1\right)\varepsilon_{u,\mathrm{i}}\right)
    \mathbf{C}_{\mathbf{h}_\mathrm{i}\mathbf{h}_\mathrm{i}}\right.\\
    &\left.+\sigma_w^2\mathbf{I}_M\right)^{-1}\mathbf{C}_{\mathbf{h}_\mathrm{i}\mathbf{h}_\mathrm{i}},
\end{align}
\begin{align}\label{Proof_D_2_2}
    \notag\mathbf{C}_{\check{\mathbf{h}}_\mathrm{i}\check{\mathbf{h}}_\mathrm{i}}
    \notag=&\mathbf{C}_{\mathbf{h}_\mathrm{i}\mathbf{h}_\mathrm{i}}
    -K\rho_\mathrm{i}\varepsilon_\mathbf{v}\varepsilon_{u,\mathrm{i}}
    \mathbf{C}_{\mathbf{h}_\mathrm{i}\mathbf{h}_\mathrm{i}}
    \left(\rho_\mathrm{i}\varepsilon_\mathbf{v}\left(1+\left(K-1\right)\varepsilon_{u,\mathrm{i}}\right)\cdot\right.\\
    &\left.\mathbf{C}_{\mathbf{h}_\mathrm{i}\mathbf{h}_\mathrm{i}}+\sigma_w^2\mathbf{I}_M\right)^{-1}\mathbf{C}_{\mathbf{h}_\mathrm{i}\mathbf{h}_\mathrm{i}}.
\end{align}
According to (\ref{Channel_Model_15}), (\ref{Proof_D_2_1}) and (\ref{Proof_D_2_2}), we can arrive at $\frac{\partial\mathbf{C}_{\hat{\mathbf{h}}_\mathrm{i}\hat{\mathbf{h}}_\mathrm{i}}}
{\partial{\left\|\overline{\mathbf{a}}^{(\mathrm{IOS})\mathrm{H}}_\mathrm{i}
\mathbf{\Theta}_\mathrm{i}\overline{\mathbf{g}}_\mathrm{i}\right\|^2}}
=\mathbf{O}_{{M}\times{M}}$ and $\frac{\partial\mathbf{C}_{\check{\mathbf{h}}_\mathrm{i}
\check{\mathbf{h}}_\mathrm{i}}}{\partial{\|\overline{\mathbf{a}}^{(\mathrm{IOS})
\mathrm{H}}_\mathrm{i}\mathbf{\Theta}_\mathrm{i}\overline{\mathbf{g}}_\mathrm{i}\|^2}}
=\mathbf{O}_{{M}\times{M}}$.

The partial derivative of $\ddot{\mathcal{R}}_{\mathrm{erg},\mathrm{r}}$ with respect to $\|\overline{\mathbf{a}}^{(\mathrm{IOS})\mathrm{H}}_\mathrm{r}\mathbf{\Theta}_\mathrm{r}
\overline{\mathbf{g}}_\mathrm{r}\|^2$ is given by
\begin{align}\label{Proof_D_4}
    \notag&\frac{\partial{\ddot{\mathcal{R}}_{\mathrm{erg},\mathrm{r}}}}
    {\partial{\left\|\overline{\mathbf{a}}^{(\mathrm{IOS})\mathrm{H}}_\mathrm{r}
    \mathbf{\Theta}_\mathrm{r}\overline{\mathbf{g}}_\mathrm{r}\right\|^2}}\\
    \notag=&\frac{\log_2\mathrm{e}\cdot\rho_\mathrm{r}\varepsilon_\mathbf{v}
    \varepsilon_{u,\mathrm{r}}}{\left(1+\rho_\mathrm{r}\varepsilon_\mathbf{v}\ddot{\zeta}_\mathrm{r}\right)
    \left(1+\rho_\mathrm{r}\varepsilon_\mathbf{v}\left(1-\varepsilon_{u,\mathrm{r}}\right)
    \ddot{\zeta}_\mathrm{r}\right)}\cdot\\
    \notag&\left(\frac{\varrho_{\mathbf{A}_\mathrm{r}}\varrho_{\mathbf{g}_\mathrm{r}}
    \kappa_{\mathbf{A}_\mathrm{r}}\kappa_{\mathbf{g}_\mathrm{r}}}
    {\left(1+\kappa_{\mathbf{A}_\mathrm{r}}\right)\left(1+\kappa_{\mathbf{g}_\mathrm{r}}\right)}\cdot
    \text{Tr}\left[\mathbf{R}^{-1}\overline{\mathbf{a}}^{(\mathrm{AP})}_\mathrm{r}
    \overline{\mathbf{a}}^{(\mathrm{AP})\mathrm{H}}_\mathrm{r}\right]\right.\\
    \notag&\left.+\frac{\varrho_{\mathbf{A}_\mathrm{r}}\varrho_{\mathbf{g}_\mathrm{r}}
    \kappa_{\mathbf{A}_\mathrm{r}}\kappa_{\mathbf{g}_\mathrm{r}}
    \left\|\overline{\mathbf{a}}^{(\mathrm{IOS})\mathrm{H}}_\mathrm{r}
    \mathbf{\Theta}_\mathrm{r}\overline{\mathbf{g}}_\mathrm{r}\right\|^2}
    {\left(1+\kappa_{\mathbf{A}_\mathrm{r}}\right)\left(1+\kappa_{\mathbf{g}_\mathrm{r}}\right)}\cdot
    \frac{\partial\text{Tr}\left[\mathbf{R}^{-1}\overline{\mathbf{a}}^{(\mathrm{AP})}_\mathrm{r}
    \overline{\mathbf{a}}^{(\mathrm{AP})\mathrm{H}}_\mathrm{r}\right]}
    {\partial{\left\|\overline{\mathbf{a}}^{(\mathrm{IOS})\mathrm{H}}_\mathrm{r}
    \mathbf{\Theta}_\mathrm{r}\overline{\mathbf{g}}_\mathrm{r}\right\|^2}}\right.\\
    &\left.+\frac{\partial\text{Tr}\left[\mathbf{R}^{-1}
    \mathbf{C}_{\hat{\mathbf{h}}_\mathrm{r}\hat{\mathbf{h}}_\mathrm{r}}\right]}
    {\partial{\left\|\overline{\mathbf{a}}^{(\mathrm{IOS})\mathrm{H}}_\mathrm{r}
    \mathbf{\Theta}_\mathrm{r}\overline{\mathbf{g}}_\mathrm{r}\right\|^2}}\right),
\end{align}
where $\frac{\partial\text{Tr}\left[\mathbf{R}^{-1}\overline{\mathbf{a}}^{(\mathrm{AP})}_\mathrm{r}
\overline{\mathbf{a}}^{(\mathrm{AP})\mathrm{H}}_\mathrm{r}\right]}{\partial{\left\|\overline{\mathbf{a}}
^{(\mathrm{IOS})\mathrm{H}}_\mathrm{r}\mathbf{\Theta}_\mathrm{r}
\overline{\mathbf{g}}_\mathrm{r}\right\|^2}}$ can be further derived as
\begin{align}\label{Proof_D_5}
    \notag\frac{\partial\text{Tr}\left[\mathbf{R}^{-1}\overline{\mathbf{a}}^{(\mathrm{AP})}_\mathrm{r}
    \overline{\mathbf{a}}^{(\mathrm{AP})\mathrm{H}}_\mathrm{r}\right]}
    {\partial{\|\overline{\mathbf{a}}^{(\mathrm{IOS})\mathrm{H}}_\mathrm{r}
    \mathbf{\Theta}_\mathrm{r}\overline{\mathbf{g}}_\mathrm{r}\|^2}}
    =&-\frac{\rho_{\mathrm{r}}\left(1-\varepsilon_\mathbf{v}\right)
    \varrho_{\mathbf{A}_\mathrm{r}}\varrho_{\mathbf{g}_\mathrm{r}}
    \kappa_{\mathbf{A}_\mathrm{r}}\kappa_{\mathbf{g}_\mathrm{r}}}
    {\left(1+\kappa_{\mathbf{A}_\mathrm{r}}\right)\left(1+\kappa_{\mathbf{g}_\mathrm{r}}\right)}\cdot\\
    &\text{Tr}\left[\mathbf{R}^{-1}\overline{\mathbf{a}}^{(\mathrm{AP})}_\mathrm{r}
    \overline{\mathbf{a}}^{(\mathrm{AP})\mathrm{H}}_\mathrm{r}\mathbf{R}^{-1}\right],
\end{align}
and $\frac{\partial\text{Tr}\left[\mathbf{R}^{-1}\mathbf{C}_{\hat{\mathbf{h}}_\mathrm{r}
\hat{\mathbf{h}}_\mathrm{r}}\right]}{\partial{\left\|\overline{\mathbf{a}}^{(\mathrm{IOS})
\mathrm{H}}_\mathrm{r}\mathbf{\Theta}_\mathrm{r}\overline{\mathbf{g}}_\mathrm{r}\right\|^2}}$ can be further rewritten as
\begin{align}\label{Proof_D_6}
    \notag\frac{\partial\text{Tr}\left[\mathbf{R}^{-1}\mathbf{C}_{\hat{\mathbf{h}}_\mathrm{r}
    \hat{\mathbf{h}}_\mathrm{r}}\right]}{\partial{\|\overline{\mathbf{a}}^{(\mathrm{IOS})
    \mathrm{H}}_\mathrm{r}\mathbf{\Theta}_\mathrm{r}\overline{\mathbf{g}}_\mathrm{r}\|^2}}
    =&-\frac{\rho_{\mathrm{r}}\left(1-\varepsilon_\mathbf{v}\right)\varrho_{\mathbf{A}_\mathrm{r}}
    \varrho_{\mathbf{g}_\mathrm{r}}\kappa_{\mathbf{A}_\mathrm{r}}\kappa_{\mathbf{g}_\mathrm{r}}}
    {\left(1+\kappa_{\mathbf{A}_\mathrm{r}}\right)\left(1+\kappa_{\mathbf{g}_\mathrm{r}}\right)}\cdot\\
    &\text{Tr}\left[\mathbf{R}^{-1}\mathbf{C}_{\hat{\mathbf{h}}_\mathrm{r}
    \hat{\mathbf{h}}_\mathrm{r}}\mathbf{R}^{-1}\right].
\end{align}
Substituting (\ref{Proof_D_5}) and (\ref{Proof_D_6}) into (\ref{Proof_D_4}), we can get
\begin{align}\label{Proof_D_7}
    \notag\frac{\partial{\ddot{\mathcal{R}}_{\mathrm{erg},\mathrm{r}}}}
    {\partial{\left\|\overline{\mathbf{a}}^{(\mathrm{IOS})\mathrm{H}}_\mathrm{r}
    \mathbf{\Theta}_\mathrm{r}\overline{\mathbf{g}}_\mathrm{r}\right\|^2}}
    =&\frac{\log_2\mathrm{e}\cdot\rho_\mathrm{r}\varepsilon_\mathbf{v}\varepsilon_{u,\mathrm{r}}
    \cdot\frac{\varrho_{\mathbf{A}_\mathrm{r}}\varrho_{\mathbf{g}_\mathrm{r}}
    \kappa_{\mathbf{A}_\mathrm{r}}\kappa_{\mathbf{g}_\mathrm{r}}}
    {\left(1+\kappa_{\mathbf{A}_\mathrm{r}}\right)\left(1+\kappa_{\mathbf{g}_\mathrm{r}}\right)}}
    {\left(1+\rho_\mathrm{r}\varepsilon_\mathbf{v}\ddot{\zeta}_\mathrm{r}\right)
    \left(1+\rho_\mathrm{r}
    \varepsilon_\mathbf{v}\left(1-\varepsilon_{u,\mathrm{r}}\right)\ddot{\zeta}_\mathrm{r}\right)}\cdot\\
    &\text{Tr}\left[\mathbf{R}^{-1}\mathbf{\Xi}\mathbf{R}^{-1}\right],
\end{align}
where
\begin{align}
    \notag&\mathbf{\Xi}=\overline{\mathbf{a}}^{(\mathrm{AP})}_\mathrm{r}
    \overline{\mathbf{a}}^{(\mathrm{AP})\mathrm{H}}_\mathrm{r}
    \left(\rho_\mathrm{r}\left(1-\varepsilon_{\mathbf{v}}\varepsilon_{u,\mathrm{r}}\right)
    \mathbf{C}_{\check{\mathbf{h}}_\mathrm{r}\check{\mathbf{h}}_\mathrm{r}}
    +\rho_\mathrm{t}\varepsilon_{\mathbf{v}}\left(1-\varepsilon_{u,\mathrm{t}}\right)\cdot\right.\\
    \notag&\left.\mathbf{C}_{\check{\mathbf{h}}_\mathrm{t}\check{\mathbf{h}}_\mathrm{t}}
    +\rho_\mathrm{t}\left(1-\varepsilon_\mathbf{v}\right)
    \left(\frac{\varrho_{\mathbf{A}_\mathrm{t}}\varrho_{\mathbf{g}_\mathrm{t}}
    \left(1+\kappa_{\mathbf{g}_\mathrm{t}}+\kappa_{\mathbf{A}_\mathrm{t}}\right)N_\mathrm{t}}
    {\left(1+\kappa_{\mathbf{A}_\mathrm{t}}\right)\left(1+\kappa_{\mathbf{g}_\mathrm{t}}\right)}
    +\varrho_{\mathbf{b}_\mathrm{t}}\right.\right.\\
    \notag&\left.\left.+\frac{\varrho_{\mathbf{A}_\mathrm{t}}\varrho_{\mathbf{g}_\mathrm{t}}
    \kappa_{\mathbf{A}_\mathrm{t}}\kappa_{\mathbf{g}_\mathrm{t}}
    \left\|\overline{\mathbf{a}}_\mathrm{t}^{(\mathrm{IOS})\mathrm{H}}
    \mathbf{\Theta}_\mathrm{t}\overline{\mathbf{g}}_\mathrm{t}\right\|^2}
    {\left(1+\kappa_{\mathbf{A}_\mathrm{t}}\right)\left(1+\kappa_{\mathbf{g}_\mathrm{t}}\right)}\right)\mathbf{I}_M
    +\sigma_w^2\mathbf{I}_M\right)\\
    \notag&+\rho_\mathrm{r}\left(1-\varepsilon_\mathbf{v}\right)
    \left(\frac{\varrho_{\mathbf{A}_\mathrm{r}}\varrho_{\mathbf{g}_\mathrm{r}}
    \left(1+\kappa_{\mathbf{g}_\mathrm{r}}\right)N_\mathrm{r}}
    {\left(1+\kappa_{\mathbf{A}_\mathrm{r}}\right)\left(1+\kappa_{\mathbf{g}_\mathrm{r}}\right)}
    +\varrho_{\mathbf{b}_\mathrm{r}}\right)\cdot\\
    &\left(\overline{\mathbf{a}}^{(\mathrm{AP})}_\mathrm{r}
    \overline{\mathbf{a}}^{(\mathrm{AP})\mathrm{H}}_\mathrm{r}-\mathbf{I}_M\right).
\end{align}
Since $\mathbf{C}_{\check{\mathbf{h}}_\mathrm{r}\check{\mathbf{h}}_\mathrm{r}}\succ\mathbf{0}$, $\mathbf{C}_{\check{\mathbf{h}}_\mathrm{t}\check{\mathbf{h}}_\mathrm{t}}\succ\mathbf{0}$, $\overline{\mathbf{a}}^{(\mathrm{AP})}_\mathrm{r}\overline{\mathbf{a}}^{(\mathrm{AP})\mathrm{H}}
_\mathrm{r}\succeq\mathbf{0}$ and $\mathbf{R}\succ\mathbf{0}$, we can write:
\begin{align}\label{Proof_D_9}
    \notag\text{Tr}\left[\mathbf{R}^{-1}\mathbf{\Xi}\mathbf{R}^{-1}\right]>
    &\rho_\mathrm{r}\left(1-\varepsilon_\mathbf{v}\right)
    \left(\frac{\varrho_{\mathbf{A}_\mathrm{r}}\varrho_{\mathbf{g}_\mathrm{r}}
    \left(1+\kappa_{\mathbf{g}_\mathrm{r}}\right)N_\mathrm{r}}
    {\left(1+\kappa_{\mathbf{A}_\mathrm{r}}\right)\left(1+\kappa_{\mathbf{g}_\mathrm{r}}\right)}
    +\varrho_{\mathbf{b}_\mathrm{r}}\right)\cdot\\
    &\text{Tr}\left[\mathbf{R}^{-1}\left(\overline{\mathbf{a}}^{(\mathrm{AP})}_\mathrm{r}
    \overline{\mathbf{a}}^{(\mathrm{AP})\mathrm{H}}_\mathrm{r}-\mathbf{I}_M\right)
    \mathbf{R}^{-1}\right].
\end{align}
Furthermore, according to (\ref{Proof_D_2_2}), we have
\begin{align}\label{Proof_D_10}
    \notag&\rho_\mathrm{r}\varepsilon_\mathbf{v}\left(1-\varepsilon_{u,\mathrm{r}}\right)
    \mathbf{C}_{\check{\mathbf{h}}_\mathrm{r}\check{\mathbf{h}}_\mathrm{r}}
    +\rho_\mathrm{t}\varepsilon_\mathbf{v}\left(1-\varepsilon_{u,\mathrm{t}}\right)
    \mathbf{C}_{\check{\mathbf{h}}_\mathrm{t}\check{\mathbf{h}}_\mathrm{t}}+\sigma_w^2\mathbf{I}_M\\
    \notag=&\mathbf{C}_{\mathbf{h}_\mathrm{r}\mathbf{h}_\mathrm{r}}
    -K\rho_\mathrm{r}\varepsilon_\mathbf{v}\varepsilon_{u,\mathrm{r}}
    \mathbf{C}_{\mathbf{h}_\mathrm{r}\mathbf{h}_\mathrm{r}}\left(\rho_\mathrm{r}\left(1+\left(K-1\right)
    \varepsilon_\mathbf{v}\varepsilon_{u,\mathrm{r}}\right)
    \mathbf{C}_{\mathbf{h}_\mathrm{r}\mathbf{h}_\mathrm{r}}\right.\\
    \notag&\left.+\sigma_w^2\mathbf{I}_M\right)^{-1}\mathbf{C}_{\mathbf{h}_\mathrm{r}\mathbf{h}_\mathrm{r}}
    +\mathbf{C}_{\mathbf{h}_\mathrm{t}\mathbf{h}_\mathrm{t}}
    -K\rho_\mathrm{t}\varepsilon_\mathbf{v}\varepsilon_{u,\mathrm{t}}
    \mathbf{C}_{\mathbf{h}_\mathrm{t}\mathbf{h}_\mathrm{t}}\cdot\\
    \notag&\left(\rho_\mathrm{t}\left(1+\left(K-1\right)\varepsilon_\mathbf{v}\varepsilon_{u,\mathrm{t}}\right)
    \mathbf{C}_{\mathbf{h}_\mathrm{t}\mathbf{h}_\mathrm{t}}+\sigma_w^2\mathbf{I}_M\right)^{-1}
    \mathbf{C}_{\mathbf{h}_\mathrm{t}\mathbf{h}_\mathrm{t}}+\sigma_w^2\mathbf{I}_M\\
    \overset{(\text{a})}\approx&\sigma_w^2\mathbf{I}_M,
\end{align}
where (a) is based on $\frac{\varepsilon_{u,\mathrm{r}}}{1-\varepsilon_{u,\mathrm{r}}}\gg1$, $\frac{\varepsilon_{u,\mathrm{t}}}{1-\varepsilon_{u,\mathrm{t}}}\gg1$ and $\frac{\varepsilon_\mathbf{v}}{1-\varepsilon_\mathbf{v}}\gg1$. Therefore, we can write:
\begin{align}\label{Proof_D_11}
    \notag\mathbf{R}=&\left(\rho_\mathrm{r}\left(1-\varepsilon_\mathbf{v}\right)
    \left(\frac{\varrho_{\mathbf{A}_\mathrm{r}}\varrho_{\mathbf{g}_\mathrm{r}}
    \left(1+\kappa_{\mathbf{g}_\mathrm{r}}+\kappa_{\mathbf{A}_\mathrm{r}}\right)N_\mathrm{r}}
    {\left(1+\kappa_{\mathbf{A}_\mathrm{r}}\right)\left(1+\kappa_{\mathbf{g}_\mathrm{r}}\right)}
    +\varrho_{\mathbf{b}_\mathrm{r}}\right.\right.\\
    \notag&\left.\left.+\varrho_{\mathbf{A}_\mathrm{r}}\varrho_{\mathbf{g}_\mathrm{r}}
    \frac{\kappa_{\mathbf{A}_\mathrm{r}}\kappa_{\mathbf{g}_\mathrm{r}}
    \left\|\overline{\mathbf{a}}_\mathrm{r}^{(\mathrm{IOS})\mathrm{H}}
    \mathbf{\Theta}_\mathrm{r}\overline{\mathbf{g}}_\mathrm{r}\right\|^2}
    {\left(1+\kappa_{\mathbf{A}_\mathrm{r}}\right)
    \left(1+\kappa_{\mathbf{g}_\mathrm{r}}\right)}\right)\right.\\
    \notag&\left.+\rho_\mathrm{t}\left(1-\varepsilon_\mathbf{v}\right)
    \left(\frac{\varrho_{\mathbf{A}_\mathrm{t}}\varrho_{\mathbf{g}_\mathrm{t}}
    \left(1+\kappa_{\mathbf{g}_\mathrm{t}}+\kappa_{\mathbf{A}_\mathrm{t}}\right)N_\mathrm{t}}
    {\left(1+\kappa_{\mathbf{A}_\mathrm{t}}\right)\left(1+\kappa_{\mathbf{g}_\mathrm{t}}\right)}
    +\varrho_{\mathbf{b}_\mathrm{t}}\right.\right.\\
    &\left.\left.+\varrho_{\mathbf{A}_\mathrm{t}}\varrho_{\mathbf{g}_\mathrm{t}}
    \frac{\kappa_{\mathbf{A}_\mathrm{t}}\kappa_{\mathbf{g}_\mathrm{t}}
    \left\|\overline{\mathbf{a}}_\mathrm{t}^{(\mathrm{IOS})\mathrm{H}}
    \mathbf{\Theta}_\mathrm{t}\overline{\mathbf{g}}_\mathrm{t}\right\|^2}
    {\left(1+\kappa_{\mathbf{A}_\mathrm{t}}\right)\left(1+\kappa_{\mathbf{g}_\mathrm{t}}\right)}\right)
    +\sigma_w^2\right)\mathbf{I}_M,
\end{align}
which is a diagonal matrix. According to (\ref{Proof_D_9}), we have:
\begin{align}\label{Proof_D_12}
    \notag\text{Tr}\left[\mathbf{R}^{-1}\mathbf{\Xi}\mathbf{R}^{-1}\right]>
    \notag&\rho_\mathrm{r}\left(1-\varepsilon_\mathbf{v}\right)
    \left(\frac{\varrho_{\mathbf{A}_\mathrm{r}}\varrho_{\mathbf{g}_\mathrm{r}}
    \left(1+\kappa_{\mathbf{g}_\mathrm{r}}\right)N_\mathrm{r}}
    {\left(1+\kappa_{\mathbf{A}_\mathrm{r}}\right)\left(1+\kappa_{\mathbf{g}_\mathrm{r}}\right)}
    +\varrho_{\mathbf{b}_\mathrm{r}}\right)\cdot\\
    \notag&\text{Tr}[\mathbf{R}^{-2}]
    \cdot\text{Tr}\left[\overline{\mathbf{a}}^{(\mathrm{AP})}_\mathrm{r}
    \overline{\mathbf{a}}^{(\mathrm{AP})\mathrm{H}}_\mathrm{r}-\mathbf{I}_M\right]\\
    =&0.
\end{align}
Based on (\ref{Proof_D_7}) and (\ref{Proof_D_12}), we can get $\frac{\partial{\ddot{\mathcal{R}}_{\mathrm{erg},\mathrm{r}}}}
{\partial{\|\overline{\mathbf{a}}^{(\mathrm{IOS})\mathrm{H}}_\mathrm{r}
\mathbf{\Theta}_\mathrm{r}\overline{\mathbf{g}}_\mathrm{r}\|^2}}>0$. Similarly, we can get $\frac{\partial{\ddot{\mathcal{R}}_{\mathrm{erg},\mathrm{t}}}}
{\partial{\|\overline{\mathbf{a}}^{(\mathrm{IOS})\mathrm{H}}_\mathrm{t}
\mathbf{\Theta}_\mathrm{t}\overline{\mathbf{g}}_\mathrm{t}\|^2}}>0$. Therefore, $\ddot{\mathcal{R}}_{\mathrm{erg},\mathrm{r}}$ and $\ddot{\mathcal{R}}_{\mathrm{erg},\mathrm{t}}$ are monotonically increasing functions with respect to $\left\|\overline{\mathbf{a}}^{(\mathrm{IOS})\mathrm{H}}_\mathrm{r}
\mathbf{\Theta}_\mathrm{r}\overline{\mathbf{g}}_\mathrm{r}\right\|^2$ and $\left\|\overline{\mathbf{a}}^{(\mathrm{IOS})\mathrm{H}}_\mathrm{t}
\mathbf{\Theta}_\mathrm{t}\overline{\mathbf{g}}_\mathrm{t}\right\|^2$, respectively. Hence, problem (P1) is equivalent to problem (P2).

\section{Proof of Theorem~\ref{theorem_5}}\label{Appendix_F}
In (\ref{Beamforming_Design_18}), $\ddot{\zeta}_\mathrm{i}$ is given by
\begin{align}\label{Proof_E_1}
    \notag\ddot{\zeta}_\mathrm{i}
    =&\frac{\varrho_{\mathbf{A}_\mathrm{i}}\varrho_{\mathbf{g}_\mathrm{i}}
    \kappa_{\mathbf{A}_\mathrm{i}}\kappa_{\mathbf{g}_\mathrm{i}}N_\mathrm{i}^2}
    {\left(1+\kappa_{\mathbf{A}_\mathrm{i}}\right)\left(1+\kappa_{\mathbf{g}_\mathrm{i}}\right)}
    \overline{\mathbf{a}}^{(\mathrm{AP})\mathrm{H}}_\mathrm{i}\mathbf{R}^{-1}
    \overline{\mathbf{a}}^{(\mathrm{AP})}_\mathrm{i}
    +\text{Tr}\left[\mathbf{R}^{-1}\mathbf{C}_{\hat{\mathbf{h}}_\mathrm{i}\hat{\mathbf{h}}_\mathrm{i}}\right]\\
    \notag\overset{(\text{a})}\leq&\frac{\frac{\varrho_{\mathbf{A}_\mathrm{i}}
    \varrho_{\mathbf{g}_\mathrm{i}}
    \kappa_{\mathbf{A}_\mathrm{i}}\kappa_{\mathbf{g}_\mathrm{i}}N_\mathrm{i}^2}
    {\left(1+\kappa_{\mathbf{A}_\mathrm{i}}\right)\left(1+\kappa_{\mathbf{g}_\mathrm{i}}\right)}
    \text{Tr}\left[\overline{\mathbf{a}}^{(\mathrm{AP})}_\mathrm{i}
    \overline{\mathbf{a}}^{(\mathrm{AP})\mathrm{H}}_\mathrm{i}\right]
    +\text{Tr}\left[\mathbf{C}_{\mathbf{h}_\mathrm{i}\mathbf{h}_\mathrm{i}}\right]}
    {\rho_\mathrm{r}\left(1-\varepsilon_{\mathbf{v}}\right)\eta_\mathrm{r}
    +\rho_\mathrm{t}\left(1-\varepsilon_{\mathbf{v}}\right)\eta_\mathrm{t}+\sigma_w^2}\\
    \overset{(\text{b})}=&\frac{M\eta_\mathrm{i}}
    {\rho_\mathrm{r}\left(1-\varepsilon_{\mathbf{v}}\right)\eta_\mathrm{r}
    +\rho_\mathrm{t}\left(1-\varepsilon_{\mathbf{v}}\right)\eta_\mathrm{t}+\sigma_w^2},
\end{align}
where (a) is based on (\ref{Proof_D_11}) as well as on $\mathbf{C}_{\mathbf{h}_\mathrm{i}\mathbf{h}_\mathrm{i}}
=\mathbf{C}_{\hat{\mathbf{h}}_\mathrm{i}\hat{\mathbf{h}}_\mathrm{i}}
+\mathbf{C}_{\check{\mathbf{h}}_\mathrm{i}\check{\mathbf{h}}_\mathrm{i}}$, $\mathbf{C}_{\check{\mathbf{h}}_\mathrm{i}\check{\mathbf{h}}_\mathrm{i}}\succ\mathbf{0}$ and $\mathbf{R}\succ\mathbf{0}$, while (b) is based on (\ref{Channel_Model_15}). Upon substituting (\ref{Proof_E_1}) into (\ref{Beamforming_Design_9}), we arrive at (\ref{Beamforming_Design_19}).

\bibliographystyle{IEEEtran}
\bibliography{IEEEabrv,TAMS}

\begin{thebibliography}{10}
\providecommand{\url}[1]{#1}
\csname url@samestyle\endcsname
\providecommand{\newblock}{\relax}
\providecommand{\bibinfo}[2]{#2}
\providecommand{\BIBentrySTDinterwordspacing}{\spaceskip=0pt\relax}
\providecommand{\BIBentryALTinterwordstretchfactor}{4}
\providecommand{\BIBentryALTinterwordspacing}{\spaceskip=\fontdimen2\font plus
\BIBentryALTinterwordstretchfactor\fontdimen3\font minus
  \fontdimen4\font\relax}
\providecommand{\BIBforeignlanguage}[2]{{%
\expandafter\ifx\csname l@#1\endcsname\relax
\typeout{** WARNING: IEEEtran.bst: No hyphenation pattern has been}%
\typeout{** loaded for the language `#1'. Using the pattern for}%
\typeout{** the default language instead.}%
\else
\language=\csname l@#1\endcsname
\fi
#2}}
\providecommand{\BIBdecl}{\relax}
\BIBdecl

\bibitem{li2021analog}
Y.~Li, F.~Wang, M.~El-Hajjar, and L.~Hanzo, ``Analog radio-over-fiber-aided
  optical-domain {MIMO} signal processing for high-performance low-cost radio
  access networks,'' \emph{IEEE Commun. Mag.}, vol.~59, no.~1, pp. 126--132,
  2021.

\bibitem{basar2021present}
E.~Basar and H.~V. Poor, ``Present and future of reconfigurable intelligent
  surface-empowered communications [perspectives],'' \emph{IEEE Sign. Proces.
  Mag.}, vol.~38, no.~6, pp. 146--152, 2021.

\bibitem{pan2021reconfigurable}
C.~Pan, H.~Ren, K.~Wang, J.~F. Kolb, M.~Elkashlan, M.~Chen, M.~Di~Renzo,
  Y.~Hao, J.~Wang, A.~L. Swindlehurst \emph{et~al.}, ``Reconfigurable
  intelligent surfaces for {6G} systems: Principles, applications, and research
  directions,'' \emph{IEEE Commun. Mag.}, vol.~59, no.~6, pp. 14--20, 2021.

\bibitem{li2022reconfigurable_iot}
Q.~Li, M.~El-Hajjar, I.~Hemadeh, D.~Jagyasi, A.~Shojaeifard, E.~Basar, and
  L.~Hanzo, ``The reconfigurable intelligent surface-aided multi-node {IoT}
  downlink: Beamforming design and performance analysis,'' \emph{IEEE Internet
  Things J.}, vol.~10, no.~7, pp. 6400--6414, 2022.

\bibitem{pan2022overview}
C.~Pan, G.~Zhou, K.~Zhi, S.~Hong, T.~Wu, Y.~Pan, H.~Ren, M.~Di~Renzo, A.~L.
  Swindlehurst, R.~Zhang \emph{et~al.}, ``An overview of signal processing
  techniques for {RIS/IRS}-aided wireless systems,'' \emph{IEEE J. Sel. Top.
  Sign. Proces.}, vol.~16, no.~5, pp. 883--917, 2022.

\bibitem{li2023reconfigurable_tvt}
Q.~Li, M.~El-Hajjar, I.~Hemadeh, A.~Shojaeifard, A.~A. Mourad, and L.~Hanzo,
  ``Reconfigurable intelligent surface aided amplitude-and phase-modulated
  downlink transmission,'' \emph{IEEE Trans. Veh. Technol.}, vol.~72, no.~6,
  pp. 8146--8151, 2023.

\bibitem{wu2019towards}
Q.~Wu and R.~Zhang, ``Towards smart and reconfigurable environment: Intelligent
  reflecting surface aided wireless network,'' \emph{IEEE Commun. Mag.},
  vol.~58, no.~1, pp. 106--112, 2019.

\bibitem{gong2020toward}
S.~Gong, X.~Lu, D.~T. Hoang, D.~Niyato, L.~Shu, D.~I. Kim, and Y.-C. Liang,
  ``Toward smart wireless communications via intelligent reflecting surfaces: A
  contemporary survey,'' \emph{IEEE Commun. Surv. \& Tut.}, vol.~22, no.~4, pp.
  2283--2314, 2020.

\bibitem{wu2021intelligent}
Q.~Wu, S.~Zhang, B.~Zheng, C.~You, and R.~Zhang, ``Intelligent reflecting
  surface-aided wireless communications: A tutorial,'' \emph{IEEE Trans.
  Commun.}, vol.~69, no.~5, pp. 3313--3351, 2021.

\bibitem{bjornson2022reconfigurable}
E.~Bj{\"o}rnson, H.~Wymeersch, B.~Matthiesen, P.~Popovski, L.~Sanguinetti, and
  E.~de~Carvalho, ``Reconfigurable intelligent surfaces: A signal processing
  perspective with wireless applications,'' \emph{IEEE Sign. Proces. Mag.},
  vol.~39, no.~2, pp. 135--158, 2022.

\bibitem{zheng2022survey}
B.~Zheng, C.~You, W.~Mei, and R.~Zhang, ``A survey on channel estimation and
  practical passive beamforming design for intelligent reflecting surface aided
  wireless communications,'' \emph{IEEE Commun. Surv. \& Tut.}, vol.~24, no.~2,
  pp. 1035--1071, 2022.

\bibitem{li2022reconfigurable_tvt}
Q.~Li, M.~El-Hajjar, I.~A. Hemadeh, A.~Shojaeifard, A.~Mourad, B.~Clerckx, and
  L.~Hanzo, ``Reconfigurable intelligent surfaces relying on non-diagonal phase
  shift matrices,'' \emph{IEEE Trans. Veh. Technol.}, vol.~71, no.~6, pp.
  6367--6383, 2022.

\bibitem{li2023achievable_tcom}
Q.~Li, M.~El-Hajjar, Y.~Sun, I.~Hemadeh, A.~Shojaeifard, Y.~Liu, and L.~Hanzo,
  ``Achievable rate analysis of the {STAR-RIS} aided {NOMA} uplink in the face
  of imperfect {CSI} and hardware impairments,'' \emph{IEEE Trans. Commun.},
  vol.~71, no.~10, pp. 6100--6114, 2023.

\bibitem{jiang2023reconfigurable}
Y.~Jiang, F.~Gao, M.~Jian, S.~Zhang, and W.~Zhang, ``Reconfigurable intelligent
  surface for near field communications: Beamforming and sensing,'' \emph{IEEE
  Trans. Wireless Commun.}, vol.~22, no.~5, pp. 3447--3459, 2023.

\bibitem{zhang2023channel}
H.~Zhang, N.~Shlezinger, G.~C. Alexandropoulos, A.~Shultzman, I.~Alamzadeh,
  M.~F. Imani, and Y.~C. Eldar, ``Channel estimation with hybrid reconfigurable
  intelligent metasurfaces,'' \emph{IEEE Trans. Commun.}, vol.~71, no.~4, pp.
  2441--2456, 2023.

\bibitem{li2023low}
Q.~Li, M.~El-Hajjar, I.~Hemadeh, A.~Shojaeifard, and L.~Hanzo, ``Low-overhead
  channel estimation for {RIS}-aided multi-cell networks in the presence of
  phase quantization errors,'' \emph{IEEE Trans. Veh. Technol.}, 2023, {E}arly
  {A}ccess.

\bibitem{xie2023performance}
K.~Xie, G.~Cai, G.~Kaddoum, and J.~He, ``Performance analysis and resource
  allocation of {STAR-RIS} aided wireless-powered {NOMA} system,'' \emph{IEEE
  Trans. Commun.}, vol.~71, no.~10, pp. 5740--5755, 2023.

\bibitem{li2023performance_tvt}
Q.~Li, M.~El-Hajjar, I.~Hemadeh, D.~Jagyasi, A.~Shojaeifard, and L.~Hanzo,
  ``Performance analysis of active {RIS}-aided systems in the face of imperfect
  {CSI} and phase shift noise,'' \emph{IEEE Trans. Veh. Technol.}, vol.~72,
  no.~6, pp. 8140--8145, 2023.

\bibitem{yang2020outage}
L.~Yang, Y.~Yang, D.~B. da~Costa, and I.~Trigui, ``Outage probability and
  capacity scaling law of multiple {RIS}-aided networks,'' \emph{IEEE Wireless
  Commun. Lett.}, vol.~10, no.~2, pp. 256--260, 2020.

\bibitem{yang2023covert}
L.~Yang, W.~Zhang, P.~S. Bithas, H.~Liu, M.~O. Hasna, T.~A. Tsiftsis, and
  D.~W.~K. Ng, ``Covert transmission and secrecy analysis of
  {RS}-{RIS}-{NOMA}-aided {6G} wireless communication systems,'' \emph{IEEE
  Trans. Veh. Technol.}, vol.~72, no.~8, pp. 10\,659--10\,670, 2023.

\bibitem{chen2023active}
P.~Chen, L.~Yang, H.~Liu, G.~Pan, Y.~Li, and Z.~Yan, ``Active {RIS-NOMA}-aided
  covert communication with hardware impairments,'' \emph{IEEE Wireless Commun.
  Lett.}, vol.~12, no.~12, pp. 2278--2282, 2023.

\bibitem{wu2019intelligent}
Q.~Wu and R.~Zhang, ``Intelligent reflecting surface enhanced wireless network
  via joint active and passive beamforming,'' \emph{IEEE Trans. Wireless
  Commun.}, vol.~18, no.~11, pp. 5394--5409, 2019.

\bibitem{li2020weighted}
Z.~Li, M.~Hua, Q.~Wang, and Q.~Song, ``Weighted sum-rate maximization for
  multi-{IRS} aided cooperative transmission,'' \emph{IEEE Wireless Commun.
  Lett.}, vol.~9, no.~10, pp. 1620--1624, 2020.

\bibitem{han2019large}
Y.~Han, W.~Tang, S.~Jin, C.-K. Wen, and X.~Ma, ``Large intelligent
  surface-assisted wireless communication exploiting statistical {CSI},''
  \emph{IEEE Trans. Veh. Technol.}, vol.~68, no.~8, pp. 8238--8242, 2019.

\bibitem{xu2022simultaneously}
J.~Xu, Y.~Liu, X.~Mu, J.~T. Zhou, L.~Song, H.~V. Poor, and L.~Hanzo,
  ``Simultaneously transmitting and reflecting intelligent omni-surfaces:
  Modeling and implementation,'' \emph{IEEE Veh. Technol. Mag.}, vol.~17,
  no.~2, pp. 46--54, 2022.

\bibitem{zhang2022intelligent}
H.~Zhang, S.~Zeng, B.~Di, Y.~Tan, M.~Di~Renzo, M.~Debbah, Z.~Han, H.~V. Poor,
  and L.~Song, ``Intelligent omni-surfaces for full-dimensional wireless
  communications: Principles, technology, and implementation,'' \emph{IEEE
  Commun. Mag.}, vol.~60, no.~2, pp. 39--45, 2022.

\bibitem{mu2021simultaneously}
X.~Mu, Y.~Liu, L.~Guo, J.~Lin, and R.~Schober, ``Simultaneously transmitting
  and reflecting ({STAR}) {RIS} aided wireless communications,'' \emph{IEEE
  Trans. Wireless Commun.}, vol.~21, no.~5, pp. 3083--3098, 2021.

\bibitem{niu2021weighted}
H.~Niu, Z.~Chu, F.~Zhou, P.~Xiao, and N.~Al-Dhahir, ``Weighted sum rate
  optimization for {STAR}-{RIS}-assisted {MIMO} system,'' \emph{IEEE Trans.
  Veh. Technol.}, vol.~71, no.~2, pp. 2122--2127, 2021.

\bibitem{zhang2020beyond}
S.~Zhang, H.~Zhang, B.~Di, Y.~Tan, Z.~Han, and L.~Song, ``Beyond intelligent
  reflecting surfaces: Reflective-transmissive metasurface aided communications
  for full-dimensional coverage extension,'' \emph{IEEE Trans. Veh. Technol.},
  vol.~69, no.~11, pp. 13\,905--13\,909, 2020.

\bibitem{zhang2021intelligent}
S.~Zhang, H.~Zhang, B.~Di, Y.~Tan, M.~Di~Renzo, Z.~Han, H.~V. Poor, and
  L.~Song, ``Intelligent omni-surfaces: Ubiquitous wireless transmission by
  reflective-refractive metasurfaces,'' \emph{IEEE Trans. Wireless Commun.},
  vol.~21, no.~1, pp. 219--233, 2021.

\bibitem{dhok2022rate}
S.~Dhok and P.~K. Sharma, ``Rate-splitting multiple access with {STAR} {RIS}
  over spatially-correlated channels,'' \emph{IEEE Trans. Commun.}, vol.~70,
  no.~10, pp. 6410--6424, 2022.

\bibitem{wu2021coverage}
C.~Wu, Y.~Liu, X.~Mu, X.~Gu, and O.~A. Dobre, ``Coverage characterization of
  {STAR}-{RIS} networks: {NOMA} and {OMA},'' \emph{IEEE Commun. Lett.},
  vol.~25, no.~9, pp. 3036--3040, 2021.

\bibitem{wu2022resource}
C.~Wu, X.~Mu, Y.~Liu, X.~Gu, and X.~Wang, ``Resource allocation in
  {STAR-RIS}-aided networks: {OMA} and {NOMA},'' \emph{IEEE Trans. Wireless
  Commun.}, vol.~21, no.~9, pp. 7653--7667, 2022.

\bibitem{zhao2022ergodic}
B.~Zhao, C.~Zhang, W.~Yi, and Y.~Liu, ``Ergodic rate analysis of {STAR-RIS}
  aided {NOMA} systems,'' \emph{IEEE Commun. Lett.}, vol.~26, no.~10, pp.
  2297--2301, 2022.

\bibitem{papazafeiropoulos2022coverage}
A.~Papazafeiropoulos, Z.~Abdullah, P.~Kourtessis, S.~Kisseleff, and
  I.~Krikidis, ``Coverage probability of {STAR}-{RIS} assisted massive {MIMO}
  systems with correlation and phase errors,'' \emph{IEEE Wireless Commun.
  Lett.}, vol.~11, no.~8, pp. 1738--1742, 2022.

\bibitem{wu2023two}
C.~Wu, C.~You, Y.~Liu, S.~Han, and M.~Di~Renzo, ``Two-timescale design for
  {STAR-RIS} aided {NOMA} systems,'' \emph{IEEE Trans. Commun.}, vol.~72,
  no.~1, pp. 585--600, 2024.

\bibitem{aldababsa2023simultaneous}
M.~Aldababsa, A.~Khaleel, and E.~Basar, ``Simultaneous transmitting and
  reflecting reconfigurable intelligent surfaces-empowered {NOMA} networks,''
  \emph{IEEE Syst. J.}, vol.~17, no.~4, pp. 5441--5451, 2023.

\bibitem{liu2021star}
Y.~Liu, X.~Mu, J.~Xu, R.~Schober, Y.~Hao, H.~V. Poor, and L.~Hanzo, ``{STAR}:
  Simultaneous transmission and reflection for 360$^{\circ}$ coverage by
  intelligent surfaces,'' \emph{IEEE Wireless Commun.}, vol.~28, no.~6, pp.
  102--109, 2021.

\bibitem{yang2022performance}
L.~Yang, P.~Li, F.~Meng, and S.~Yu, ``Performance analysis of {RIS}-assisted
  {UAV} communication systems,'' \emph{IEEE Trans. Veh. Technol.}, vol.~71,
  no.~8, pp. 9078--9082, 2022.

\bibitem{an2022joint}
J.~An, C.~Xu, L.~Wang, Y.~Liu, L.~Gan, and L.~Hanzo, ``Joint training of the
  superimposed direct and reflected links in reconfigurable intelligent surface
  assisted multiuser communications,'' \emph{IEEE Trans. Green Commun. Netw.},
  vol.~6, no.~2, pp. 739--754, 2022.

\bibitem{zhang2021reconfigurable}
Q.~Zhang, Y.-C. Liang, and H.~V. Poor, ``Reconfigurable intelligent surface
  assisted {MIMO} symbiotic radio networks,'' \emph{IEEE Trans. Commun.},
  vol.~69, no.~7, pp. 4832--4846, 2021.

\bibitem{bjornson2017massive}
E.~Bj{\"o}rnson, J.~Hoydis, L.~Sanguinetti \emph{et~al.}, ``Massive {MIMO}
  networks: Spectral, energy, and hardware efficiency,'' \emph{Foundations and
  Trends{\textregistered} in Signal Processing}, vol.~11, no. 3-4, pp.
  154--655, 2017.

\bibitem{bjornson2014massive}
E.~Bj{\"o}rnson, J.~Hoydis, M.~Kountouris, and M.~Debbah, ``Massive {MIMO}
  systems with non-ideal hardware: Energy efficiency, estimation, and capacity
  limits,'' \emph{IEEE Trans. Inf. Theory}, vol.~60, no.~11, pp. 7112--7139,
  2014.

\bibitem{kang2023active}
X.~Kang, H.~Lei, L.~Yang, G.~Pan, T.~A. Tsiftsis, and H.~Liu,
  ``Active-{RIS}-aided covert communications in {NOMA} systems with cooperative
  jamming,'' \emph{IEEE Trans. Veh. Technol.}, 2023.

\bibitem{zhang2017matrix}
X.~Zhang, \emph{Matrix analysis and applications}.\hskip 1em plus 0.5em minus
  0.4em\relax Cambridge University Press, 2017.

\end{thebibliography}
\end{document}